\documentclass[traditabstract]{aa} 
\usepackage{epsfig,amsmath}
\usepackage[dvips]{color}
\usepackage{verbatim}
\usepackage{graphicx}
\usepackage{txfonts}
\usepackage{natbib}
\bibpunct{(}{)}{;}{a}{}{,}
\usepackage{hyperref}
\usepackage{rotating}
\usepackage{longtable}

\definecolor{Black}{named}{Black}
\definecolor{Red}{named}{Red}
\definecolor{Green}{named}{Green}
\definecolor{Blue}{named}{Blue}

\begin{document}

 \title{The extinction curves of star-forming regions from $z=0.1$ to $6.7$ using GRB afterglow spectroscopy\thanks{Based on observations collected at the European Southern Observatory (ESO) utilizing the 8.2m Very Large Telescope (VLT), Chile, under programs 075.D-0270 (PI: Fynbo), 077.D-0661 (PI: Vreeswijk), 077.D-0805 (PI: Tagliaferri), 078.D-0416 (PI: Vreeswijk), 079.D-0429 (PI: Vreeswijk), 080.D-0526 (PI: Vreeswijk), 081.A-0135 (PI: Greiner), and 281.D-5002 (PI: Della Valle).}}
\authorrunning{Zafar et al.}
\author{Tayyaba Zafar\inst{1}
 \and Darach Watson\inst{1}
    \and Johan P. U. Fynbo\inst{1}
    \and Daniele Malesani\inst{1}
        \and P\'all Jakobsson\inst{2}
    \and Antonio de Ugarte Postigo\inst{1}} 

\institute{Dark Cosmology Centre, Niels Bohr Institute, University of Copenhagen,
Juliane Maries Vej 30, DK-2100 Copenhagen \O, Denmark.
\and Centre for Astrophysics and Cosmology, Science Institute, University of Iceland, Dunhagi 5, IS-107 Reykjav\'ik, Iceland.}

\offprints{tayyaba@dark-cosmology.dk}

\date{Received  / Accepted }

\abstract
{Studies of extinction curves provide insights into the properties of
interstellar dust. Until recently, however, very few extinction curves
existed outside the Local Group. GRB afterglows are well suited to
extinction studies due to their brightness, simple power-law spectra
and their occurrence in distant star forming galaxies. In this
paper we present results from the SED analysis of a sample of 41 GRB
afterglows, from X-ray to NIR wavelengths. The sample is based on
spectra from VLT-FORS, with additional data primarily from
\emph{Swift}. This is the largest sample of extinction curves outside
the Local Group and, to date, the only extragalactic sample of
absolute extinction curves based on spectroscopy. Estimates of the
distribution of restframe visual extinctions, the extinction curves,
and the intrinsic spectral shapes of GRB afterglows are obtained.
Their correlation with $\ion{H}{i}$ column density as well as total
and gas-phase metal column density are examined. The line-of-sight
gas-to-dust and metals-to-dust ratios are determined and examined as a
function of total column density, ISM metallicity and redshift. The
intrinsic SEDs of the afterglows show that approximately half the
sample require a cooling break between the optical and X-ray ranges.
The broken power-law SEDs show an average change in the spectral index of
$\Delta\beta=0.51$ with a very small standard deviation of 0.02
(excluding the outlier GRB\,080210). This is consistent with the
expectations from a simple synchrotron model. Such a close convergence
of values suggests that the X-ray afterglows of GRBs may be used with
considerably more confidence to set the absolute flux level and intrinsic
spectral indices in the optical and UV. Of the sample, 63\% are well
described by a featureless (SMC-type) extinction curve. Almost a
quarter of our sample is consistent with no significant extinction
(typically $A_V\lesssim0.1$). The 2175\,\AA\ extinction bump is
detected unequivocally in 7\% of our sample (3 GRBs), which all have
$A_V>1.0$, while one afterglow has a very unusual extinction curve
with a sharp UV rise. However, we can only say that the bump is
\emph{not} present in about a quarter of our sample because of low
extinction or lack of coverage of the 2200\,\AA\ region. All the afterglows
well fit with SMC type curves have moderate or low extinction, with
$A_V<0.65$. This suggests that the SMC extinction curve is not as
nearly-universal as previously believed and that extinction curves
more similar to those found in the Galaxy and the LMC may be quite
prevalent. We find an anti-correlation between gas-to-dust ratio and
metallicity consistent with the Local Group relation; we find,
however, no correlation between the metals-to-dust ratios and the
metallicities, redshift and visual extinction; we find no strong
correlation of the extinction column with metallicity either. Our
metals-to-dust ratios derived from the soft X-ray absorption are
always larger (3--30 times) than the Local Group value, which may mean
that GRB hosts may be less efficient at turning their metals into
dust. However, we find that gas, dust, and metal column densities are
all likely to be influenced by photo-ionization and dust destruction
effects from the GRB to differing extents and caution must be used in
extrapolating the ratios of dust and gas-derived properties from GRB
afterglows to the star-forming population in general.} 
\maketitle{}

\section{Introduction}

Extinction of light is the scattering and absorption of photons along
the travel path from an object to the observer. Dust grains are the
major contributor to extinction in the optical and ultraviolet (UV)
range. More than 30\% of the UV and optical light from stars in the
Universe may be absorbed and re-radiated thermally at infrared and
millimeter wavelengths by the obscuring dust
\citep{lagache,franceschini}. Extinction curves are the standard tool
to study dust in the optical and UV energy bands, revealing
information about the dust grain size and composition by considering the
amount of light lost due to scattering and absorption. The extinction
curves of the Milky Way (MW) and the Large and Small Magellanic
Clouds (LMC and SMC) have been modeled in a variety of ways
\citep{fm,ccm,pei}. These extinction laws vary significantly, largely
due to the presence and relative strength of the 2175\,\AA{} extinction
feature which is believed to be mainly due to absorption by graphite
grains \citep{stecher,draine03}. Extinction curves are typically
obtained by comparing pairs of stars, however, outside the Local Group of
galaxies, the method of estimating extinction curves from stellar
atmosphere models is not applicable because individual stars are
difficult or impossible to observe.

Long duration gamma-ray bursts (GRBs) are signposts of star formation due to their
association with the deaths of short-lived, massive stars
\citep[e.g.,][]{woosley, galama98, hjorth, stanek, della03,
malesani,campana06,starling10}. GRBs are excellent probes to study extinction
curves in distant galaxies due to their intrinsic brightness, simple
spectra and their occurrence in dense, star forming environments. GRB
afterglows typically have intrinsically featureless piecewise
power-law spectra from the X-ray to the optical/NIR and radio
wavelengths. Theoretically, the most prominent explanation of GRB
afterglows is the fireball model
\citep[e.g.][]{rees92,meszaros,galama,sari,piran99,granot,zhang06} which provides 
some notable successes in explaining the observations. According to
the fireball model the GRB afterglow originates from synchrotron
radiation produced by the interaction between the ultra-relativistic
jet and the surrounding interstellar medium (ISM). Cooling of electrons produces a 
break in the synchrotron spectrum, which can sometimes be located between 
the X-ray and optical/NIR. The cooling frequency will lead to a well-defined change in
spectral slope of $\Delta\beta=\beta_2-\beta_1=0.5$ resulting in a
softer spectrum on the high energy side. Therefore, the X-ray--optical
SED is expected to be a single or broken power-law. The simple spectral
shape of GRBs and the lack of absorption above $\sim3$\,keV allow us to
place constraints not only on the reddening due to dust, but on the
absolute extinction and thereby obtain absolute extinction curves for
sources over a wide range of redshifts ($z=0.0086-8.2$).

In the era of the \emph{Swift} satellite \citep{gehrels} and fast
responding ground-based telescopes, better-sampled GRB afterglow
lightcurves and SEDs enable us to get more complete extinction curves
and even samples of extinction curves. To date, SMC-type extinction
(i.e.\ without a 2175\,\AA{} bump), has provided the best representation
of GRB afterglow extinction
\citep{bloom98,jensen01,fynbo01smc,price01,savaglio04,jakobsson04,kann06,starling07,kann10,schady}.
The 2175\,\AA\ feature has been firmly detected in only two GRB afterglows
so far: GRB\,070802 \citep{ardis,kruhler08} and GRB\,080607 \citep{prochaska09,perley09}. Grey/wavelength-independent extinction
laws have also been claimed for a few GRB afterglows
\citep{galama03,nardini08,perley08}. Dust with an unusual extinction
curve, claimed to be due to supernova-origin dust, has been suggested
in two cases, GRB\,071025 \citep{perley10} and GRB\,050904
\citep{stratta}, but ruled out for the latter case with revised
multi-epoch SED analysis \citep{zafar}.

In this paper, we present the SED analysis of a sample of forty one
GRB afterglows observed mainly with the Very Large Telescope (VLT).
This is the largest study to date of extinction curves outside the
Local Group of galaxies. It is also unique as the first spectroscopic
sample of absolute extinction curves
\citep[cf.][]{kann06,starling07,schady072,schady,kann10,kruhler10,gallerani2}. Our
main objectives are: $i)$ to examine the extinction curves of
star-forming galaxies at all redshifts, $ii)$ to understand the
properties of the dust and their relationship to the gas and metals in
these galaxies, $iii)$ to infer information about the typical spectral
shapes of GRB afterglows.

To the best of our knowledge, this paper presents the first set of absolute extinction curves based on spectroscopy. Using spectroscopy in addition to photometry has significant benefits in such an analysis: \emph{i)} the actual shape of the continuum and of the extinction curve can be determined, including the width, strength and central wavelength of the 2175\,\AA{} bump where it is present \emph{ii)} the effects of gas absorption on the spectrum can be removed, including Ly$\alpha$ at $z\gtrsim2$ and the Ly$\alpha$ forest at higher redshifts, \emph{iii)} precise redshifts can be obtained from the spectrum, \emph{iv)} the potential influence of intervening absorbers can be determined.

 In \S2 we
describe the X-ray, UV/optical and NIR data reduction. In \S3 we
describe the different models used to fit the data. The results of our
modeling are presented in \S4. A discussion on the implications of our
results is given in \S5, and our conclusions are summarized in \S6. A
cosmology where $H_0=72$\,km\,s$^{-1}$\,Mpc$^{-1}$, $\Omega_\Lambda =
0.73$ and $\Omega_{\rm m}=0.27$ \citep{spergel} is assumed. Throughout spectral
indices are denoted by $\beta$ where $\beta_1$ is the low energy
(typically the optical/NIR) slope and $\beta_2$ is the high energy
(typically X-ray) slope. The temporal indices are denoted by $\alpha$: $F_\nu(\nu,t)\propto t^{-\alpha}\nu^{-\beta}$. All errors are $1\sigma$ unless otherwise
specified. For the cases where there is no detection, $2\sigma$ upper
limits are reported.
\section{Multi-wavelength observations and data reduction}
\subsection{Ensemble selection criteria}
The sample consists of the flux-calibrated spectra from
\citet[][hereafter F09]{fynbo} taken with the Focal Reducer and
low dispersion Spectrographs (FORS1 and FORS2) on the VLT with photometric observations in at least $R$ or $I$-bands. Forty one GRB afterglows fulfilled our selection criteria (see Table \ref{ebv}). The spectra were obtained through target-of-opportunity (ToO) programs.

\subsection{Optical spectroscopy}
With the aim of constructing composite SEDs, we collected flux calibrated spectroscopic data for each GRB  afterglow listed in Table \ref{ebv} (see also F09). The sample includes GRB afterglows from April 2005 to September 2008. The longslit spectra were obtained with either the FORS1 or FORS2 spectrographs \citep{appenzeller} using the 300V grism for all bursts except for GRB\,080319B (600B) and GRB\,080913 (600z). This yields a spectral resolution of $R=\lambda/\Delta\lambda= 440$, 780 and 1390 in the 300V, 600B and 600z grisms, respectively, with a slit of 1\arcsec. Further details of the optical spectroscopy are available in F09. The data were reduced using standard tasks in IRAF and corrected for cosmic rays using the method developed by \citet{van}. All the spectra were flux calibrated using spectrophotometric standard stars observed on the same night of the afterglow spectra. All FORS1/2 spectra from F09 are used except GRB\,060526, GRB\,060719, GRB\,060807, GRB\,070411, GRB\,070508, GRB\,070306, GRB\,080411, GRB\,080413B and GRB\,080523. There were no suitable standard star spectra obtained near the time of observations of GRB\,060526, GRB\,080411 and GRB\,080413B. The spectra of GRB\,060719, GRB\,070411, GRB\,070508 and GRB\,080523 have extremely low S/N ratios. GRB\,060807 has no redshift estimation. The spectrum of GRB\,070306 was dominated by the host galaxy. The spectrum of GRB\,080607 was obtained with Keck using the Low Resolution Imaging Spectrometer \citep[LRIS;][]{oke} and is therefore formally outside our sample. We include it here as a comparison object, since it is one of only two known GRB afterglows with a confirmed 2175\,\AA\ bump in the host galaxy prior to this sample. We used photometry from acquisition camera imaging, mostly performed using the $R$ filter, to set the absolute flux level of the spectra. The time of the acquisition camera image was then used as the mean time of the SED. The magnitudes of the afterglows are obtained using ESO zero-points from the night of the observation.

To avoid contamination from absorption caused by the Ly$\alpha$ forest, data (both spectroscopic and photometric) with restframe wavelength $\lambda<1216$ \AA{} were excluded from the SED fitting. Furthermore, emission and absorption lines arising from different metal species and atmospheric molecular absorption lines were removed from the optical spectra. 

\subsection{Photometry}\label{sec:photometry}
The 30\,cm Ultraviolet Optical Telescope (UVOT) onboard \emph{Swift}, has six UV/optical filters: $v$, $b$, $u$, $uvw1$, $uvm2$ and $uvw2$ \citep{roming05}. The data are made available at the UK \emph{Swift} Science Data Centre. The UVOT filter lightcurves were extracted using the UVOT imaging data tool \texttt{UVOTMAGHIST}. The fluxes in each band were obtained by using the standard UVOT photometric calibration techniques defined by \citet{poole}. For the ground-based optical and NIR data, lightcurves for each filter were constructed using data either from refereed publications or from GRB Coordinates Network (GCN) circulars. For ease of access to the data published in GCN circulars for each GRB, the information was obtained via the GRBlog\footnote {\url{http://grblog.org/grblog.php}} database \citep{quimby04}. To generate SEDs at instantaneous epochs, the magnitudes in the relevant filters were obtained by interpolating or extrapolating the lightcurves to the time of interest. For this purpose, usually the $R$-band lightcurve decay index is used for extrapolation and interpolation in other bands since it is typically the most complete.

The optical spectra and optical/NIR photometry have been corrected for Galactic dust
extinction using the \citet{schlegel} dust extinction maps. The $E(B-V)$ values along the line of sight for each burst are given in Table ~\ref{ebv} and were retrieved from the NASA Extragalactic database
(NED\footnote{\url{http://nedwww.ipac.caltech.edu/}}) web-calculator. Several works have attempted to verify the \citet{schlegel} results (e.g. \citealt{dutra,schlafly}), finding overall a good agreement at low values of $A_V$. Some uncertainty is still present, e.g.\ recently \citet{schlafly} suggested that these values might be systematically overestimated by $\sim 14$\% on the average. For our analysis we corrected the Galactic $E(B-V)$ values of \citet{schlegel} for the 14\% overestimation (see Table \ref{ebv}). However, extinction at high redshift has a much greater effect for a given dust column on the observed spectrum since we are observing dust effects in the restframe UV of the GRB host. For example, for a GRB host at $z=2$ with SMC extinction, the effective observer frame $A_R$ is
$\sim4$ times larger than the equivalent Galactic $A_R$.  Discrepancies between the \citet{schlafly} and
\citet{schlegel} values in the Galactic extinction are therefore small compared to the typical uncertainty on the
measured restframe absorption and extinction values.

\begin{table}
\begin{minipage}[t]{\columnwidth}
\caption{Basic data on the flux calibrated spectroscopic sample. The details of the sample are provided in the columns as (1) burst name, (2) Galactic \ion{H}{i} column density, (3) time of SED, time since GRB trigger, (4) Galactic $E(B-V)$ from the \citet{schlegel} extinction maps corrected for 14\% overestimation, (5) redshift, and (6) magnitude of the afterglow from the acquisition image.}      
\label{ebv} 
\centering
\renewcommand{\footnoterule}{}  
\setlength{\tabcolsep}{2pt}
\begin{tabular*}{\columnwidth}{@{\extracolsep{\fill}}l c c c c c c}\hline\hline                       
GRB & $N_{\ion{H},{\rm{Gal}}}$ & $\Delta t$ & $E(B-V)_{\rm{Gal}}$ & $z$ & $\rm{Mag_{acq}}$\\
	& $10^{20}$ cm$^{-2}$ 	&	hr	&		mag	&  & mag\\ 
\hline
050401 & 4.40 & 14.7 & 0.056 & 2.8983 & $23.27\pm0.09$ \\
050730 & 2.99 & 4.10 & 0.044 & 3.9693  & $17.80\pm0.04$ \\
050824 & 3.72 & 9.50 & 0.030 & 0.8278 & $20.60\pm0.06$ \\
060115 & 9.48 & 8.90 &  0.114 &  3.5328  & $22.00\pm0.07$ \\
060512 & 1.53 & 8.80 & 0.015 & 2.1000 & $20.40\pm0.20$\\
060614 & 1.87 & 21.1 & 0.019 & 0.1257 & $19.80\pm0.05$ \\
060707 & 1.44 & 34.4 & 0.019 & 3.4240  & $22.40\pm0.06$ \\
060708 & 2.10 & 43.0 & 0.009 & 1.9200  &  $22.90\pm0.07$\\
060714 & 6.05 & 10.1 & 0.066 & 2.7108  & $20.90\pm0.06$\\
060729 & 4.49 & 13.2 & 0.046 & 0.5428  & $17.50\pm0.05$ \\
060904B & 11.3 & 5.10 & 0.149 & 0.7029 & $19.90\pm0.05$ \\
060906 & 9.81 & 1.00 & 0.048 & 3.6856  & $20.00\pm0.06$ \\
060926 & 7.58 & 7.70 & 0.135 & 3.2086  & $23.00\pm0.06$ \\
060927 & 4.60 & 12.5 & 0.053 & 5.4636  &  $24.00\pm0.09$ \\
061007 & 1.77 & 17.4 &  0.018 & 1.2622  & $21.50\pm0.06$ \\
061021 & 4.52 & 16.5 & 0.049 &  0.3463  & $20.50\pm0.04$ \\
061110A & 4.26 & 15.0 & 0.077 &  0.7578  &  $23.10\pm0.05$ \\
061110B & 3.35 & 2.50 & 0.055 &  3.4344  & $22.50\pm0.05$ \\
070110 & 1.61 & 17.6 & 0.012 & 2.3521  & $20.80\pm0.04$ \\
070125 & 4.31 & 31.5 & 0.045 & 1.5471  &$18.80\pm0.20$\\
070129 & 6.96 & 2.20 & 0.118 & 2.3380  & $21.30\pm0.05$ \\ 
070318 & 1.44 & 16.7 & 0.015 & 0.8397  & $20.20\pm0.10$ \\
070506 & 3.80 & 4.00 & 0.034 & 2.3090 & $21.00\pm0.04$ \\
070611 & 1.38 & 7.70 & 0.010 & 2.0394  & $21.00\pm0.04$ \\
070721B & 2.33 & 21.6 & 0.027 &  3.6298  & $24.30\pm0.07$ \\
070802 & 2.90 & 2.00 & 0.023 &  2.4541  & $22.60\pm0.06$ \\
071020 & 5.12 & 5.12 & 0.052 &  2.1462  & $20.33\pm0.04$ \\
071031 & 1.22 &  1.20 & 0.010 & 2.6918 & $18.90\pm0.04$ \\
071112C & 7.44 &  9.50 & 0.101 & 0.8227 & $22.20\pm0.11$ \\
071117 & 2.33 & 9.00 & 0.021 &  1.3308  & $23.58\pm0.09$ \\
080210 & 5.47 &  1.69 & 0.071 & 2.6419  & $19.57\pm0.05$ \\
080319B & 1.12 & 26.0 & 0.009 & 0.9382  & $20.36\pm0.04$ \\
080520 & 6.82 & 7.30 &  0.071 &  1.5457  & $23.00\pm0.06$ \\
080605 & 6.67 & 1.74 & 0.117 & 1.6403  & $19.81\pm0.05$ \\
080707 & 6.99 & 1.10 & 0.086 & 1.2322 & $19.60\pm0.04$ \\
080721 & 6.94 & 10.2 & 0.086  & 2.5914  & $20.20\pm0.12$\\
080805 & 3.46 & 1.00 & 0.037 &  1.5042  & $21.50\pm0.05$ \\
080905B & 3.50 & 8.30 & 0.040 & 2.3739  & $20.20\pm0.05$ \\
080913 & 3.17 & 1.88 & 0.037 & 6.6950 & $23.36\pm0.13$\\
080916A & 1.84 & 17.1 & 0.016 & 0.6887 & $22.30\pm0.05$\\
080928 & 5.59 & 15.5 & 0.058 & 1.6919  & $21.07\pm0.07$ \\[5pt]
080607 \footnote{Keck spectrum.} &1.69 & 0.08 & 0.060 & 3.0368  & $17.97\pm0.07$ \\
\hline
\end{tabular*}
\end{minipage}
\end{table}

 \subsection{X-ray data}
 \label{xraysection}
The GRB afterglows present in our sample have been all detected by \emph{Swift}. The X-Ray Telescope \citep[XRT;][]{burrows} onboard \emph{Swift} performed the X-ray observations in each case and the X-ray data for each GRB have been obtained from the UK  \emph{Swift} science data center\footnote{\url{http://www.swift.ac.uk/swift_portal/archive.php}}. The X-ray spectrum for each GRB afterglow was reduced in the 0.3--10.0 keV energy range using the HEAsoft software (version 6.10). For each afterglow, the X-ray lightcurves were produced from the XRT data \citep{evans,evans2}. To get a better estimate of the X-ray flux, we fitted a decay model \citep{beuermann} to the X-ray lightcurves. For X-ray spectral analysis we used photon counting (PC) mode data and source counts were extracted within a circular region centered on the source, and background counts from a circular region $\sim10$ times greater than the source region. We used \texttt{XSELECT} (v2.4) to extract spectral files from the event data in the 0.3--10 keV energy band. To avoid pile-up all X-ray spectra were taken from time intervals where the measured count rate is less than 0.6 counts s$^{-1}$. The X-ray spectral files were grouped to 12--20 counts per energy channel. Effective area files for each spectrum were created using the \texttt{FTOOLS} \texttt{XRTMKARF} recipe and bad columns were corrected by using the exposure maps. Response matrices  (v11) from the \emph {Swift} XRT calibration files were used. The X-ray spectra were obtained over an interval as close as possible to the mid-time of the optical spectra. The flux of the final X-ray spectrum was corrected to the SED time by taking the ratio of the X-ray lightcurve model fits at the photon weighted mean time (PWMT) of the X-ray spectrum and the SED time. The spectra were fitted within \texttt{XSPEC} \citep[v12.6.0;][]{arnaud} with a Galactic-absorbed power-law, with absorption from the Galactic neutral hydrogen column density taken from \citet[][using the nH FTOOL]{kalberla}, and additional absorption from the GRB host galaxy. The Galactic column density value used for each GRB is given in Table~\ref{nhx}. The equivalent neutral hydrogen column density from the host galaxy of each GRB was estimated from the soft X-ray absorption and is denoted here as $N_{\rm{H,X}}$. $N_{\rm{H,X}}$ is modeled with \texttt{XSPEC} where solar abundances were assumed from \citet{asplund}. $N_{\rm{H,X}}$, presented in units of equivalent hydrogen column density for a solar abundance of the elements, is essentially a direct measure of the metal column density.

\begin{table}
\begin{minipage}[t]{\columnwidth}
\caption{Results of fits to the SEDs. The columns give the burst name, and then for each model the $\chi_\nu^2$, the number of degrees of freedom (dof) and the null hypothesis probability (NHP) for the fit are provided. The F-test probability is computed for each SED comparing the single (PL) and broken power-law (BPL) models. The best fit models are denoted $\dagger$.}      
\label{fitresult} 
\centering
\setlength{\tabcolsep}{3.5pt}
\renewcommand{\footnoterule}{}  
\begin{tabular}{l l l l l l}   
\hline\hline       
	& \multicolumn{4}{c}{SMC}\\           
	& \multicolumn{2}{c}{PL} & \multicolumn{2}{c}{BPL}  \\ 
GRB	& $\chi_\nu^2$ (dof) & NHP\% & $\chi_\nu^2$ (dof) & NHP\% & F-test prob.  \\
\hline
050401 & 1.09 (1011) & 2.0 & 1.08 (1009)$^\dagger$ & 2.6 & 0.02  \\
050730 & 1.23 (659) & 0.003 & 0.98 (657)$^\dagger$ & 60 & $<0.01$  \\
050824 & 1.13 (958) & 0.2 & 0.97 (956)$^\dagger$ & 71 & $<0.01$ \\
060115 & 1.15 (1014)$^\dagger$ & 0.06 &  1.15 (1012)  & 0.05 & 1.00 \\
060512 & 1.02 (1452) & 33 & 0.93 (1450)$^\dagger$ & 96 & $<0.01$  \\
060614 & 1.01 (1254) & 37 & 0.82 (1252)$^\dagger$ & 100 & $<0.01$ \\
060707 & 0.98 (1088)$^\dagger$ & 65 & 0.98 (1086) & 64 & 1.00 \\
060708 & 0.86 (867) & 100 & 0.85 (865)$^\dagger$ & 100 & $<0.01$   \\
060714 & 1.02 (1043) & 35 & 0.97 (1041)$^\dagger$ & 72 & $<0.01$  \\
060729 & 0.99 (1140)$^\dagger$ & 54 &  0.99 (1138) & 53 & 1.00 \\
060904B & 0.80 (1198)$^\dagger$  & 100 & 0.80 (1196) & 100 & 1.00  \\
060906 & 1.01 (786)  & 42 & 0.94 (784)$^\dagger$ & 87 & $<0.01$ \\
060926 & 0.95 (907)$^\dagger$ & 86 & 0.95 (905) & 85 & 1.00  \\
060927 & 1.03 (274)$^\dagger$ & 35 & 1.03 (272) & 32  & 1.00 \\
061007 & 0.97 (1341)$^\dagger$ &  78 & 0.97 (1339) & 77 & 1.00   \\
061021 & 1.26 (1703) & 0.0 & 0.98 (1701)$^\dagger$ & 70 & $<0.01$  \\
061110A & 0.98 (1093)$^\dagger$ & 63 & 0.98 (1091) & 61 & 1.00   \\
061110B & 0.95 (896) & 85 & 0.94 (894)$^\dagger$ & 88 & 0.02   \\
070110 & 1.09 (1087) & 0.17 & 1.04 (1085)$^\dagger$ & 18 & $<0.01$  \\
070125 & 1.03 (790) & 25 & 0.97 (788)$^\dagger$ & 72 & $<0.01$ \\
070129 & 0.93 (1017) & 94 & 0.91 (1015)$^\dagger$ & 98 & $<0.01$ \\ 
070506 & 0.99 (775) & 57 & 0.98 (773)$^\dagger$ & 65 & $<0.01$ \\ 
070611 & 0.77 (1248)$^\dagger$ & 100 & 0.77 (1246) & 100 & 0.77  \\
070721B & 1.00 (830)$^\dagger$ & 47 & 1.00 (828) & 46 & 0.76   \\
071020 & 1.09 (1020) & 2.1 &  1.06 (1018)$^\dagger$ & 10 & $<0.01$  \\
071031 & 2.13 (614) & 0 & 1.05 (612)$^\dagger$ & 17  & $<0.01$ \\
071112C & 1.67 (1287) & 0 & 1.00 (1285)$^\dagger$ & 48 & $<0.01$  \\
071117 & 0.99 (1171)  & 56 & 0.98 (1169)$^\dagger$ & 71 & $<0.01$   \\
080210 & 1.10 (964) & 1.7 & 1.00 (962)$^\dagger$  & 45  & $<0.01$  \\
080319B & 0.83 (1468)$^\dagger$ & 100 & 0.83 (1466) & 100 & 1.00 \\
080520 & 1.00 (1028)$^\dagger$ & 54 & 1.00 (1026) & 51 & 1.00  \\
080707 & 1.07 (743) & 9.3 & 1.03 (741)$^\dagger$ & 30 & $<0.01$  \\
080721 & 0.96 (731)$^\dagger$ & 76 & 0.96 (729) & 74 & 1.00 \\
080905B & 0.98 (1029)$^\dagger$ & 68 & 0.98 (1027) & 66 & 1.00  \\
080913 & 0.58 (81)$^\dagger$ & 100 & 0.58 (79) & 100 & 1.00 \\
080916A & 0.93 (1170) & 96 & 0.91 (1168)$^\dagger$ & 99 & $<0.01$   \\
080928 & 0.99 (1157)$^\dagger$ & 57 & 0.99 (1155) & 55 & $1.00$ \\
\hline\hline 
	 & \multicolumn{4}{c}{FM}\\           
	& \multicolumn{2}{c}{PL} & \multicolumn{2}{c}{BPL}  \\ 
\hline
070318 &  1.15 (1392)& $0.001$ &  0.88 (1390)$^\dagger$  & 100 & $<0.01$\\
070802 &  0.54 (1381)$^\dagger$ & 100 &  0.54 (1379) & 100 & 1.00  \\
080605 & 0.86 (859)$^\dagger$ & 100 & 0.86 (857) & 100 & 1.00 \\
080805 & 0.93 (1391)$^\dagger$ & 98 & 0.93 (1389) & 97 & 1.00   \\[5pt]
080607 \footnote{Keck spectrum.}& 1.79 (39)$^\dagger$ & 0.19  & 1.79 (37) & 0.17 & 1.00 \\
\hline
\end{tabular}
\end{minipage}
\end{table}

\section{Data analysis}
We describe the afterglow continuum emission with single or broken
power-law models. For each burst, we tried both laws to fit the
observed data, where the spectral break must be in the observing
window ($10^{14}$\,Hz$\,\lesssim\nu_{\rm{break}}\lesssim10^{18}$\,Hz). On top
of the power-law fits, we added empirical extinction laws
corresponding to the SMC \citep{pei}, MW \citep{fm,ccm}, or LMC
\citep{pei} (see \S\ref{dustmodels} for details), in order to account for the 
effect to dust in the GRB local environment. The X-ray to optical/NIR GRB
afterglow SEDs were generated at specific epochs (the time of the
acquisition image of the spectrum), and modeled in \texttt{IDL}. To
get errors on each parameter we used 1000 Monte Carlo (MC)
realizations of the data, with the mean value of each datapoint set to
the observed value and a gaussian distribution with a width
corresponding to the $1\sigma$ measured error on that datapoint. For
each simulated set we fitted the function and estimated the best fit
parameters. The error for each parameter was calculated using the
standard deviation of this distribution. To determine whether the use
of a broken power-law was required over a single power-law,  we used
an F-test probability $<5$\% as the threshold. The results from the spectral analysis of 
our spectroscopic GRB sample are provided in Table
\ref{fitresult}. The number of degrees of freedom in Table \ref{fitresult} is calculated by taking
the number of pixels in the optical spectra. Most of the spectra are taken with FORS1/2 grism 300V 
yielding four pixels per resolution element.

\subsection{Dust models}\label{dustmodels}
The dust extinction was modeled with empirical extinction laws
\citep{fm,ccm,pei} to determine the absolute extinction and the
characteristics of the extinction. From the \citet{pei} extinction
laws we used only the SMC extinction law. Hereafter, we will refer to
these three extinction models as FM \citep[][]{fm}, CCM \citep[][]{ccm}
and SMC \citep[][]{pei}. The important characteristic of the SMC law
used here is the lack of the graphite 2175\,\AA{} bump and the even steeper
rise in the far UV, rising faster than $1/\lambda$
\citep{prevot,gordon}.  \citet{ccm} showed that Galactic sightlines
could typically be well-fit with a parameterization of the extinction
curve that relied only on a single parameter, the ratio of absolute
to selective extinction, $R_V=A_V/E(B-V)$. Comparable to CCM, FM provides greater
freedom in reproducing the extinction curves. To investigate restframe
visual extinction, we fit all three extinction laws for each case and
found that the best fitting results and null hypothesis probabilities for
SMC and FM were similar in most cases. We therefore selected the SMC as by far the simpler model and used it to characterize the extinction in 37 cases (see Table \ref{fitresult}) where the null hypothesis probability for the fit with FM was not significantly better. The FM parameterization was used for the cases where the SMC could not provide a good fit because of the presence of the 2175\,\AA{} bump (see Table \ref{fitresult}). The CCM curve yielded a poorer fit to the observed data in every case (see below)

The continuum emission from an afterglow is believed to be dominated by synchrotron radiation \citep{sari} described by a power-law given as
\begin{equation}
F_\nu = F_0\nu^{-\beta}
\end{equation}
where $F_0$ is the flux normalization, $\nu$ is the frequency and $\beta$ is the intrinsic spectral slope. The observed flux is extinguished due to dust in the Milky Way, and in objects between our Galaxy and the GRB, the latter typically dominated by the GRB host galaxy (though see GRB\,060418; \citealt{ellison}). As explained above (Sect.~\ref{sec:photometry}), we corrected the optical/NIR data for Galactic foreground extinction maps \citep{schlegel}. With knowledge of the unextinguished flux from X-rays and the redshift, the extinction corrected spectral slope and restframe $A_V$ can be obtained. Therefore the extinguished flux will be given as

\begin{equation}
F_{\nu}^{\rm{obs}} = F_\nu10^{-0.4A_\lambda}
\end{equation}
Here $A_\lambda$ is the extinction in the host galaxy of the GRB as a function of wavelength $\lambda$. The wavelength dependence on dust extinction observed in SMC, LMC and MW environments is well reproduced by the standard dust models \citep{fm,ccm,pei}. For SMC, LMC and MW the extinction law introduced by \citet{pei} is given as
\begin{equation}
A_\lambda = A_V\left(\frac{1}{R_V}+1\right) \sum_{i=1}^6 \frac{a_i}{(\lambda/\lambda_i)^{n_i} + (\lambda_i/\lambda)^{n_i} + b_i}
\end{equation}
For SED modeling we used the SMC parameterization where $R_V=2.93$ for the SMC extinction law \citep{pei}. The parameters $a_i$, $b_i$ and $n_i$ are different for each extinction model (see \citealt{pei} for the parameter list). A parameterization for extinction in the Milky Way was advocated by \citet{ccm}. It provides a better estimation of the far UV curvature with fixed bump strength and varying $R_V$. It is defined as 
\begin{equation}
A_\lambda = A_V\left(a(x) + \frac{b(x)}{R_V}\right) \quad \mbox{ where $x=\lambda^{-1}$ }
\end{equation}
 $a(x)$ and $b(x)$ are wavelength dependent polynomials controlling the UV, optical and NIR regimes of the extinction curve. Finally, we fitted the data with an FM extinction model. The parameterization for the FM extinction curve in the UV range is given as
\begin{equation}
A_\lambda = A_V \left(\frac{1}{R_V}k(\lambda-V) + 1\right)
\end{equation}
where 
\[k(\lambda-V) = \left\{ 
\begin{array}{l l}
  c_1+c_2x+c_3 D(x,x_0,\gamma) & \quad \mbox{ $x\leq c_5$ }\\
  c_1+c_2x+c_3 D(x,x_0,\gamma)+c_4(x-c_5)^2 & \quad \mbox{ $x >c_5$ }\\ \end{array} \right. \]
 and the Lorentzian-like Drude profile, which represents the 2175\,\AA{} bump, is given as
 \begin{eqnarray}
 D(x,x_0,\gamma) = \frac{x^2}{(x^2-x_0^2)^2+x^2\gamma^2} \nonumber
 \end{eqnarray}
The underlying UV linear component is specified by the intercept, $c_1$, and the slope, $c_2$. The $\sim2175$\,\AA{} absorption bump is controlled by $c_3$ (bump strength), $x_0$ (central wave number), and $\gamma$ (the width of the bump). The parameters $c_4$ and $c_5$ give the far-UV curvature. The extinction properties in the NIR and optical are determined using spline interpolation points \citep[see][]{fm3}. The FM parameterization provides greater freedom for fitting the extinction curve; in particular, it allows characterization of the curve, for example in the strength, width and position of the 2175\,\AA\ bump. 

\section{Results}
\subsection{The SED of the afterglow}\label{spectralchange}
GRB afterglows are well described by a coherent synchrotron emission from accelerated electrons with a power-law distribution of energies promptly behind the shock. These electrons then cool both adiabatically and by emitting synchrotron and inverse Compton radiation. At late times, the electron population accelerated by shock mechanisms is expected to radiate in the slow cooling regime \citep[see][]{sari}. The GRB spectrum in that regime is defined by a broken power-law model. The transfer to the slow cooling regime causes a cooling break in the intrinsic spectrum, typically occurring between the X-ray and the optical/NIR wavelengths. The broken power-law model is given as 
\begin{equation}
F_\nu = \left\{ 
\begin{array}{l l}
 F_0 \nu^{-\beta_{1}} & \quad \mbox{ $\nu \leq \nu_{\rm{break}}$ }\\
 F_0 \nu^{\beta_{2}-\beta_{1}}_{\rm{break}} \nu^{-\beta_2} & \quad \mbox{ $\nu \geq \nu_{\rm{break}}$ } \end{array} \right. 
 \end{equation}
where $\beta_1$ and $\beta_2$ are the slopes of the optical and X-ray segments in our model respectively. The parameter $\nu_{\rm{break}}$ is the cooling break frequency joining the two power-law segments. In accordance with the theory outlined by \citet{sari}, this additional break in the spectrum will give a spectral change of $\beta_1=\beta_2-\Delta\beta$, where $\Delta\beta=0.5$. In our SED modeling we treat $\Delta\beta$ as a free parameter by allowing both slopes to vary  independently in order to test the compatibility of the synchrotron model with our observed SED sample. From our sample, 21 SEDs have a spectral break between the X-ray and optical wavelengths. We find that $95\%$ of the spectral breaks have a spectral change of $\Delta\beta\sim0.5$, in agreement with the synchrotron emission model. The mean $\Delta\beta$ of all 21 GRBs is 0.54 with a standard deviation of 0.13. This large standard deviation is dominated by a single outlier, GRB\,080210, which has a $\Delta\beta=1.1\pm0.1$. Of the 21 GRBs, GRB\,080210 is the only case that deviates from $\Delta\beta=0.5$ \citep[see][for a detailed analysis of the SED of GRB\,080210]{decia}. Considering only the remaining SEDs, $\langle\Delta\beta\rangle=0.51$ with a standard deviation of 0.02. This agreement with the prediction of \citet{sari} is remarkable, and indicates that for the vast majority of SEDs taken at similar epochs, it is valid to assume that the amplitude of the spectral break is equal to exactly 0.5. 
In the future, for broad-band SED analysis, this should remove a degree of freedom in deriving extinction curves using GRBs, making the process more robust.

This result provides strong support that the afterglow emission is
indeed synchrotron radiation, and that both the optical and X-ray
emission are produced by the same mechanism in the vast majority of
cases. While this is a key element of the fireball external shock
model, our findings are not in complete agreement with it: we checked
whether our data satisfy the closure relations between the spectral
and temporal indices as defined by \citet{sari}. In most cases,
we found that the observed temporal indices were \textit{not}
consistent with those expected from the fireball model,  even though
the specific prediction of $\Delta\beta = 0.5$ does hold. This is not
surprising. The synchrotron emission process is only one of the
elements of the fireball model. The closure relations depend not only
on the spectral shape, but also on the temporal evolution of the
fireball properties (radius, Lorentz factor, magnetic field), on the
assumption that the microphysics parameters $\varepsilon_{\rm e}$ and
$\varepsilon_{\rm B}$ do not vary in time, and on several other
details. The shape of the synchrotron spectrum, on the contrary, is
not sensitive to these additional parameters, and therefore the
prediction that $\Delta \beta = 0.5$ is much more robust. Indeed, our
analysis is not dependent on the fireball dynamics, but it just
assumes (and confirms) that the optical and X-ray emission are due to
synchrotron emission.

      \begin{figure}
   \centering
   {\includegraphics[width=\columnwidth,clip=]{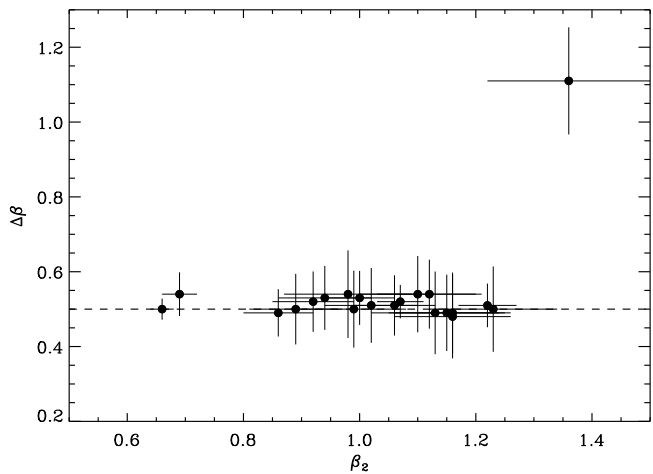}}
      \caption{$\Delta\beta$ against the X-ray spectral slope. The dashed line corresponds to $\Delta\beta=0.5$. GRB\,080210 is an outlier and has a softer X-ray slope compared to the rest of the sample.}
          \label{slope2}
   \end{figure}

\subsection{Dust extinction}
The restframe visual extinction, $A_V$, is determined for the forty one afterglows in our sample, plus GRB\,080607. Approximately two thirds (26/41), have low/moderate extinction values with $A_V$ ranging from $0.06\pm0.02$ to $0.65\pm0.04$, and are well-fit with SMC-type extinction. The mean visual extinction for these 26 GRB afterglows is $\langle A_{V}\rangle=0.24$ with a standard deviation of 0.14.  Roughly one quarter of the sample (11/41) is consistent with no dust reddening, with a mean $2\sigma$ upper limit $\langle A_{V}\rangle<0.10$ assuming an SMC type dust extinction curve. More than 7\% of the full sample ($\sim9$\% of the extinguished afterglows) are found to have a well-defined bump consistent with the 2175\,\AA\ feature. Of these bursts, GRB\,070802 was known before \citep{ardis,kruhler08}, while GRB\,080605 and GRB\,080805 are new (but see F09), doubling the number of such bumps unequivocally known in GRB host galaxies. We also include the other definite case of a bump in a GRB host extinction curve, GRB\,080607, in our analysis as a comparison extinction curve. All of these cases have $A_V\gtrsim1$, and are by far the most extinguished bursts in our sample, with a mean extinction more than five times the average extinction of extinguished bursts in the sample. Finally, GRB\,070318 was not well fit by the SMC curve and does not have a 2175\,\AA\ bump. Its extinction curve seems very unusual and is the subject of a separate investigation (Watson et al.\ in prep.). The \citet{maiolino} extinction curve associated with dust from supernovae was not used, but we infer from the good fit obtained with the SMC extinction curve in all cases without a 2175\,\AA\ bump, that this extinction curve is not required to fit the data.

\begin{table}
\caption{Parameters of the best fit with the SMC extinction curve ($R_V=2.93$) of \citet{pei}. The values of $\beta_2$ and $\nu_{\rm{break}}$ are provided for the cases where a broken power-law is the best fit.}
\label{pei} 
\centering
\begin{tabular*}{\columnwidth}{@{\extracolsep{\fill}}l c c c c }\hline\hline   
\centering
GRB	& $\beta_1$ & $\beta_2$ &  $\log\nu_{\rm{break}}$ & $A_V$  \\
	&			&		&	Hz		& mag\\
\hline
050401 & $0.39\pm0.05$ & $0.89\pm0.08$ & $16.32\pm0.89$ & $0.65\pm0.04$ \\
050730 & $0.16\pm0.02$ & $0.66\pm0.02$ & $16.00\pm0.17$ & $0.12\pm0.02$ \\
050824 & $0.40\pm0.04$ &  $0.92\pm0.07$  & $15.20\pm0.33$  & $0.15\pm0.03$ \\
060115 & $0.77\pm0.03$ & \mbox{\ldots} & \mbox{\ldots} & $0.10\pm0.02$  \\
060512 & $0.68\pm0.05$ & $1.16\pm0.10$ & $15.37\pm0.45$ & $<0.12$ \\
060614 & $0.47\pm0.04$ & $1.00\pm0.05$ & $16.06\pm0.21$ & $0.11\pm0.03$ \\
060707 & $0.59\pm0.02$ & \mbox{\ldots} & \mbox{\ldots} & $0.08\pm0.02$  \\
060708 & $0.67\pm0.04$ & $1.16\pm0.11$ & $16.92\pm1.04$ & $0.14\pm0.02$ \\
060714 & $0.44\pm0.04$ & $0.98\pm0.11$ & $15.86\pm0.53$ & $0.21\pm0.02$ \\
060729 & $0.78\pm0.03$ & \mbox{\ldots} & \mbox{\ldots} & $0.07\pm0.02$ \\
060904B & $0.90\pm0.02$ &  \mbox{\ldots} & \mbox{\ldots} & $0.34\pm0.03$ \\
060906 & $0.56\pm0.02$ & $1.10\pm0.10$ & $17.10\pm0.14$ & $<0.09$ \\
060926 & $0.82\pm0.01$ & \mbox{\ldots} & \mbox{\ldots} & $0.32\pm0.02$ \\
060927 &	$0.86\pm0.03$  &  \mbox{\ldots} & \mbox{\ldots} & $<0.12$ \\
061007 & $0.78\pm0.02$ & \mbox{\ldots} & \mbox{\ldots} & $0.34\pm0.03$  \\
061021 & $0.55\pm0.02$ & $1.07\pm0.04$ & $16.93\pm0.24$ & $<0.10$ \\
061110A	& $0.79\pm0.02$ &  \mbox{\ldots} & \mbox{\ldots} & $<0.10$ \\
061110B & $0.73\pm0.03$ & $1.23\pm0.11$ & $16.90\pm0.15$ &  $0.23\pm0.03$  \\
070110 & $0.66\pm0.02$ & $1.15\pm0.10$ & $16.84\pm0.18$ & $<0.10$ \\
070125 & $0.55\pm0.04$ & $1.06\pm0.07$ & $14.95\pm0.48$ & $0.30\pm0.04$ \\
070129 & $0.58\pm0.02$ & $1.12\pm0.09$ & $16.98\pm0.13$ & $0.28\pm0.02$ \\ 
070506 & $0.51\pm0.06$ & $1.02\pm0.08$ & $16.00\pm0.69$ & $0.44\pm0.05$ \\
070611 & $0.95\pm0.02$ & \mbox{\ldots} & \mbox{\ldots} & $0.06\pm0.02$ \\
070721B & $0.88\pm0.02$ & \mbox{\ldots} & \mbox{\ldots} &  $0.20\pm0.02$  \\
071020 & $0.15\pm0.05$ &  $0.69\pm0.03$ & $14.79\pm0.25$  &  $0.43\pm0.04$ \\
071031 & $0.64\pm0.01$ & $1.13\pm0.11$ & $15.35\pm0.55$ & $<0.07$ \\
071112C & $0.37\pm0.02$ & $0.86\pm0.06$ & $15.67\pm0.22$ & $<0.08$ \\
071117 & $0.41\pm0.03$ & $0.94\pm0.08$ & $16.95\pm0.92$ & $0.28\pm0.02$ \\
080210 & $0.25\pm0.03$ & $1.36\pm0.14$ & $16.03\pm0.15$ & $0.33\pm0.03$ \\
080319B &$0.77\pm0.02$ &  \mbox{\ldots} & \mbox{\ldots} & $<0.11$ \\
080520 & $0.78\pm0.02$ & \mbox{\ldots} & \mbox{\ldots}  & $0.22\pm0.02$\\
080707 & $0.71\pm0.03$ & $1.22\pm0.05$ & $16.90\pm0.16$  & $<0.12$ \\
080721& $0.68\pm0.02$ & \mbox{\ldots} & \mbox{\ldots} & $<0.12$ \\
080905B & $0.80\pm0.07$ & \mbox{\ldots} & \mbox{\ldots} & $0.42\pm0.03$  \\
080913 & $0.79\pm0.03$ & \mbox{\ldots} & \mbox{\ldots} & $0.12\pm0.03$ \\
080916A	 & $0.49\pm0.05$ & $0.99\pm0.09$ & $17.00\pm0.88$ & $0.15\pm0.04$ \\
080928 & $1.08\pm0.02$ & \mbox{\ldots} & \mbox{\ldots} & $0.29\pm0.03$ \\
\hline
\end{tabular*}
\end{table}
   
   \begin{figure}
   \centering
   {\includegraphics[width=\columnwidth,clip=]{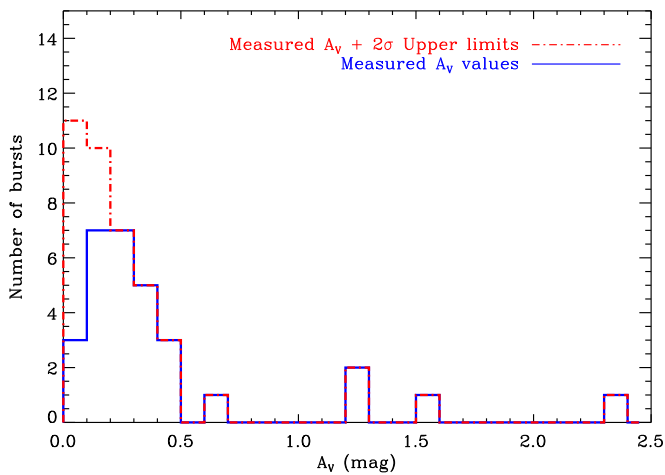}}
      \caption{Distribution of the restframe visual extinction, $A_V$, of our  flux calibrated spectroscopic sample. The solid curve corresponds to detections; the dot-dashed curve to detections and $2\sigma$ upper limits.}
          \label{avhist}
   \end{figure}

\begin{table*}
\caption{Best fit extinction curve parameters for the afterglow cases where FM is the best model (see Table.~\ref{fitresult}). GRB\,080607 was observed with Keck telescope and is not a part of our spectroscopic sample.}      
\label{fmresult} 
\centering
\begin{tabular}{@{} l c c c c @{}}   
\hline\hline       
Parameter  & 070802 & 080605& 080805 &  $\>\>\>\>\>\>\>$080607 \\
\hline
 $c_1$  				& $0.02\pm0.25$ 			& $-0.81\pm0.16$		& $0.26\pm0.15$\, 			& $\>\>\>\>\>\>\>$$1.19\pm0.43$ \\
 $c_2$ ($\mu$m) 				&  $0.82\pm0.16$			& $0.86\pm0.07$ 		& $0.57\pm0.06$\,			&  $\>\>\>\>\>\>\>$$0.37\pm0.10$ \\
 $c_3$ 				& $1.46\pm0.29$ 			&  $0.35\pm0.04$ 		& $0.89\pm0.11$\,			&  $\>\>\>\>\>\>\>$$2.96\pm0.66$ \\
 $c_4$ ($\mu$m$^{2}$)				& $0.26\pm0.08$			& $0.26\pm0.10$ 		& $0.15\pm0.14$\,			&   $\>\>\>\>\>\>\>$$0.17\pm0.10$\\
 $c_5$ ($\mu$m$^{-1}$)			 	&  $5.03\pm0.31$			& $6.30\pm0.41$ 		&  $6.50\pm0.14$ \,			&  $\>\>\>\>\>\>\>$$5.88\pm0.34$  \\
 $\gamma$ ($\mu$m$^{-1}$)			& $0.99\pm0.01$  			& $0.90\pm0.02$ 		&  $0.90\pm0.02$\,			&   $\>\>\>\>\>\>\>$$1.20\pm0.04$  \\
 $R_V$ 			 	& $2.81^{+0.67}_{-0.68}$ 		& $2.93^{-0.17}_{-0.20}$ 	 & $2.48\pm0.39$\,			&  $\>\>\>\>\>\>\>$$3.75^{+1.02}_{-1.06}$ \\
 $x_0$ ($\mu$m$^{-1}$)				&  $4.64\pm0.01$ 			&  $4.55\pm0.01$ 		 & $4.57\pm0.01$\, 			&  $\>\>\>\>\>\>\>$$4.52\pm0.02$ \\
 $\beta_1$ 		 	& $0.88\pm0.06$ 			& $0.76\pm0.03$ 		& $0.81\pm0.02$\,  			& $\>\>\>\>\>\>\>$ $0.96^{+0.05}_{-0.06}$  \\
  $A_V$  (mag)			 	& $1.19\pm0.15$ 			& $1.20^{+0.09}_{-0.10}$ 	 & $1.53\pm0.13$\,			&  $\>\>\>\>\>\>\>$$2.33^{+0.46}_{-0.43}$ \\ 
\hline
\end{tabular}
\end{table*}

\section{Discussion}

\subsection{The 2175\,\AA\ feature}
The 2175\,\AA\ absorption feature, as mentioned above, has been robustly detected in the optical spectra of four GRB afterglows (see Table \ref{fmresult}). The feature is a broad dip centered at 2175\,\AA\ due to excess extinction and was first discovered by \citet{stecher65}. Notable characteristics are the width ($\sim 1~\mu{\rm m}^{-1}$) and its variability ($\sim 20\%$), contrasted to the tight observed central wavelength of the bump (to within $\sim10$\,\AA, i.e. $<1$\%). Several candidates have been considered to explain the 2175\,\AA\ bump, ranging from carbonaceous materials \citep{henard} to iron poor silicate grains in the form of partially hydrogenated amorphous Mg$_2$SiO$_4$ particles \citep{steel}. The dominant hypothesis is some form of carbonaceous material in small grains ($\lesssim50$\,\AA), with polycyclic aromatic hydrocarbons \citep{duley98,duley06,cecchi08,duley09,cecchi10} and graphite being popular choices \citep{stecher,mathis77,draine84,draine89,sorrell90,mathis94,rouleau97,will,andersen03,clayton03,fm3}. The absence or weakness of the bump in SMC and LMC sightlines can be explained by a difference in the relative abundances of graphite and silicate grains \citep{pei,prevot}. The presence of the 2175\,\AA{} bump in GRB afterglow SEDs indicates that these GRBs occur in more evolved and massive galaxies and have dust properties similar to our local disk galaxies. Beyond our Local Group these distant GRBs can provide information on the graphite content of small grains in the surrounding circumburst environments. 

 It is notable that less than 9\% of the GRB extinction curves show an unequivocal 2175\,\AA\ bump. This is a small fraction, consistent with the prevailing notion that strongly star-forming galaxies, and in particular GRB hosts, have featureless dust extinction curves like the SMC \citep[e.g.][]{pei,calzetti,schady,kann10,kruhler10,greiner10}. However, there are only five known extinction curves toward the SMC, and one in fact does have a 2175\,\AA\ bump \citep{pei}. Furthermore, the starburst sample of \citet{calzetti} with SMC-like attenuation curves is very small, containing only six objects. Therefore we examined our sample to see in how many SEDs a bump could have been detected. We found that for 12 cases the bump region is not significantly covered by our spectra or photometry and in a few more the signal-to-noise ratio is very low. This is approximately 40--50\% of the \emph{extinguished} sample. It could be argued therefore, that in only about a quarter of the sample are we reasonably certain that no bump is present. Furthermore, there is very clear evidence in this sample that afterglows with a 2175\,\AA\ bump are preferentially heavily extinguished ($A_V\gtrsim1$).
 
 It is apparent that our spectroscopic sample is heavily reliant on optically bright afterglows; the bulk of bursts in the sample have an average magnitude of $21.1$ with a standard deviation of 1.7 magnitudes. But a MW style extinction gives 2--3 magnitudes of extinction in the $BVRI$ bands for $A_V=1$ at $z=2$. A restframe $A_V\gtrsim1$ would make spectroscopy unfeasible for the overwhelmingly majority of these bursts and a redshift would almost certainly not have been found for any except the brightest few (apart from emission lines from the host galaxy). It is obvious that we are heavily biased against these high extinction sightlines. F09 report a dark burst fraction of 25--42\% for that sample. If most of those dark bursts are similar to the most heavily extinguished bursts in our sample, many, if not all, may have a 2175\,\AA\ bump in their extinction curves. Our conclusion is that the apparent prevalence of the SMC extinction curve in GRB hosts is very likely an artifact of the bias in favour of low extinction sightlines and the frequent lack of good coverage of the bump region. We can only say that the bump is not present in about a quarter of our sample, and that it is present in 7\%, and that we know there is a clear preference for the bump to be detected in the most extinguished sightlines. Currently it is therefore not possible to say whether a featureless extinction curve or one with a bump is more common in GRB host galaxies. But it seems likely that the 2175\,\AA\ bump is far more prevalent than previously believed.

\begin{table}
\begin{minipage}[t]{\columnwidth}
\caption{Table listing the spectroscopic GRB sample with their name, Galactic column density, neutral hydrogen column density when available from the literature, equivalent column density ($N_\ion{H}{,X}$) determined from the X-ray spectral fits and the $N_{H,X}/A_V$ ratio. We reported $2\sigma$ upper limits for $N_\ion{H}{,X}$ where detection is $<2\sigma$ significant.}      
\label{nhx} 
\centering
\renewcommand{\footnoterule}{}  
\setlength{\tabcolsep}{3pt}
\begin{tabular}{l c c c c}   
\hline\hline                        
GRB & $A_V$ & log $N_\ion{H}{i}$ & $N_\ion{H}{,X}$ & $N_\ion{H}{,X}/A_V$\\
	&  mag	&	cm$^{-2}$		& $10^{22}$ cm$^{-2}$ & $10^{22}$ cm$^{-2}$ mag$^{-1}$\\
\hline
050401 &  $0.65\pm0.04$ & $22.60\pm0.30$  & $ 1.56^{+0.19}_{-0.18}$ & $2.40^{+0.33}_{-0.32}$ \\
050730 &  $0.12\pm0.02$ & $22.10\pm0.10$  & $<0.70$ & $<5.83$ \\
050824 &  $0.15\pm0.03$ & \mbox{\ldots}  & $ <0.20$ & $<1.33$ \\
060115 &  $0.10\pm0.02$ & $21.50\pm0.10$ & $<0.84$ & $<8.40$ \\
060512 & $<0.12$ & \mbox{\ldots}& $<0.07$ &  \mbox{\ldots} \\
060614 &  $0.11\pm0.03$ &  \mbox{\ldots} & $<0.05$ & $<0.45$ \\
060707 &  $0.08\pm0.02$ & $21.00\pm0.20$ & $<1.03$  & $<12.9$ \\
060708 & $0.14\pm0.02$ & \mbox{\ldots} & $<0.46$ & $<3.29$ \\
060714 &  $0.21\pm0.02$ & $21.80\pm0.10$ &  $1.39^{+0.20}_{-0.19}$ & $6.62^{+1.14}_{-1.10}$ \\
060729 & $0.07\pm0.02$ & \mbox{\ldots} & $0.13\pm0.02$ & $1.89\pm0.61$  \\
060904B &  $0.34\pm0.03$ &  \mbox{\ldots} & $0.47\pm0.04$ & $1.38\pm0.17$  \\
060906 &  $<0.09$  & $21.85\pm0.10$ & $<5.3$ &  \mbox{\ldots} \\
060926 &  $0.32\pm0.02$ & $22.70\pm0.10$ & $<8.00$ & $<25.0$ \\
060927 &  $<0.12$ & $22.50\pm0.15$ & $<5.7$ &  \mbox{\ldots} \\
061007 & $0.34\pm0.03$ & \mbox{\ldots} & $0.45\pm0.03$ & $1.29\pm0.14$ \\
061021 &  $<0.10$ & \mbox{\ldots} & $0.07^{+0.02}_{-0.01}$ & $>0.70$ \\
061110A &  $<0.10$ & \mbox{\ldots} & $<0.03$ &  \mbox{\ldots} \\
061110B &  $0.23\pm0.03$ & $22.35\pm0.10$  & $<8.40$ & $<36.52$ \\
070110 &  $<0.10$ & $21.70\pm0.10$ & $<0.34$ &  \mbox{\ldots} \\
070125 & $0.30\pm0.04$ & \mbox{\ldots} & $0.39^{+0.20}_{-0.17}$  & $1.30^{+0.69}_{-0.59}$ \\
070129 & $0.28\pm0.02$ & \mbox{\ldots} & $0.78^{+0.22}_{-0.20} $ & $2.79^{+0.81}_{-0.74}$  \\ 
070318 & \mbox{\ldots} & \mbox{\ldots} & $0.45\pm0.05$ & \mbox{\ldots} \\
070506 &  $0.44\pm0.05$ & $22.00\pm0.30$  & $<1.26$  & $<2.86$ \\
070611 &  $0.06\pm0.02$ & $21.30\pm0.20$ & $<0.94$ & $<15.7$ \\
070721B &  $0.20\pm0.02$ & $21.50\pm0.20$ & $<1.57$ & $<7.85$ \\
070802 & $1.19\pm0.15$ & $21.50\pm0.20$ & $<2.90$ & $<2.44$ \\
071020 & $0.43\pm0.04$ & $<20.30$ & $0.46\pm0.02$ & $1.07\pm0.11$ \\
071031 &  $<0.07$ & $22.15\pm0.05$ & $<1.20$ &  \mbox{\ldots} \\
071112C & $<0.08$ & \mbox{\ldots} & $<0.16$ &  \mbox{\ldots} \\
071117 & $0.28\pm0.02$ & \mbox{\ldots} & $1.18^{+0.20}_{-0.30}$ & $4.21^{+0.77}_{-1.11}$ \\
080210 &  $0.33\pm0.03$ & $21.90\pm0.10$ &  $2.15^{+0.62}_{-0.54}$ & $6.52^{+1.97}_{-1.74}$ \\
080319B & $<0.11$ & \mbox{\ldots} & $0.12\pm0.01$ &  $>1.10$ \\
080520 &  $0.22\pm0.02$ & \mbox{\ldots} & $1.38^{+0.40}_{-0.59}$ & $6.27^{+1.90}_{-2.74}$ \\
080605 & $1.20^{+0.09}_{-0.10}$ & \mbox{\ldots} & $0.71\pm0.08$ & $0.59\pm0.08$ \\
080707 & $<0.12$ & \mbox{\ldots} & $0.38\pm0.19$ & $>3.17$ \\
080721 &  $<0.12$ & $21.60\pm0.10$  &  $0.86\pm0.02$ & $>7.17$ \\
080805 &  $1.53\pm0.13$ & \mbox{\ldots}  &  $1.22^{+0.35}_{-0.45}$ & $0.80^{+0.24}_{-0.30}$ \\
080905B & $0.42\pm0.03$ & $<22.15$ & $2.38^{+0.51}_{-0.43}$ & $5.67^{+1.28}_{-1.10}$ \\
080913 & $0.12\pm0.03$ & $<21.14$ & $<9.30$ & $<77.5$ \\
080916A & $0.15\pm0.04$ & \mbox{\ldots} & $0.68\pm0.07$ & $4.53\pm1.29$ \\
080928 & $0.29\pm0.03$ & \mbox{\ldots} & $0.30^{+0.13}_{-0.12}$ & $1.03^{+0.46}_{-0.43}$ \\[5pt]
080607\footnote{Keck spectrum.} &$2.33^{+0.46}_{-0.43}$  & $22.70\pm0.15$ & $3.77^{+0.23}_{-0.22}$ & $1.62^{+0.33}_{-0.31}$ \\
\hline
\end{tabular}
\end{minipage}
\end{table}

   \begin{figure}
   \centering
   {\includegraphics[width=\columnwidth,clip=]{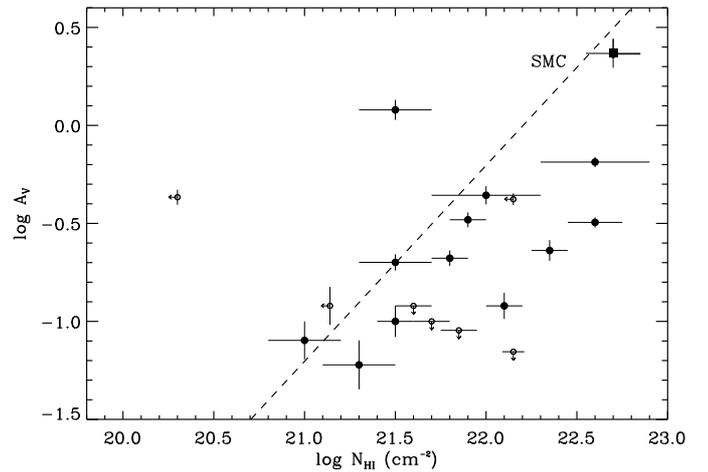}}
      \caption{$A_V$ against spectroscopically measured $\ion{H}{i}$ column density in log-log space. $N_{\ion{H}{i}}$ values are obtained from the literature and given in Table \ref{nhx}. The dashed line corresponds to the $N_{\ion{H}{i}}/A_V$ ratio for the SMC environment given by \citet{weingartner01}.}
          \label{nhav}
   \end{figure}

\subsection{Constraints on $R_V$}
We attempted to constrain $R_V$ specifically for the extinction curves
that were well fit with SMC-type extinction. We fitted the data again
with the FM model, this time fixing the FM parameters to the SMC
average extinction curve values given in \citet{gordon} and fixing
$\Delta\beta$ to 0.5 for the 20 relevant spectral break cases (see
\S\ref{spectralchange}), but allowing $R_V$ to vary freely. In most
cases $R_V$ is degenerate with $A_V$, so that we cannot constrain
$R_V$ in any useful sense. For GRB\,071020 and 080928 we can constrain
$R_V$ to $3.0^{+0.7}_{-0.8}$ and $2.0^{+0.5}_{-0.7}$ respectively
($1\sigma$ errors for two parameters of interest). From these cases,
we can determine that it is the strong constraint on the absolute
optical flux level that allows the constraint on $R_V$. In the former
case it is because the spectral break is very close to the optical,
and in the latter because the X-ray and optical fluxes are so similar.
In both cases the extinction is relatively high -- $A_V=0.43\pm0.04$
and $0.29\pm0.03$ respectively -- and the X-ray spectrum fairly
well-defined. To fix $R_V$ in future observations will require that
the unextinguished power-law is well-constrained, either through
infrared spectroscopy or better X-ray observations. The ideal way 
to do this is to use mid-infrared data in concert with good X-ray data.
To date this has been published for only one GRB afterglow using
observations with \emph{Spitzer} \citep{kevin}, albeit with a low S/N result. While more mid-infrared observations are now unlikely, and better X-ray
observations will not be routinely available in the foreseeable future, near-infrared spectra are now regularly obtained with
the X-shooter spectrograph on the VLT and may allow us to obtain
reasonably well-constrained $R_V$ values for GRB extinction sightlines
in the next few years.
   
\subsection{Gas-to-dust ratios} \label{nhavsec}
\begin{figure}
   \centering
   {\includegraphics[width=\columnwidth,clip=]{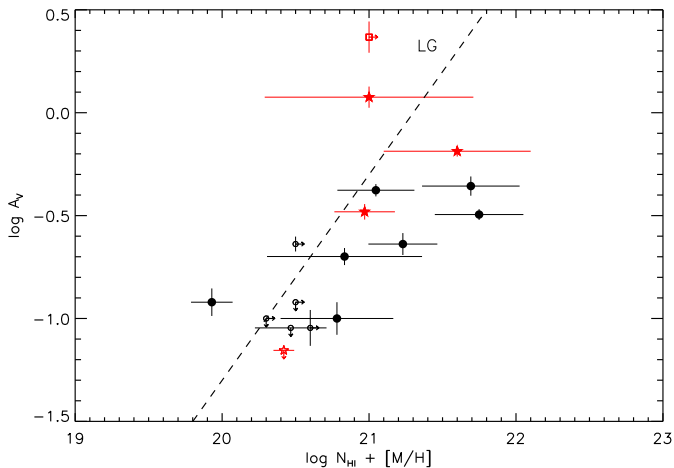}}
      \caption{Visual extinction against UV metal abundances in solar units. The metallicities obtained from the literature based on Zn element are marked with red stars. The rest of the data points are based on Si element metallicities (see Table \ref{metallicities}). GRB\,080607 is indicated with a red square. The metallicity or $A_V$ upper limits are represented by small circles or stars. The dashed line indicates metals-to-dust relation for the Local Group (indicated as LG).}
          \label{metalsav}
   \end{figure}

The $\ion{H}{i}$ column densities for these lines of sight have been measured from the 1216\,\AA\ Ly$\alpha$ absorption line in F09. The ratio of $N_{\ion{H}{i}}$ to $A_V$ gives, in principle, an estimate of the gas-to-dust ratio in the host galaxy of the GRB (Fig.~\ref{nhav}). However, the potential ionizing and destructive effect of the GRB and its afterglow may substantially complicate this picture. The vast majority of data points lie at values greater than the SMC gas-to-dust ratio -- the dashed line in Fig. \ref{nhav} corresponds to the gas-to-dust ratio of the SMC environment with a value of $N_{\ion{H}{i}}/A_V=1.6\times10^{22}$ cm$^{-2}$ mag$^{-1}$ \citep{weingartner01}. The only clear outlier is  GRB\,071021, with a very small neutral hydrogen column density quoted in F09. However, the burst is at $z=2.1$, and the spectrum is cut off in the blue. On close re-inspection of the 2D spectrum, it seems unlikely that such a strong limit can indeed be placed on the Ly$\alpha$ line in the afterglow of this burst. Indeed, it seems possible that this burst could have a large enough neutral column density in hydrogen to make it perfectly consistent with the SMC value. Therefore, the only burst in our sample that securely yields a gas-to-dust ratio lower than the SMC is GRB\,070802, with a gas-to-dust ratio consistent with Galactic sightline values. As pointed out by \citet{ardis}, this afterglow sightline is highly unusual among GRBs in this regard, a fact strengthened by our results. However, it is clearly not a property of all sightlines with the 2175\,\AA\ bump, since GRB\,080607 has a gas-to-dust ratio a factor of ten higher. Due to the low redshift, we do not have $\ion{H}{i}$ column density measurements for GRB\,080605 and GRB\,080805. The overall distribution suggests that gas-to-dust ratios in GRB environments are typically SMC-like or higher. An obvious idea is that metallicities that are SMC-like or lower are responsible for the high gas-to-dust ratios. However two arguments point against such a conclusion: first, the metals-to-dust ratio, where the metal column density is derived from soft X-ray absorption is typically an order of magnitude or more above the values from the Local Group \citep[see below and][]{watson07,schady}; second, when we correct the $\ion{H}{i}$ column for the metallicity in Fig.~\ref{metalsav}, we show that, even corrected for metallicity, most GRB sightlines have a super--Local-Group gas-to-dust ratio.

\begin{figure}
   \centering
   {\includegraphics[width=\columnwidth,clip=]{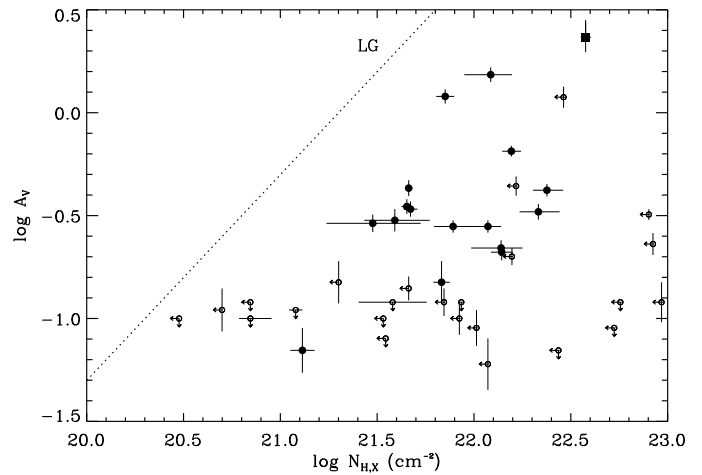}}
      \caption{$A_V$ measurements versus $N_{\rm{H,X}}$ for the spectroscopic GRB sample determined from X-ray spectral analysis and SED fitting. The dotted curve represents the metals-to-dust ratio for the Local Group environments (indicated as LG) assuming solar abundances. }
          \label{nhxav}
   \end{figure}

In Fig.~\ref{metalsav} we check if metallicity effects alone could be responsible for the high gas-to-dust ratios without reference to the X-ray absorption. We derived a metal column density in solar units using log $(N_{\ion{H}{i}}/\rm{cm}^{-2})+\rm{[M/H]}$ (equivalent to $N_{\rm{H,X}}$). These are metal column densities in the gas phase, derived from low ionization species, based on Zn and Si. The issue of depletion of the elements into dust grains is important. Zn and S are non-refractory elements and are the best tracers of the metallicity. We have used Zn metallicities wherever possible, however in many cases we have used Si to measure metallicity. Si is affected by depletion of the metal out of the gas phase, but less than strongly refractory elements like Fe and Cr \citep[e.g.,][]{ledoux02,draine03,fynbomet}. The $\rm{[M/H]}$ values are given in Table \ref{metallicities} and the $N_{\ion{H}{i}}$ values for the corresponding GRBs are taken from Table \ref{nhx}. In Fig.~\ref{metalsav} we see that the low ionization metals-to-dust ratios are closer to the Local Group value than the gas-to-dust ratios, but are still typically higher.  Conclusions on the gas-to-dust ratio of the host galaxies of GRBs are inevitably complicated by the fact that we know a majority of the gas along the line of sight to a GRB is ionized close to the burst \citep{watson07,schady,prochaska07,vreeswijk07,ledoux09}, and it is suspected that any dust near the GRB will be either sublimated or cracked by the GRB \citep[e.g.][]{fruchter,perna02}. The X-ray absorption, by contrast, is far less affected by photo-ionization, and so GRBs show much larger metals-to-dust ratios (Fig.~\ref{nhxav}). The most probable interpretation is that the column has been strongly affected by the GRB/afterglow, and the ionization distance and the dust sublimation distance are different. Therefore the metals-to-dust and gas-to-dust ratios derived from the optical/UV are not representative of the intrinsic ratios for the GRB host, but instead are due to the interaction of the distribution of the gas and dust along the column and the luminosity and spectrum of the GRB and its afterglow. Thus, the SMC-like and higher gas-to-dust ratios found in GRB afterglows, even after accounting for metallicity, seem to be due to the competing effects of gas ionization and dust destruction. The ratios found are closer to the values in the Local Group than are the metals-to-dust ratios derived from X-rays. An obvious explanation for this is that since we know that hydrogen ionization certainly occurs in the surroundings of the GRB, the $\ion{H}{i}$ column densities are simply lower limits to the actual gas column. So far we have no unequivocal evidence for dust destruction in GRB surroundings, but it seems probable that it occurs; what we can at least say from these results is that hydrogen ionization in the GRB surroundings is a more efficient process than dust-destruction, drawing the gas values closer to the Local Group ratio. 
   
\subsection{Metals-to-dust ratios} \label{nhxavsec}
Metal column densities ($N_{\rm{H,X}}$) from soft X-ray absorption in excess of the Galactic absorption \citep[e.g.][]{galama01,watson02,stratta04,starling05,starling07,gendre06,butler07,watson07,schady,campana10,evans} have been detected in most GRB afterglows; a rather surprising fact, since the absorption becomes harder to detect with increasing redshift. The effective X-ray absorption is dominated by various species, mostly oxygen and silicon K-shell absorption and including helium and iron L-shell \citep{morrison}. At higher redshifts (i.e.\ $z>2.5$) the soft X-ray absorption becomes increasingly dominated by the heavier elements such as iron, silicon and sulphur \citep{kallman,gou05,butler072,hurkett}. It is important to note that the X-ray column densities are typically reported in units of equivalent hydrogen column density assuming solar abundances. The upper limits and measured $N_{\rm{H,X}}$ values of our spectroscopic GRB sample are given in Table~\ref{nhxav}. Significant excess X-ray absorption was detected for 21 GRBs in our sample. The X-ray column densities are effectively metal column densities in both gas and dust, and would have to be corrected by the metallicity to obtain the gas column density. However, this is not enough. Due to a lack of information on the precise ionization state and structure of the gas, the absorption is invariably fit with a neutral gas model. We are well aware that the gas surrounding the GRB that causes the majority of the absorption is highly ionized \citep{watson07,schady11,prochaska07}. The metal column densities derived from the fits using neutral models are therefore strictly lower limits to the actual metal column densities. How close they are to the real value of the total column density depends strongly on the ionization state of the gas. It is reasonable to assume we are not underestimating the column densities by a factor of more than ten, since in the cases with the highest column densities, Compton scattering effects would become significant \citep{watsonandlaursen,campana11}, and we would not detect some bursts in X-rays. 

In Fig.~\ref{nhxav} we plot the dust extinction and soft X-ray--derived metal column densities for the sample. It is immediately apparent that all of the dust extinctions lie far below the expectation for the Local Group metals-to-dust ratios.
       \begin{figure}
   \centering
   {\includegraphics[width=\columnwidth,clip=]{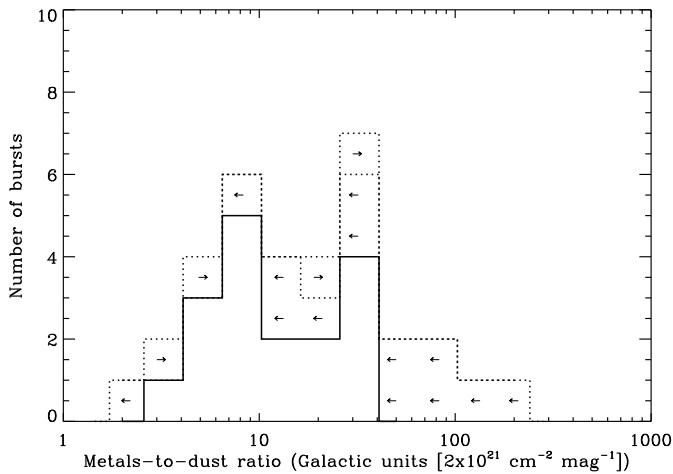}}
      \caption{Metals-to-dust ratio ($N_{\rm{H,X}}/A_V$) distribution of our spectroscopic GRB sample in Galactic units. The solid line show the cases where both $N_{\rm{H,X}}$ and $A_V$ are significantly measured. The upper and lower limits to the metals-to-dust ratios are shown by the dotted histogram and the limits in each bin are marked by left or right arrows.}
          \label{nhxhist}
   \end{figure}  
For the Galaxy, $N_{\rm{H,X}}/A_V$ has been measured for many sightlines and using different techniques \citep[e.g.][]{guver09,predehl,gorenstein}. For the LMC and SMC, the gas-to-dust ratio ($N_{\ion{H}{i}}/A_V$) can be converted to the same units ($N_{\rm{H,X}}/A_V$) by correcting for the metallicity. The $N_{\ion{H}{i}}/A_V$ relation for the MW varies in the range $(1.7-2.2)\times10^{21}$\,cm$^{-2}$\,mag$^{-1}$. Correcting for metallicities, the LMC \citep{fitzpatrick85} and SMC \citep{weingartner01} values are consistent with this range \citep[see for a review][]{draine03}. Therefore, we use a canonical value of  $N_{H,X}/A_V\approx0.2\times10^{22}$\,cm$^{-2}$\,mag$^{-1}$ for the metals-to-dust ratio of the Local Group of galaxies for comparison to our sample. \citet{daiandkochanek} showed that in lensing galaxies, the metals-to-dust ratio is also consistent with values in the Local Group. Together, those results hint that there may be an almost universal metals-to-dust ratio, at least in galaxies that are somewhat evolved.

In apparent contrast to this, our sample shows a spread of metals-to-dust ratios that covers a range from approximately three to thirty times the Local Group value (Fig.~\ref{nhxhist}). The fact that none of our objects have a metals-to-dust ratio below the Local Group value suggests that the Local Group ratio may at least be a minimum. Indeed, our result here, which confirm much previous work \citep[e.g.,][]{galama01,schady,schady11,stratta}, may be consistent with a universal metals-to-dust ratio if dust destruction by the GRB and afterglow is responsible for the lack of dust that leads to these large metals-to-dust ratios. Under this hypothesis, the GRB would have to destroy the equivalent of between 1 and 10 magnitudes of extinction ($A_V$) of dust within a few pc of the burst \citep{watson07}, which seems feasible \citep{fruchter,waxmananddraine,perna03}. As pointed out by \citet{fruchter}, \citet{waxmananddraine}, and \citet{perna03}, observations within the first few seconds of a burst should, for most GRBs, show a colour change related to the destruction of the dust. Based on these data and the assumption of a universal metal-to-dust ratio, such a colour change would be very large; however, given that we rely on bright restframe UV emission for detection, and it is precisely the bright UV emission that is primarily responsible for the dust destruction, it seems unlikely that we will be able to detect such a colour change before it is complete. 

It is worth noting here again that neither ionization nor metallicity can resolve the dust-poor nature of these absorptions. In the former case we know the gas must already be highly ionized; the dust would not survive in such an environment. As to the latter, we are directly observing the metal absorption, changes to the overall metallicity therefore have no effect on this ratio.

It is worth noting in this discussion that because our sample is almost certainly biased towards low extinction sightlines due to the requirement to have an optical spectrum, we may be preferentially selecting those sightlines with the highest gas-to-dust and metals-to-dust ratios without needing to
invoke dust destruction. It is possible that we might simply be missing many extinguished GRB afterglows with low metals-to-dust and gas-to-dust ratios. However, in this scenario, one would have to accept that sightlines in a significant number of GRB hosts exist with metals-to-dust ratios tens of times higher than in the Local Group. Such higher metals-to-dust ratios are also found in AGN \citep[e.g.,][]{maiolino01,willott} using similar techniques to
those used here. In those cases, however, most of the high X-ray absorbing column is attributed to the AGN and is assumed to be essentially dust-free. Sightlines throughout the Local Group from the SMC to the Milky Way, on the other hand, have metals-to-dust ratios that do not seem to vary by more than some tens of percent.

\begin{figure}
   \centering
   {\includegraphics[width=\columnwidth,clip=]{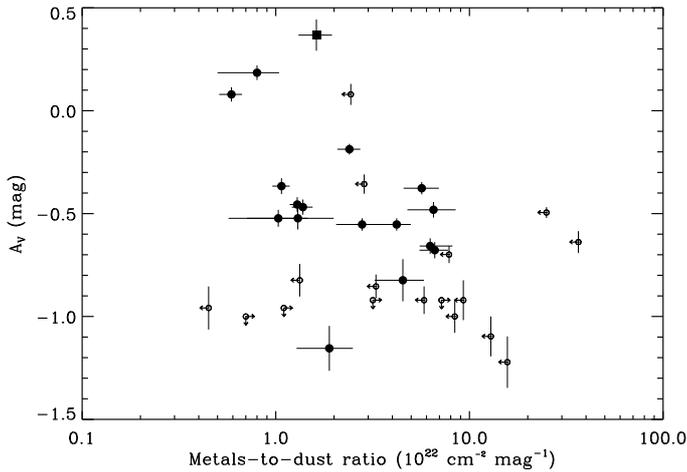}}
      \caption{Restframe $A_V$ and metals-to-dust ratio ($N_{\rm{H,X}}/A_V$) for our GRB sample. The filled circles are for the cases where $N_{\rm{H,X}}$ and $A_V$ are $\ge2\sigma$ significant. The open circles represent upper limits either obtained for $N_{\rm{H,X}}$ or $A_V$. The filled square corresponds to GRB\,080607 observed with the Keck telescope.}
          \label{nhxavav}
   \end{figure}

We examined the possibility of a relationship between the metals-to-dust ratio and the line-of-sight extinction (Fig.~\ref{nhxavav}). While it might be anticipated that GRB hosts that possessed high extinction sightlines might be dust-enriched and have low metals-to-dust ratios, we see no clear correlation here either. Even the highly extinguished events with a 2175\,\AA\ bump have metals-to-dust ratios similar to the rest of the sample.

We briefly compared the metals-to-dust ratios of our spectroscopic sample with the burst redshift. We see no obvious trend of the ratio with redshift except what might be expected from the detectability limits of the X-ray absorption increasing with redshift and while the limit for $A_V$ essentially decreases with redshift.

\subsection{The effect of metallicity} \label{meta}

Metallicities measured in GRBs vary from less than 1/100 to nearly solar values, and are usually higher than the values found in QSO-DLA sightlines \citep[see][]{fynbo06,prochaska07,fynbo083,ledoux09}. This trend supports the idea that GRBs occur in low mass or dwarf galaxies, representing the largest fraction of galaxies in the local Universe with large star formation rates. This favours the currently accepted scenario of long duration GRBs, the collapse of massive, rotating Wolf-Rayet star \citep{woosley,macfadyen}. The metallicities are obtained from Voigt-profile fitting to mildly saturated lines arising from non-refractory elements using mid/high resolution spectroscopy. In our metallicity sub-sample we obtained metallicities for the GRBs from the literature (see Table \ref{metallicities}). The metallicities based on Zn are given for GRB\,050401, GRB\,070802, GRB\,071031 and GRB\,080210. In addition, we derived Si column densities, and hence, metallicities, for GRB\,060707, GRB\,060714, GRB\,070110, using the optically thin limit approximation \citep[e.g.][]{petitjean98} with oscillator strengths from \citet{morton}. For these three bursts we used the equivalent width of the weak and hence likely only mildly saturated \ion{Si}{ii} (1808\,\AA) line to estimate the \ion{Si}{ii} column density. For a few more GRBs there is no detection of \ion{Si}{ii} (1808\,\AA) whereas \ion{Si}{ii} (1526\,\AA) is well-detected (see F09). We inferred metallicities for eight GRB hosts using the restframe equivalent widths of \ion{Si}{ii} (1526\,\AA{}). The equivalent widths for all these GRBs are obtained from F09. The metallicities are derived in these cases using the metallicity--EW$_{1526\,\AA}$ correlation for QSO-DLAs \citep[Eq. 1]{prochaska1526}. Other independent QSO-DLA studies also confirm this tight trend between EW$_{1526\,\AA}$ and $[\rm{M/H}]$ \citep{kaplan1526}; GRB-DLAs also exhibit the same trend \citep{prochaska1526}. For GRB\,080607 the absorption line features are highly saturated, therefore only lower limits are obtained \citep{prochaska09}. The metallicity for this burst could be solar or even super solar. The reported metallicity here is a lower limit based on Zn.  No metallicity measurements are available for $60\%$ of our sample, mainly due to a lack of high- or medium-resolution spectroscopy or a redshift $\lesssim2$, placing the Ly$\alpha$ line outside the observed wavelength range. 

\begin{figure}
   \centering
   {\includegraphics[width=\columnwidth,clip=]{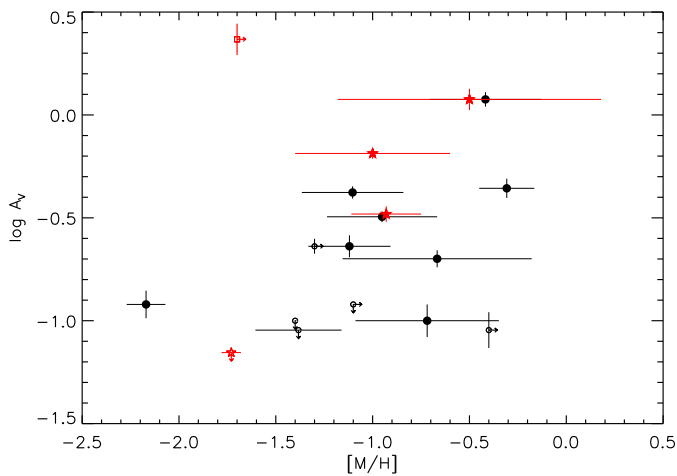}}
      \caption{Restframe $A_V$ measurements versus GRB metallicity. The metallicities obtained from the literature based on Zn are marked with red stars. The remaining data points are based on Si element metallicities (see Table \ref{metallicities}). GRB\,080607 is shown as an open square. The  small circles or stars represent metallicity or $A_V$ upper limits.}
          \label{metav}
   \end{figure}

In Fig.~\ref{metav} we plot metallicities for our GRB sub-sample versus the visual dust extinction, $A_V$. Fig.~\ref{metav} shows that there may be a slight trend between the metallicity and dust extinction. We see somewhat higher dust with increasing metallicity, suggesting that more metal rich systems may be more dusty. However we do not claim this correlation here with any confidence. 

\begin{figure}
   \centering
   {\includegraphics[width=\columnwidth,clip=]{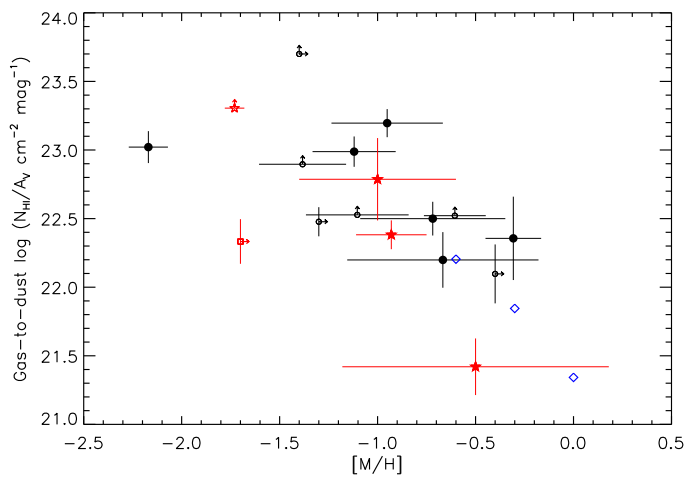}}
      \caption{Gas-to-dust ratio versus metallicity for GRB sub-sample. The metallicities obtained from the literature based on Zn are marked with red stars. The remaining data points are based on Si (see Table \ref{metallicities}). GRB\,080607 is indicated with a red square. The small circles and stars denote metallicity or gas-to-dust ratio upper limits. MW, LMC and SMC environments are denoted with blue diamonds from right to left.}
          \label{metnhav}
   \end{figure}

\begin{figure}
   \centering
   {\includegraphics[width=\columnwidth,clip=]{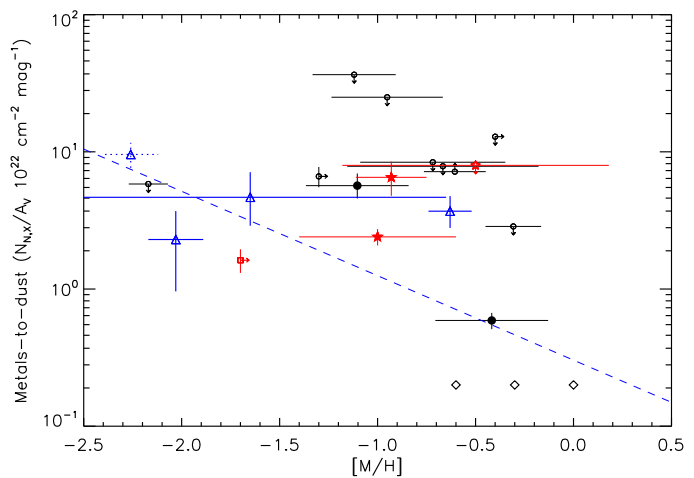}}
      \caption{Metals-to-dust ratio versus metallicity for the GRB sub-sample. The metallicities based on Zn are marked with red stars. The black circles represent Si-derived metallicity  (see Table \ref{metallicities}). GRB\,080607 is marked by a red square. The small circles and stars illustrate metallicity or metals-to-dust ratio upper limits. MW, LMC and SMC environments are denoted with diamonds from right to left. The blue triangles correspond to four data points taken from \citet{schady}. The blue dashed line represents the best fit to the three Local Group environments and the four GRBs from \citet{schady}. The blue triangle with dotted error bars is GRB\,050730 which is also included in both \citet{schady} and our sample.}
          \label{metnhx}
   \end{figure}

\citet{schady} also found high X-ray--derived metals-to-dust ratios and proposed that these ratios could be related to the metallicity of the GRB environment. They suggested that there might be an anti-correlation between the cold, gas-phase metallicity and the metals-to-dust ratios; i.e.\ metal poor galaxies forming dust less efficiently from their metals than the metal rich ones. Fig.~\ref{metnhx} shows the relation using the data in our sample. We see no obvious trend between the metals-to-dust ratios and the gas-phase metallicities and cannot confirm the relation. For the Spitzer Infrared Nearby Galaxies Survey (SINGS) sample \citet{draine07} found that the \emph{gas}-to-dust ratio was anti-correlated with metallicity. The anti-correlation was also found for dwarf galaxies \citep{lisenfeld}. Such a correlation would not be terribly surprising since the dust is composed of metals. In Fig. \ref{metnhav} we plot the gas-to-dust ratios of our GRB sub-sample versus metallicity. There is an anti-correlation between the two values; a Spearman rank test for a random origin for such an anti-correlation yields a $<1$\% probability. So, while there is some evidence of an anti-correlation between the gas-to-dust ratio and metallicity in GRB host galaxy sightlines, we find no evidence for a relationship between the metals-to-dust ratio and the metallicity. GRB environments have consistently higher metals-to-dust ratios than the Local Group but at the same time their metallicities cover the range as observed in the Local Group with a large scatter. Fig.~\ref{metnhx} in fact shows the large gap between the Local Group and GRB hosts in metals-to-dust ratios.

For most of the GRB afterglows we used Si based metallicity. In any case, the possible depletion of Si does not affect our conclusions above, since if the Si metallicities are affected by dust depletion, they will be underestimated. Specifically, high metals-to-dust ratios would therefore be higher if corrected for dust depletion (Fig. \ref{metalsav}), depletion is likely to be higher in systems with lower gas-to-dust ratios, thus making the correlation in Fig. \ref{metnhav} potentially stronger, and correcting for possible depletion in Fig. \ref{metnhx} is only likely to make the lack of correlation between metals-to-dust ratio and metallicity even clearer.

\begin{table}
\begin{minipage}[t]{\columnwidth}
\caption{GRB sub-sample metallicity measurements taken either from the literature or based on the \ion{Si}{ii} lines strength. The \ion{Si}{ii} ($\lambda$ = 1808 \AA{}, 1526 \AA{}) absorption lines are used to obtain metallicity by using the optically thin approximation and the \citet{prochaska1526} correlation. The metallicity for GRB\,080607 is based on Zn element. The metallicity could be solar or even super-solar for this burst \citep{prochaska09}.}      
\label{metallicities} 
\centering
\renewcommand{\footnoterule}{}  
\begin{tabular}{@{\extracolsep{\fill}}l c c }\hline\hline   
GRB	& [M/H] & Reference  \\
\hline
050401	& $-1.00\pm0.40$ & \citet{watson06} \\
050730	& $-2.17\pm0.10$ & \citet{ledoux09} \\
060115	& $-0.72\pm0.37$ & \citet{prochaska1526} correlation  \\
060707	& $>-0.40$ &  From $\ion{Si}{ii}$(1808) \\
060714	& $>-1.30$ &  From $\ion{Si}{ii}$(1808) \\
060906	& $-1.38\pm0.22$   & \citet{prochaska1526} correlation\\
060926	& $-0.95\pm0.28$  & \citet{prochaska1526} correlation\\
061110B 	& $-1.12\pm0.21$   & \citet{prochaska1526} correlation\\
070110 	& $>-1.40$ &  From $\ion{Si}{ii}$(1808)  \\
070506 	& $-0.31\pm0.14$  & \citet{prochaska1526} correlation\\
070721B 	& $-0.67\pm0.49$  & \citet{prochaska1526} correlation\\
070802 	& $-0.50\pm0.68$ & \citet{ardis} \\
071031 & 	$-1.73\pm0.05$ & 	\citet{ledoux09} \\
080210 	& $-0.93\pm0.18$ & \citet{decia} \\
080605\footnote{The metallicity for GRB\,080605 is based on the $\ion{Si}{ii}$ line EW. Being at $z<2$, ${\ion{H}{i}}$ is not detected for this burst. Therefore it is not included in Fig. \ref{metalsav} and \ref{metnhav}.} 	& $-0.42\pm0.29$   & \citet{prochaska1526} correlation\\
080607	& $>-1.70$ & 	\citet{prochaska09} \\
080721	& $>-1.10$  & \citet{starling09}\\
080905B	& $-1.10\pm0.26$  & \citet{prochaska1526} correlation\\
\hline
\end{tabular}
\end{minipage}
\end{table}


\section{Conclusions}
In this work we have presented the results of our analysis of 41 GRB afterglows from the X-ray to the NIR. This is not only the first spectroscopic study of \emph{absolute} extinction curves in distant galaxies, but also the largest extinction sample outside the Local Group. The SEDs are almost universally well-described by a power-law or broken power-law with absorption in soft X-rays and extinction in the optical/UV. Our analysis shows that a break occurs between optical and X-ray in about half of all SEDs at times typically between a few hours and a day after the burst. The mean spectral change excluding one outlier, is $\Delta\beta=\beta_2-\beta_1=0.51$, with a very small scatter (0.02), suggesting that in future afterglow SED analyses, the spectral change can typically be fixed to 0.5.

$63\%$ (26/41) of the bursts in the sample are well-fit with an SMC extinction curve, with 27\% (11/41) of the afterglows found to have no significant extinction. Of the remaining 4, 3 have a 2175\,\AA\ bump and one has a very unusual extinction curve. The bump cases comprise 7\% of our spectroscopic sample. Most bursts of our sample have moderate or low extinction, with $A_V<0.65$. However cases with a 2175\,\AA{} bump all have $A_V>1.0$. The predominance of the bump in the extinction curves of high-extinction sightlines allows us to claim that since our sample is heavily biased against high extinction sightlines, missing between a quarter and a half of all bursts, the percentages of extinction curves listed above are not at all representative of GRB hosts as a whole. Previous conclusions suggesting that the bump is a rare feature in extragalactic sightlines are also likely to be incorrect. An important conclusion from this work is that the most extinguished systems seem to correlate with the systems with a 2175\,\AA\ bump and possibly with high metallicity. Both facts point to the most extinguished systems being the most evolved, and therefore likely more massive. This is consistent with reports on work by \citet{perley} based on Spitzer data suggesting that dark bursts are more likely to have red, dusty host galaxies. Since a very substantial fraction ($\sim25-42\%$) of GRB afterglows are dark and likely extinguished (F09; \citealt{greiner10}), and all four known 2175\,\AA\ systems have restframe $A_V>1$, corresponding to $>$2--3 magnitudes of extinction in the observed $BVRI$ bands, it is premature to claim unequivocally that GRBs reside in small, sub-luminous, blue galaxies, or even that they follow star-formation.

The gas-to-dust ratios of GRBs are typically higher than the SMC. Correcting for gas-phase metallicity does not provide an explanation for these high ratios, and we find that gas-to-dust ratios from optical/UV measures are likely to be dominated by ionization and dust-destruction effects by the GRB, and not indicative of the intrinsic ratios of the GRB host galaxies. Soft X-ray--derived metals-to-dust ratios are 3--30 times the Local Group values. This is not due to ionization or metallicity effects. GRB hosts may however be consistent with a universal metals-to-dust ratio if dust destruction by the GRB is responsible for the apparent paucity of dust. We find no correlation between the metals-to-dust ratios in GRBs and the cold, gas-phase metallicity, the redshift or the extinction.

\begin{acknowledgements}
The Dark Cosmology Centre is funded by the Danish National Research Foundation. This work made use of data supplied by the UK \emph{Swift} Science Data Centre at the University of Leicester. We are thankful to the GRB community for observing these remarkable events and providing well sampled data. We are thankful to Daniel A. Perley, Marco Nardini and Annalisa De Cia for their help and support. PJ acknowledges support by a Marie Curie European Reintegration Grant within the 7th European Community Framework Program, and a Grant of Excellence from the Icelandic Research Fund.

\end{acknowledgements}

\bibliographystyle{aa}
\bibliography{grb.bib}{}

\begin{appendix}
\section{Details on the spectroscopic GRB afterglow sample}
\subsection{GRB\,050401}
The optical spectrum of the afterglow of GRB\,050401 was secured with the VLT/FORS2 instrument (\citealt{watson06}; F09). The NIR photometry in the $J$, $H$ and $K$ bands from \citet{watson06} is interpolated to the time of the SED using the $R$-band decay rate ($\alpha=0.86\pm0.1$). We find that the SED of the afterglow can be well reproduced by a broken power-law with substantial amount of SMC-type dust extinction with a value of $A_V=0.65\pm0.04$\,mag. The best-fit parameters are given in Table \ref{pei}. 

GRB\,050401 at $z=2.8983$ is categorized as a ``dark burst" as defined by \citet{jakobsson}. Previously \citet{watson06} found the afterglow SED is compatible with a broken power-law model and SMC type dust reddening with $A_V=0.62\pm0.06$, in agreement with our results. This burst has remarkably one of the highest $\ion{H}{i}$ column densities with log $N_\ion{H}{i}$/cm$^{-2}=22.60\pm0.3$ \citep{watson06}.
\subsection{GRB\,050730}
The optical spectrum of the afterglow of GRB\,050730 ($z=3.9693$) presented in this paper was carried out with the VLT/FORS2 (F09). The afterglow SED is constructed at 4.1 hr after the burst. The SED can be explained by a broken power-law, consistent with a $\Delta\beta\sim0.5$, plus a small amount of SMC type extinction with $A_V=0.12\pm0.02$\,mag.

Optical spectra of GRB\,050730 were also acquired by the Magellan Inamori Kyocera Echelle (MIKE) echelle spectrograph on the Magellan Telescopes \citep{chen05}, the Intermediate-dispersion Spectroscopic and Imaging System (ISIS) on the William Herschel Telescope \citep[WHT;][]{starling05}  and VLT UV-visual Echelle Spectrograph \citep[UVES;][]{delia07,ledoux09}. The derived \ion{H}{i} column density is log $N_\ion{H}{i}$/cm$^{-2}=22.1\pm0.1$ \citep{starling05,delia07,ledoux09}. A foreground QSO is also detected in the 2D spectrum at an impact parameter of 20 arcsec (F09). Previously \citep{oates09} derived $A_V=0.15$ from a fit of UVOT and X-ray data. \citet{kann10} find $A_V=0.10\pm0.02$ from their optical/NIR SED fitting. Both results are consistent with our findings. \citet{starling05} find $A_V=0.01$ from spectral fitting. \citet{schady} find that no extinction model is capable to fit of the NIR through X-ray SED, and derive $A_V\approx0.16-0.23$ for the SMC model using single and broken power-laws.
\subsection{GRB\,050824}
The optical spectrum of the afterglow of GRB\,050824 ($z=0.8278$) was taken with the VLT/FORS2 instrument (\citealt{sollerman07}; F09). GRB\,050824 is an X-ray flash (XRF) \citep{sollerman07}. The optical and X-ray spectrum are normalized to the level of acquisition image photometry taken at 9.5 hr after the burst trigger. The SED of the afterglow of GRB\,050824 is well explained by a broken power-law and SMC extinction with $A_V=0.14\pm0.04$\,mag. 

\citet{schady072} find SMC dust fits the data best, but they are not able to discriminate among the single and broken power-law models. For both power-law models \citet{schady} find $A_V\approx0.12-0.16$, in excellent agreement with our values. \citet{kann10} also find that SMC model well fits the optical/NIR SED with an insignificant amount of dust ($A_V=0.14\pm0.13$).

\subsection{GRB\,060115}
The optical spectrum and the $R$-band photometry of the afterglow of GRB\,060115 ($z=3.5328$) were carried out with the VLT/FORS1 (\citealt{piranomonte06}; F09). The SED is generated at $t_0+8.9$ hr after the burst. The afterglow SED of GRB\,060115 is described by a single power-law and SMC type extinction with a very small amount of dust ($A_V=0.10\pm0.02$\,mag). The optical spectrum has very low S/N and this can be seen clearly in the extinction curve (see Fig. \ref{extinction1}). F09 derived an \ion{H}{i} column density of log $N_\ion{H}{i}$/cm$^{-2}=21.5\pm0.1$.
\subsection{GRB\,060512}
The spectrum of the afterglow of GRB\,060512 was carried out with the VLT/FORS2 (\citealt{starlinggcn}; F09). For this afterglow there is uncertainty about the redshift in the GCN circulars. The redshift of $z=2.1$ comes from a single absorption line detected in the Telescopio Nazionale Galileo \citep[TNG;][]{starling062} and the FORS1 spectra  (F09). This redshift is also consistent with the spectral break detected by UVOT \citep{depasquale,oates09}. The $R$-band afterglow light curve is constructed by using magnitudes given by \citet{mundell06,cenko062,milne06,cenko063}. The $K_s$-band photometry is interpolated at the time of the SED by using observations given in \citet{hearty06,tanaka06}. The $J$-band observations of the afterglow are taken from \citet{sharapov06}. We reduced the UVOT data and the afterglow is clearly detected in the $v$, $b$ and $u$ bands. The interpolated lightcurve for each band is obtained by using the $R$-band lightcurve decay rate of $\alpha=1.05\pm0.1$. The SED of the afterglow of GRB\,060512 fits well with a broken power-law and no dust reddening (see Table \ref{pei}). We estimated the $2\sigma$ upper limit for extinction in the rest-frame $V$-band ($<0.08$\,mag).

Previously \citet{schady} find that no dust model is capable of fitting the NIR through X-ray data and derive high extinction using SMC model ($A_V\approx0.47-0.66$). The extinction corrected data presented in \citet{schady} do not match with their given intrinsic slope of the afterglow SED.
\subsection{GRB\,060614}
The optical spectrum of the afterglow of GRB\,060614 was obtained with the VLT/FORS2 (\citealt{della06}; F09). GRB\,060614 has a redshift of $z=0.1257$, the lowest redshift in our flux calibrated GRB sample. The redshift is estimated from the emission lines from the GRB host galaxy \citep{della06}. GRB\,060614 is a nearby, long duration GRB but not accompanied by a bright supernova \citep{fynbo06,della06,gal-yam,gehrels06}. The UVOT data were previously published in \citet{xu09} and \citet{mangano}. We re-reduced the UVOT data for $v$, $b$, $u$, $uvw1$, $uvm2$ and calibrated the afterglow fluxes in these bands using the photometric technique given in \citet{poole}. The optical to the X-ray SED of the afterglow of GRB\,060614 is well modeled with a broken power-law and low SMC type extinction with $A_V=0.1\pm0.03$\,mag.

\citet{mangano} find that SMC dust well fits the optical through X-ray SED, with $A_V=0.05\pm0.02$, values comparable to ours. \citet{kann08} find SMC dust model fits the optical data, and even higher extinction is obtained ($A_V=0.28\pm0.07$). The SED is not provided in \citet{kann08}, therefore a detailed comparison could not be made. 
\subsection{GRB\,060707}
The optical spectrum of the afterglow of GRB\,060707 ($z=3.4240$) together with $R$-band photometry are taken with the VLT/FORS1 (\citealt{jakobsson06}; F09). The afterglow SED is constructed at 34.4 hr after the burst trigger. The optical to X-ray SED of the afterglow of GRB\,060707 can be nicely reproduced with a single power-law and SMC type dust extinction ($A_V=0.08\pm0.02$\,mag). The derived \ion{H}{i} column density is log $N_\ion{H}{i}$/cm$^{-2}=21.0\pm0.2$ \citep{jakobsson06}.
\subsection{GRB\,060708}
The optical spectrum and $R$-band photometry of the afterglow of GRB\,060708 are obtained with the VLT/FORS2 (\citealt{jakobsson063}; F09). A precise redshift of the burst is not obtained due to the low S/N of the optical spectrum (F09). The reported redshift of $z=1.92$ was based on spectral break seen in UVOT data due to Ly$\alpha$ break \citep{schady06}. Together with the X-ray and optical spectrum we reduced the UVOT data in all bands obtained from the \emph{Swift} science archive facility. The data were reduced using the standard UVOT photometric calibration technique described in \citet{poole}. The afterglow is clearly detected in the UVOT $v$, $b$ and $u$ photometric bands. The optical to X-ray SED of the afterglow is well fit with a broken power-law together with SMC type dust extinction, $A_V=0.13\pm0.02$\,mag. 
\subsection{GRB\,060714}
The optical spectrum of the afterglow of GRB\,060714 ($z=2.7108$) was carried out with the VLT/FORS1. The $R$-band lightcurve is obtained from the circulars spanning from 0.04 to 4 days after the burst. The $I$ and $J$-band photometric magnitudes were obtained from the CTIO A Novel Dual Imaging CAMera (ANDICAM) instrument \citep{cobb06}. The UVOT data of the afterglow is reduced and the source is well detected in the $v$ and $b$ bands. We modeled the intrinsic SED of the afterglow and found that the afterglow can be well explained by a broken power-law together with SMC type extinction with restframe extinction of $A_V=0.22\pm0.02$\,mag.

Previously \citet{jakobsson06} obtained the redshift and the $\ion{H}{i}$ column density of log $N_\ion{H}{i}$/cm$^{-2}=21.8\pm0.1$ from the optical afterglow spectrum. Moreover extended Ly$\alpha$ emission in the center of the DLA trough is clearly detected (F09). \citet{schady} find that no dust model is able to fit the optical/X-ray data. The least bad is the SMC-type dust and power-law with dust extinction $A_V=0.46\pm0.17$, comparable to our results. The afterglow SED in \citet{schady} is constructed from the $Rvb$ band and X-ray data. While $b$ band is clearly affected by the Ly$\alpha$ absorption at $z=2.7108$ and might causes high amount of extinction.
\subsection{GRB\,060729}
The optical spectrum of the afterglow of GRB\,060729 at $z=0.5428$ is obtained with the VLT/FORS2 (\citealt{thone06}; F09). The acquisition image photometry was carried out using the $I$-band filter as reported in Table \ref{ebv}. The UVOT data of the afterglow of GRB\,060729 is previously published in \citet{grupe07}. The \emph{Swift} UVOT data of the afterglow of GRB\,060729 have been re-reduced and the source is clearly detected in the $v$, $b$, $u$, $uvw1$ and $uvm2$ filters. The optical spectrum is not of good quality and has broad undulations, therefore, the true shape of the underlying continuum cannot be seen from the spectrum. The optical spectrum has very low S/N and is affected by the high airmass at the time of the observation (F09). The afterglow SED can be fitted by a single power-law and SMC type dust extinction, and resulted in a very small amount of dust ($A_V=0.07\pm0.02$\,mag).

\citet{schady} find that SMC/power-law is the best model, resulting in low amount of extinction with $A_V=0.03\pm0.01$ and consistent with our value.

\subsection{GRB\,060904B}
The optical spectrum of the afterglow of GRB\,060904B ($z=0.7029$) was obtained with the VLT/FORS1 instrument (\citealt{fugazza}; F09). In our sample GRB\,060904B has the highest amount of foreground extinction and absorption (see Table \ref{ebv}). The SED of the afterglow is constructed at 5.1 hr after the burst. The SED fits well with a single power-law and SMC type extinction with extinction of $A_V=0.31\pm0.02$\,mag. The SED seems like a broken power-law case with softer X-ray slope. However, due to large uncertainties on the X-ray spectrum the fitting routine cannot minimize both models accurately and prefers a single power-law.

Previously \citet{schady} find no good fit for any dust model but find that single power-law is less likely, suggesting low extinction with $A_V=0.12^{+0.05}_{-0.04}$. \citet{kann10} find that their optical-NIR SED is consistent with no dust reddening using any dust model. The SED is not provided in \citet{kann10}, therefore a detailed comparison could not be made.
\subsection{GRB\,060906}
The optical spectrum of the afterglow of GRB\,060906 ($z=3.6856$) was obtained with the VLT/FORS1 spectrograph (\citealt{jakobsson06}; F09). The $I$-band observation was taken at the New Mexico Skies Observatory \citep{torii06} and corrected for the time using the $R$-band lightcurve power-law with shallow decay index $\alpha=0.56\pm0.07$ \citep{cenko065,cenko064}. The $z^\prime$ band observation is taken from the lightcurve presented in \citet{cenko09}. The optical to X-ray SED can be fitted with a broken power-law with no dust extinction. The derived $2\sigma$ extinction upper limit is $<0.09$\,mag. The X-ray to optical slope differ by $\Delta\beta\sim0.5$ which corresponds to the change in slope caused by the cooling frequency \citep{sari} lying within the observed frequency range.

The derived \ion{H}{i} column density is log $N_\ion{H}{i}$/cm$^{-2}=21.85\pm0.1$ \citep{jakobsson06}. Previously \citet{cenko09} fitted the SED and found dust extinction with $A_V=0.20^{+0.01}_{-0.12}$. \citet{kann10} fit the optical afterglow SED and find insignificant dust extinction $A_V=0.05\pm0.05$, in agreement with our results.
\subsection{GRB\,060926}
The optical afterglow of GRB\,060926 ($z=3.2086$) was observed with the VLT/FORS1 (\citealt{piranomonte062}; F09). The Ly$\alpha$ emission from the host galaxy of the GRB is detected in the trough of the DLA (F09). The estimated \ion{H}{i} column density is log $N_\ion{H}{i}$/cm$^{-2}=22.7\pm0.1$ \citep{jakobsson06}. The optical to X-ray SED is normalized to the acquisition image photometry at 7.7 hr after the burst. The optical to X-ray SED is well described by a single power-law and SMC type dust with a moderate extinction of $A_V=0.31\pm0.02$\,mag. 
\subsection{GRB\,060927}
The optical spectrum of the afterglow of GRB is carried out with the VLT/FORS1 spectrograph (\citealt{ruiz07}; F09). GRB\,060927 with a redshift of $z=5.4636$ is the second highest redshift GRB in our spectroscopic GRB sample. The redshift is based on a single $\ion{Si}{ii}$ absorption line \citep{ruiz07}. The $K$, $J$ and $I$ band observations of the afterglow were carried out with VLT filters \citep{ruiz07}. We took NIR photometry and spectroscopy to construct the composite SED at 12.5 hr after the burst trigger. The SED of the afterglow can be fitted with a single power-law without any dust extinction. The estimated $2\sigma$ upper limit for the $A_V$ is $<0.12$\,mag.

The optical spectrum of the afterglow is previously published in \citet{ruiz07} reporting the \ion{H}{i} column density of log $N_\ion{H}{i}$/cm$^{-2}=22.50\pm0.15$. \citet{kann10} fit the NIR SED of the afterglow and found $\le2.5\sigma$ significant dust with $A_V=0.21\pm0.08$ for SMC dust. Due to the large uncertainties on the NIR data, no model could be able to fit the NIR segment alone.
\subsection{GRB\,061007}
GRB\,061007 is an extremely bright burst detected by \emph{Swift} accompanied by a very luminous afterglow and similar decay rate in the X-ray and optical bands \citep{schady07}. The spectrum of the optical afterglow of GRB\,061007 at $z=1.2622$ was obtained with the VLT/FORS1 (\citealt{jakobsson064}; Paper). The $I$-band photometry is obtained from the lightcurve presented by \citet{mundell07}. We reduced the UVOT data of the GRB afterglow and detected source in the $v$, $b$ and $u$ photometric bands. The SED of the afterglow is nicely fitted with a single power-law together with SMC extinction with a value of $A_V=0.35\pm0.03$\,mag. 

\citet{oates09} and \citet{schady} find high values of extinction varying from $A_V\approx0.66-0.75$, while \citet{schady} prefer LMC dust model. The results of \citet{mundell07} with $A_V=0.48\pm0.19$ and \citet{kann10} with $A_V=0.48\pm0.10$ are comparable with our values. There is an indication of a 2175\,\AA{} bump in the blue end of the optical spectrum. The central wavelength of the bump (i.e. $\lambda_{\rm{obs}}\sim4520$\,\AA{}) does not match with the redshift of the GRB. An intervening absorber at redshift $z=1.066$ is also seen in the spectrum with strong $\ion{Mg}{ii}$ and $\ion{Ca}{ii}$ lines. This absorption feature could be associated with the intervening absorber. Apart from the optical spectrum, no other sign of the intervening absorber is seen at this redshift.

\subsection{GRB\,061021}
The spectrum of the GRB\,061021 was secured with the VLT/FORS1 spectrograph (\citealt{thone062}; F09). The redshift of the burst (i.e. $z=0.3463$) is obtained from $\ion{Mg}{ii}$ absorption lines (Hjorth et al, in prep). This is the second lowest redshift of our flux calibrated GRB afterglow sample. The UVOT photometric data have been reduced and the afterglow is well detected in the $v$, $b$, $u$, $uvw1$ and $uvm2$ filters (see Fig.~\ref{sed1}). The intrinsic SED of the afterglow can be explained by a broken power-law consistent with a cooling break as explained by \citet{sari} and with no dust extinction. The estimated $2\sigma$ extinction upper limit is $<0.1$\,mag. 

\subsection{GRB\,061110A}
The optical spectrum of the afterglow of GRB\,061110A at $z = 0.7578$ was obtained with the VLT/FORS1 (\citealt{fynbo07}; F09). The intrinsic SED of the afterglow is scaled to the acquisition image photometry. The afterglow spectrum of this burst has low S/N. The optical to X-ray SED can be modeled with a single power-law with no dust reddening. We derived the $2\sigma$ extinction upper limit of $A_V<0.1$\,mag.

\subsection{GRB\,061110B}
The VLT/FORS1 carried out the optical observation of the afterglow of GRB\,061110B at $z=3.4344$ (\citealt{fynbo063}; F09). The afterglow SED is constructed at $t_0+2.5$ hr epoch scaling with the $R$-band flux. The estimated \ion{H}{i} column density is log $N_\ion{H}{i}$/cm$^{-2}=22.35\pm0.1$ (F09). The intrinsic SED of the afterglow can be explained by a broken power-law all the way from optical to the X-ray with SMC type extinction with $A_V=0.22\pm0.03$\,mag. 

\subsection{GRB\,070110}
The optical spectrum of GRB\,070110 ($z=2.3521$) was obtained with the VLT/FORS2 spectrograph (\citealt{jaunsen}; F09). The UVOT data of this burst have been reduced by using UVOT calibration technique given by \citet{poole}. The afterglow is detected only in the $v$ and $b$ bands. The optical to X-ray SED of the afterglow is fitted with a broken power-law indicating a cooling break as suggested by \citet{sari}. We find no dust extinction in this burst. The derived $2\sigma$ extinction upper limit is $A_V<0.11$\,mag.

The estimated \ion{H}{i} column density is log $N_\ion{H}{i}$/cm$^{-2}=21.7\pm0.1$ (F09). Previously \citet{schady} find that broken power-law well fits the data, but they could not distinguish between the MW, LMC and SMC dust models, finding dust extinction of $A_V=0.23^{+0.06}_{-0.05}$ with SMC model.

\subsection{GRB\,070125}
The optical spectrum of the afterglow of GRB\,070125 ($z=1.5471$) was carried out with the VLT/FORS2 and previously published in F09. The broad undulations in the optical spectrum are due to systematics in the flux calibration. The redshift of the GRB is determined from the data obtained with Gemini North Telescope equipped with GMOS \citep{cenko08}. The $R_c$ and $I$ band observations have been collected by the MITSuME Telescope \citep{yoshida07}. The $J$, $H$ and $K_s$ band observations of the afterglow have been obtained with the robotic Peters Automatic Infrared Imaging Telescope \citep[PAIRITEL;][]{bloom07} and scaled to $R_c$-band flux level using the decay rate from a decent $R$-band lightcurve obtained from the circulars. The $R$-band lightcurve exhibits a plateau phase and a break to a steeper decay with decay indices $\alpha_1=1.4\pm0.1$ and $\alpha_2=2.55\pm0.07$ and break time at $\sim4$ days after the burst. The optical spectrum is normalized to $R_c$-band flux level. The X-ray to optical/NIR SED can be fitted with a broken power-law and SMC-type linear extinction with $A_V=0.27\pm0.03$\,mag.

Previously \citet{kann10} carried out a fit to the optical/NIR data and found that SMC dust best fits the data with little amount of dust ($A_V=0.11\pm0.04$). We find that the errors on the photometry are large and the NIR/optical X-ray joint fit suggests a cooling break in the SED and moderate extinction.

\subsection{GRB\,070129}
The optical spectrum of the afterglow of GRB\,070129 was obtained with the VLT/FORS2 spectrograph together with the $R$-band imaging (F09). We collected the $I$-band detection of the afterglow from Michigan-Dartmouth-MIT (MDM) Hiltner telescope \citep{halpern07}, which is close to the time of the SED. The redshift ($z=2.338$) reported here was estimated from the host galaxy [$\ion{O}{iii}$] emission line (Bo Milvang-Jensen private comm.). The intrinsic SED of the afterglow is scaled to the level of the $R$-band observation. The X-ray to optical SED of the afterglow is reproduced well with a broken power-law with SMC type extinction with a typical value of $A_V=0.30\pm0.02$\,mag.

\subsection{GRB\,070318}
The optical spectrum of the afterglow of GRB\,070318 ($z=0.8397$) was carried out with the VLT/FORS1 spectrograph (\citealt{jaunsen07}; F09). The spectrum shows an unusual sharp break in the optical spectrum around $\lambda_{\rm{obs}}\sim5000$ \AA{}. A spectrum simultaneously obtained at the Magellan telescope, showed the same optical break \citep{chen07}. The \emph{Swift} UVOT data have been reduced together with the NIR data from ANDICAM. The UVOT slope is consistent with the downturn in the optical spectrum. The SED of the afterglow is generated at $t_0+16.7$ hr after the burst. There is clear evidence for dust attenuation in the NIR to X-ray afterglow SED. Due to the sharp break in the optical spectrum and peculiar dust reddening the data are not consistent with SMC-origin extinction curve. We fit the SED with FM-dust induced model and found that the SED fits with a sharp break in the optical spectrum. The extinction curve is very unusual due to the sharp optical break. The GRB is not included in our extinction sample due to non-consistency with the synchrotron model. This sharp break could be due to destruction of dust grains by the GRB. A more detailed and comprehensive analysis of the SED will be given in Watson et al. (in prep).

Previously \citet{schady} implemented UVOT-XRT joint fit and found that a broken power-law is a slightly better fit to the data with dust extinction of $A_V=0.59^{+0.01}_{-0.06}$. 

\subsection{GRB\,070506}
The optical spectrum of the afterglow of GRB\,070506 ($z=2.3090$) was taken with the VLT/FORS2 instrument (\citealt{thone07}; F09). The estimated $\ion{H}{i}$ column density is log $N_\ion{H}{i}$/cm$^{-2}=22.0\pm0.3$ (F09). The optical spectrum redwards of $\lambda_{\rm{obs}}\approx7000$\,\AA\, is affected by strong fringing, therefore, this part is not included in our SED analysis. The intrinsic SED of the afterglow is modeled and we find a broken power-law to be a reasonable fit to the data. In addition to the broken power-law the data require dust using SMC type dust grains with $A_V=0.44\pm0.05$\,mag.

\subsection{GRB\,070611}
The optical spectrum of the afterglow of GRB\,070611 ($z=2.0394$) was obtained with the VLT/FORS2 (\citealt{thone072}; F09). The derived neutral hydrogen column density of the DLA is log $N_\ion{H}{i}$/cm$^{-2}=21.3\pm0.2$ (F09). The SED is scaled to the time of the $R$-band photometry at 7.7 hr. The X-ray to optical SED of the afterglow is defined by a single power-law and SMC type dust extinction ($A_V=0.06\pm0.02$\,mag). 

\subsection{GRB\,070721B}
The optical spectrum of the afterglow of GRB\,070721B ($z=3.6298$) is carried out with the VLT/FORS2 spectrograph (\citealt{malesani07}; F09). The estimated \ion{H}{i} column density is log $N_\ion{H}{i}$/cm$^{-2}=21.5\pm0.2$ (F09). The SED is scaled to the flux level of the $R$-band photometry. The intrinsic SED of the afterglow is nicely reproduced by using a broken power-law and SMC-type extinction  with $A_V=0.2\pm0.02$\,mag (see Fig.~\ref{sed1}). 

\subsection{GRB\,070802}
The optical observations of GRB\,070802 ($z=2.4541$) were obtained with the VLT/FORS2 spectrograph. The 2175\,\AA{} dust extinction feature is clearly seen in the optical spectrum of the afterglow \citep[F09;][]{ardis}. The burst is categorized as dark burst as defined by \citet{jakobsson}. The $J$, $H$ and $K$ band observations are taken at $t_0+2.0$ hr from the lightcurves presented in \citet{kruhler08} for each band. We reduced UVOT data using the calibration techniques defined by \citet{poole}. The afterglow is well detected in the $v$ and $b$ bands (see Fig.~\ref{sed1}). Due to the presence of a strong 2175\,\AA{} bump feature, the SMC extinction curve gives the worst fit to the data. The observed bump is shallower than the MW, therefore, CCM is also not a good fit to the data. We fitted the SED using a FM extinction model with both single and broken power-laws. The intrinsic SED is well described with a single power-law and a large amount of dust observed in rest-frame $V$-band ($A_V=1.19\pm0.15$\,mag) with total-to-selective extinction of $R_V=2.81\pm0.68$. The complete log of the best fit FM parameters is given in Table \ref{fmresult}.

The neutral hydrogen column density of log $N_\ion{H}{i}$/cm$^{-2}=21.5\pm0.2$ is determined from the optical spectrum \citep{ardis}. The rest-frame extinction curve of the afterglow is shown in Fig.~\ref{extinction1}. The NIR to X-ray SED has been previously fitted by \citet{ardis,kruhler08,cenko09,greiner10}, finding high dust extinction ($A_V\gtrsim1.0$) with a single power-law, similar to our results.

\subsection{GRB\,071020}
There was no PC mode data available near the time of the optical spectrum of GRB\,071020 ($z=2.1462$), therefore, we reduced early and late time PC mode X-ray data for this burst using the procedure described in \S\ref{xraysection}. We further checked that the X-ray spectral slope for early and late time are similar. The optical spectrum of the afterglow was obtained with the VLT/FORS2 instrument (\citealt{jakobsson07}; F09). The derived \ion{H}{i} column density is log $N_\ion{H}{i}$/cm$^{-2}<20.30$ (Nardini et al. in prep; F09). In our spectroscopic sample, this is the lowest column density obtained from the optical spectrum. The Ly$\alpha$ is in the very blue end of the spectrum and is not seen clearly (see F09). The$R$, $J$ and $K$ band observations are taken from Nardini et al. (in prep). The $H$-band photometry is carried out with PAIRITEL \citep{bloom072} and scaled to the time of the SED. A comprehensive analysis of the optical spectrum of GRB\,071020 will be discussed in Nardini et al. (in prep). The SED of the afterglow of GRB\,071020 can be modeled using a broken power-law with SMC type extinction with $A_V=0.40\pm0.04$\,mag. The change in slope is consistent with $\Delta\beta=0.5$ (see Table \ref{pei}). 

The best fit of \citet{kann10} implies SMC-type dust with $A_V=0.28\pm0.09$, comparable to our results within uncertainties.  

\subsection{GRB\,071031}
The acquisition image photometry and the spectroscopy of the afterglow of GRB\,071031 ($z=2.6918$) were carried out with the VLT/FORS2 instrument (\citealt{ledoux07}; F09). The $i^\prime$, $z^\prime$, $J$, $H$ and $K_s$ band photometry have been obtained from GROND \citep{kruhler09}. We also compared acquisition camera photometry to the $r^\prime$ band photometry presented in \citet{kruhler09} and taken at similar time, finding consistent values. The obtained column density of the GRB-DLA is log $N_\ion{H}{i}$/cm$^{-2}=22.15\pm0.05$ \citep{ledoux09}. The SED of the afterglow of GRB\,071031 fits well with a broken power-law and no dust reddening. The estimated $2\sigma$ reddening upper limit in the $V$-band is $<0.07$\,mag. 

Previously \citet{kann10} modeled the optical-NIR data and found that SMC model well fits the data with insignificant amount of dust $A_V=0.14\pm0.13$. The SED presented in this paper clearly indicate no dust. \citet{greiner10} find no dust extinction from the GROND-XRT joint fit, consistent with our results.

\subsection{GRB\,071112C}
The optical spectrum of the afterglow of GRB\,071112C at $z=0.8227$ was carried out with the VLT/FORS2 instrument together with the acquisition image photometry (\citealt{jakobsson072}; F09). The only available $H$ band photometry has been taken with the MAGNUM \citep{minezaki07}. The $R$-band lightcurve is retrieved from the GRBlog and fitted with a decay index of $\alpha=0.92\pm0.06$. We used the $R$-band lightcurve decay rate to obtain the NIR photometry at 10 hr after the burst. The SED and optical/NIR observations of the afterglow have been previously discussed in \citet{uehara}. The intrinsic SED of the afterglow is modeled well with a broken power-law and presents no dust extinction. The derived upper limit for dust reddening is $<0.08$\,mag.

Previously \citet{kann10} studied the optical-NIR afterglow SED and found that SMC dust best fit the data with an insignificant amount of dust ($A_V=0.23\pm0.21$). On the contrary, NIR to X-ray afterglow SED studied in this paper suggests no dust extinction (see Fig. \ref{sed1}).

\subsection{GRB\,071117}
The optical spectrum of the afterglow of GRB\,071117 ($z=1.3308$) was obtained with the VLT/FORS1 (\citealt{jakobsson073}; F09). The estimated redshift of the afterglow comes from [$\ion{O}{ii}$] emission line. The optical and X-ray spectrum are normalized to the $R$-band acquisition camera photometry at 9 hr after the burst trigger. The NIR to X-ray data fit well with a single power-law and SMC type with dust reddening of $A_V=0.26\pm0.02$\,mag.

\subsection{GRB\,080210}
The optical spectrum of the afterglow of GRB\,080210 at $z=2.6419$ was carried out using the VLT/FORS2 instrument (\citealt{jakobsson08}; \citealt{decia}; F09). The early and late time $R$-band photometry has been acquired with FORS2 \citep[see][]{decia}. We selected the $R$-band photometry at 1.69 hr which is near the mid time of the optical spectrum and scaled the afterglow SED to that level. The optical spectrum is corrected for slit losses and kindly provided by Annalisa De Cia \citep[see also][]{decia}. In addition to the optical spectrum the optical/NIR data have been obtained by using the GROND telescope \citep{greiner10}. We obtained the $g^\prime$, $r^\prime$, $i^\prime$, $z^\prime$, $J$, $H$ and $K_s$ band photometry at the time of the SED (Kr{\"u}hler, private comm.). The $i^\prime$ and $z^\prime$ data show a spectral plateau and are clearly down compared to the rest of the photometry (see Fig. \ref{sed1}), therefore, the GROND photometry is not included in our SED fitting. The optical-XRT SED is modeled nicely with a broken power-law and SMC-type dust reddening. The dust extinction is found to be $A_V=0.33\pm0.03$. The change in the spectral slope for this afterglow is $\Delta\beta=1.14\pm0.13$ (see Fig. \ref{slope2}).

The estimated \ion{H}{i} column density for the DLA is of log $N_\ion{H}{i}$/cm$^{-2}=21.90\pm0.10$ \citep{decia}. \citet{kann10} fitted the GROND SED and assumed the $i^\prime$ and $z^\prime$ data as an indication of 2175\,\AA{} bump, resulting in some deviation from the dust model and give quite large extinction ($A_V=0.7$). An independent analysis of \citet{decia} also confirms that the spectral change is around 1.0 and is inconsistent with the fireball model. This is the only outlier with twice the spectral slope change as compared to the other break frequency cases. The X-ray slope for this case is much softer than the other cases (see Fig. \ref{slope2}). We checked that the X-ray spectrum becomes gradually softer ($\Gamma=\beta+1>2.0$) in the time interval ranging from 1000 to 10000 s, suggesting that energy injection from the central engine is likely to be decreasing at these times. For SMC type extinction \citet{decia} found $A_V=0.18\pm0.03$. \citet{greiner10} employed a joint GROND-XRT fit to the data with fixed $\Delta\beta=0.5$ and find $A_V=0.24\pm0.03$.

\subsection{GRB\,080319B}
The optical spectrum of the afterglow of GRB\,080319B ($z=0.9382$) was obtained with the VLT/FORS2 spectrograph almost one day after the burst trigger (F09). The GRB is commonly referred to as the ``naked eye burst'' due to its peak visual magnitude of 5.3\,mag \citep{racusin08,wozniak09,bloom09}. The $g^\prime$, $r^\prime$, $i^\prime$ and $z^\prime$ band photometry has been obtained from the lightcurves for these filters presented in \citet{tanvir10}. The UVOT data have been reduced and the magnitudes obtained for the $v$, $b$, $u$ and $uvw1$ filters, where the afterglow is clearly detected. We generated the SED at 26 hr after the burst and modeled it using power-laws and dust extinction. We found the observed data fit very well with a single power-law all the way from optical to the X-ray and no dust reddening. The upper limit on the amount of dust is estimated to be $<0.11$\,mag.

\citet{kann10} find no dust extinction from optical-NIR SED fitting, in excellent agreement with our result.

\subsection{GRB\,080520}
The optical spectrum of the afterglow of GRB\,080520 ($z=1.5457$) was obtained with the VLT/FORS2 spectrograph at the time when the afterglow was very faint. The UVOT data have been reduced and the afterglow is only detected in the $v$ band. The SED of the afterglow is constructed and the data are well fitted with a single power-law and SMC-type dust extinction with $A_V=0.23\pm0.02$\,mag. 

\citet{greiner10} fit the XRT through optical/NIR data and find weakly constrained high amount of extinction ($A_V=0.53^{+0.40}_{-0.42}$). Our results are consistent to \citet{greiner10} within $1\sigma$.

\subsection{GRB\,080605}
The optical spectrum of the afterglow of GRB\,080605 ($z= 1.6403$) was carried out with the VLT/FORS2 in different grisms (\citealt{jakobsson083}; F09). The 2175\,\AA{} absorption dip is clearly seen in two FORS2 grism spectra. Being at redshift $z<2$ the Ly$\alpha$ is not detected for this burst due to the sensitivity range of the spectrograph. Therefore metal abundances cannot be obtained for this interesting case. We reduced the UVOT data of the afterglow and found that the source is contaminated by a nearby object. After subtraction the afterglow is not visible in any UVOT bands. The 2175\,\AA{} absorption dip is very weak and narrow. The SED is modeled with the FM extinction law. The SED is well described by a single power-law and large amount of dust extinction in restframe $V$ band of $A_V=1.2^{+0.09}_{-0.10}$\,mag. The extinction curve of the afterglow suggests $R_V$ consistent with its value in the MW (see Fig. \ref{extinction1}). The metallicity obtained from the equivalent width of $\ion{Si}{ii}$ (1526\,\AA{}) suggests a fairly high metallicity for this burst (see Table \ref{metallicities}).

\citet{greiner10} find that the MW dust model best-fit the data with moderate amount of extinction ($A_V=0.47\pm0.03$). The SED presented in this paper is not complete and lacking the NIR data, which could be in the restframe $V$-band. This particular case with other 2175\,\AA{} bump cases will be discussed in Zafar et al. (in prep).

\subsection{GRB\,080607}
The optical spectrum of the afterglow of GRB\,080607 ($z=3.0368$) was obtained with the Keck telescope and previously published in \citet{prochaska09,perley09} and F09. Because of being not observed with the VLT, this burst is not a part of our spectroscopic GRB sample. The derived neutral hydrogen column density of GRB-DLA is log $N_\ion{H}{i}$/cm$^{-2}=22.70\pm0.15$ (\citealt{prochaska09}; F09). Due to the occurrence of GRBs in star-forming molecular clouds, molecular hydrogen has been searched for many GRBs and resulted in non-detections \citep{vreeswijk04,tumlinson07,sheffer}. \citet{fynbo062} tentatively interpreted an absorption feature as H$_2$ in their spectrum of GRB\,060206. For the first time, strong unambiguous H$_2$ and CO molecular absorption lines were detected in the optical afterglow spectrum of GRB\,080607 \citep{prochaska09,perley09}. Most of the 2175\,\AA{} extinction bump is clearly detected in the optical spectrum. The remaining part of the bump is covered by using the $z$ and $i$ filters. The afterglow is detected in the NIR $J$, $H$ and $K_s$ bands with PAIRITEL. The $I$ and $V$ band observations have been performed with the Katzman Automatic Imaging Telescope (KAIT) and the $R$, $i$ and $z$ band detection are from the robotic Palomar 60\,inch telescope \citep[see Table:1][]{perley09}. This burst classifies as dark by the definition of \citet{jakobsson}. The GRB afterglow is remarkably luminous in optical and NIR wavelengths. Since the optical data have strong H$_2$ molecular absorption lines which makes it hard to see the underlying continuum, therefore, we fit with the binned spectrum where lines are already taken out \citep[see Table: 4;][]{perley09}. We fit the data with the FM parameterization and the best fit parameters are given in Table \ref{fmresult}. The intrinsic SED of the afterglow is nicely reproduced with $A_V=2.31^{+0.46}_{-0.43}$\,mag and a very high $R_V=3.75\pm1.04$. The 2175\,\AA{} is the widest compared to all other bump cases (see Table \ref{fmresult}). The host galaxy of the GRB afterglow is also extremely red in the optical and NIR bands \citep{chen10}.
 
Previously \citet{prochaska09} found $A_V\approx 3.2$ and \citet{perley09} estimated $A_V=3.3\pm0.4$ the optical/NIR data fit. Both authors found $R_V\sim4$ which is consistent with our result of a shallower extinction curve.

\subsection{GRB\,080707}
The optical spectrum of the afterglow of GRB\,080707 ($z=1.2322$) is obtained with the VLT/FORS1 under bad conditions (\citealt{fynbo08}; F09). The SED is normalized to the $R$-band observation taken at 1.1 hr after the burst. The intrinsic SED of the afterglow is fitted with a broken power-law and no dust reddening. The upper limit on the dust content is $A_V<0.13$\,mag. 

\citet{greiner10} implemented the GROND-XRT joint fit to the data and find that the data are consistent with no dust with an insignificant value of $A_V=0.11^{+0.14}_{-0.08}$\,mag.

\subsection{GRB\,080721}
The optical spectrum of the afterglow of GRB\,080721 at $z=2.5914$ was obtained with the VLT/FORS1 spectrograph (\citealt{starling09}; F09). We extracted the UVOT data and the afterglow is detected in the $v$ and $b$ band. The $I$-band photometry is obtained at 10.2 hr using the $I$-band observations given by \citet{starling09}. The neutral hydrogen column density of the DLA is log $N_\ion{H}{i}$/cm$^{-2}=21.6\pm0.1$ (F09). The intrinsic SED is scaled to the $R$-band acquisition camera photometry. The data is well described by using a single power-law and no dust reddening. The estimated upper limit for dust reddening is $<0.12$\,mag. 

Previously \citet{starling09} found no cooling break between the X-ray and optical wavelengths. Despite the large errors on the photometry, \citet{kann10} fit the optical SED and find $A_V=0.35\pm0.07$.

\subsection{GRB\,080805}
The optical spectrum of the afterglow of GRB\,080805 ($z=1.5042$) was obtained with the VLT/FORS2 (\citealt{jakobsson084}; F09). The spectrum clearly show a 2175\,\AA{} absorption feature at the redshift of the GRB. Due to the relative faintness in the optical, the afterglow is not detected in any of the UVOT filters. The spectrum shows a pretty clear absorption dip and the extinction curve resembles that of the MW. The intrinsic SED of the afterglow is generated 1.0 hr after the burst trigger. The X-ray to optical SED is reproduced nicely with a single power-law plus FM extinction curve with a large amount of dust reddening of $A_V=1.53\pm0.13$\,mag and total-to-selective extinction of $R_V=2.48\pm0.39$.

Previously \citet{greiner10} modeled the GROND-XRT data and found that MW dust model, with a 2175\,\AA{} bump, provides the best explanation to the data with $A_V=1.01^{+0.19}_{-0.08}$. The value is consistent with the results we find from our analysis.

\subsection{GRB\,080905B}
The optical spectroscopy of the afterglow of GRB\,080905B at $z=2.3739$ was taken with the VLT/FORS2 together with the $R$ band acquisition image photometry (\citealt{vreeswijk082}; F09). We implemented MIDAS package \texttt{FITLYMAN} \citep{fontana} to determine the neutral hydrogen column density. The derived $\ion{H}{i}$ column density of the afterglow of GRB\,080905B is log $N_\ion{H}{i}$/cm$^{-2}<22.15$. Since the afterglow flux is contaminated by the light from another object on the slit, we suggest the column density to be an upper limit. We modeled the intrinsic afterglow SED and the data are well fitted with a single power-law and dust extinction with $A_V=0.41\pm0.03$\,mag

\subsection{GRB\,080913}
The spectrum of the afterglow of GRB\,080913 was taken with the VLT/FORS2 (\citealt{greiner}; F09; \citealt{zafar11}). GRB\,080913 is the highest redshift GRB in our spectroscopic sample. The acquisition image was taken with $z$-Gunn filter. The magnitude reported in Table \ref{ebv} is strongly affected by Ly$\alpha$ absorption. We corrected the magnitude for Ly$\alpha$ absorption using spectro-photometric analysis (see \citealt{zafar11} for detail). A comprehensive description of the SED analysis is given in \citet{zafar11}. The neutral $\ion{H}{i}$ column density of the DLA is $N_\ion{H}{i}$/cm$^{-2}<21.14$ \citep{greiner}. 

\subsection{GRB\,080916A}
The optical spectrum of the optical afterglow of GRB\,080916A ($z=0.6887$) was carried out with the VLT/FORS2 (\citealt{fynbo082}; F09). The SED is constructed at 17.1 hr i.e. the time of $R$-band acquisition image photometry. The intrinsic SED is nicely reproduced with a broken power-law and SMC type dust extinction with $A_V=0.15\pm0.04$\,mag.

\subsection{GRB\,080928}
The optical spectrum of the afterglow of GRB\,080928 at $z=1.6919$ was obtained with the VLT/FORS2 (\citealt{vreeswijk08}; F09). We collected the optical/NIR photometry at the time of the SED taken with GROND from \citet{rossi10}. We scaled the spectrum to the level of the $r^\prime$ band photometry taken at 15.5 hr. In addition we reduced the UVOT data for this burst and fluxes are obtained for the $v$, $b$ and $u$ bands at the common SED epoch. The intrinsic SED is well fitted with a single power-law and SMC extinction. The best fit $A_V$ value is $0.3\pm0.03$\,mag. 

\citet{kann10} find that a MW dust model best fits the optical-NIR data with a small amount of extinction ($A_V=0.14\pm0.08$). We do not find any evidence for the 2175\,\AA{} bump in the optical spectrum.
\end{appendix}

\begin{appendix}
\clearpage
  \begin{figure*}
  \begin{tabular}{c c}
   {\includegraphics[width=0.9\columnwidth,clip=]{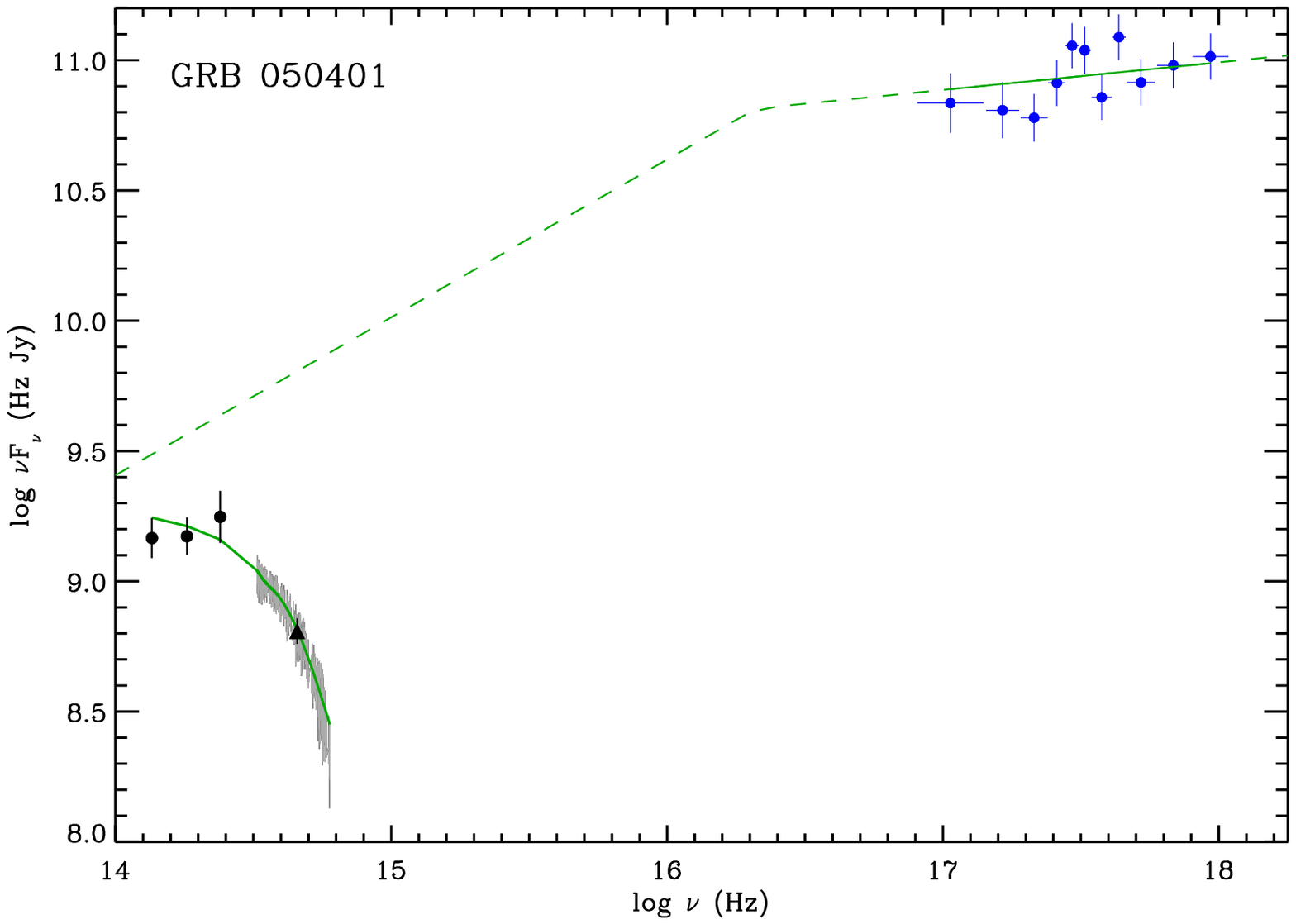}} &
   {\includegraphics[width=0.9\columnwidth,clip=]{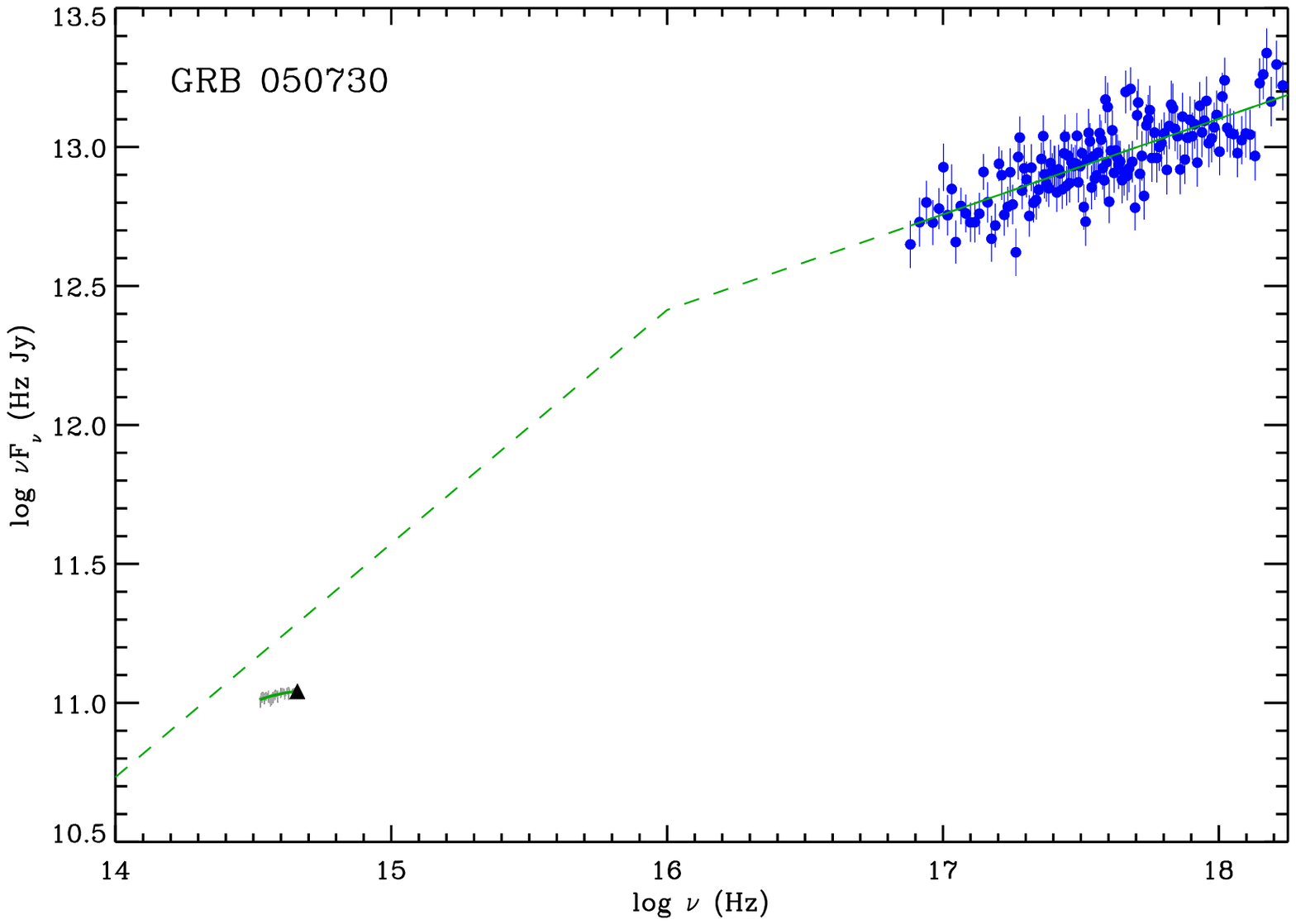}} \\
   {\includegraphics[width=0.9\columnwidth,clip=]{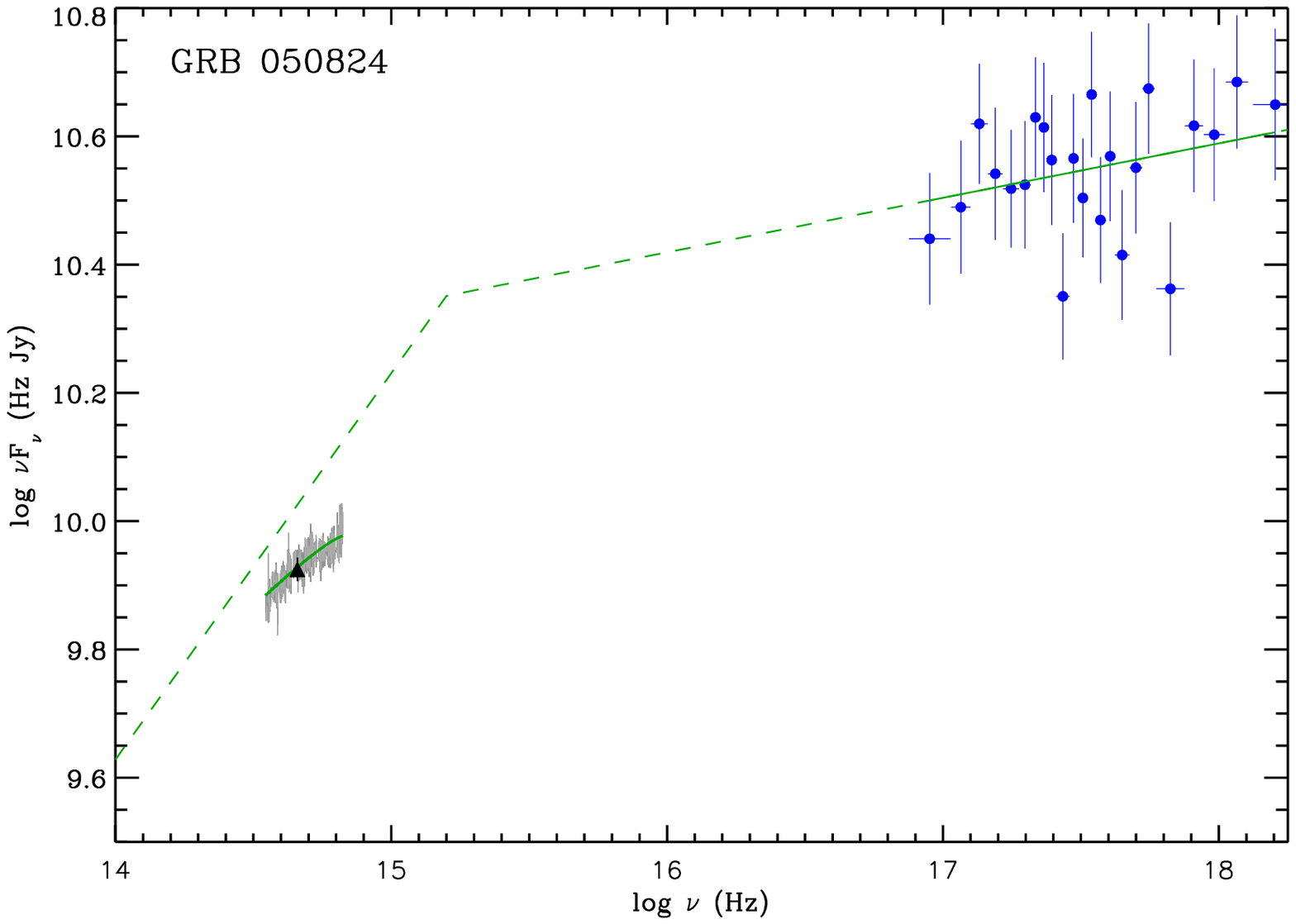}} &
         {\includegraphics[width=0.9\columnwidth,clip=]{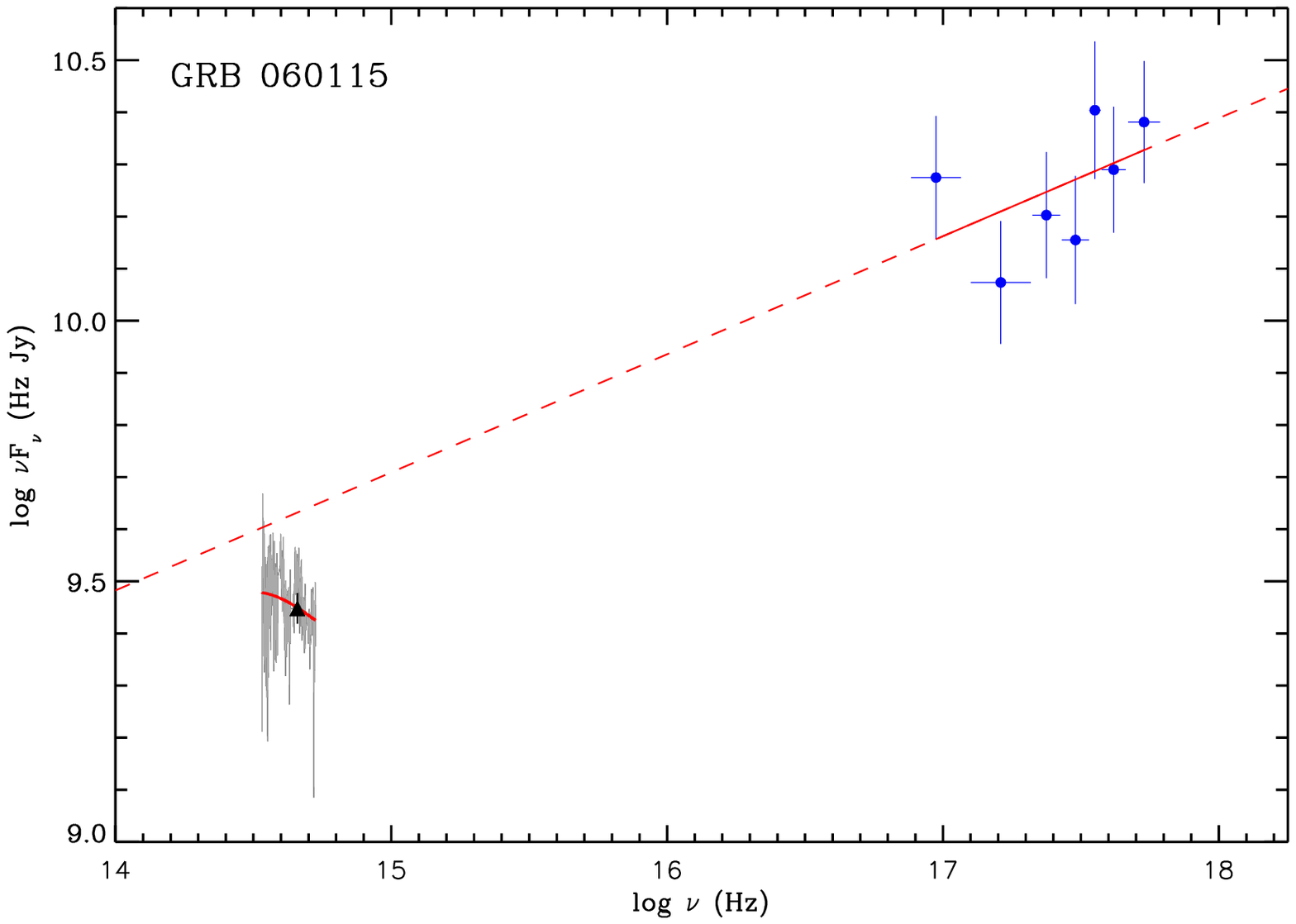}} \\
     {\includegraphics[width=0.9\columnwidth,clip=]{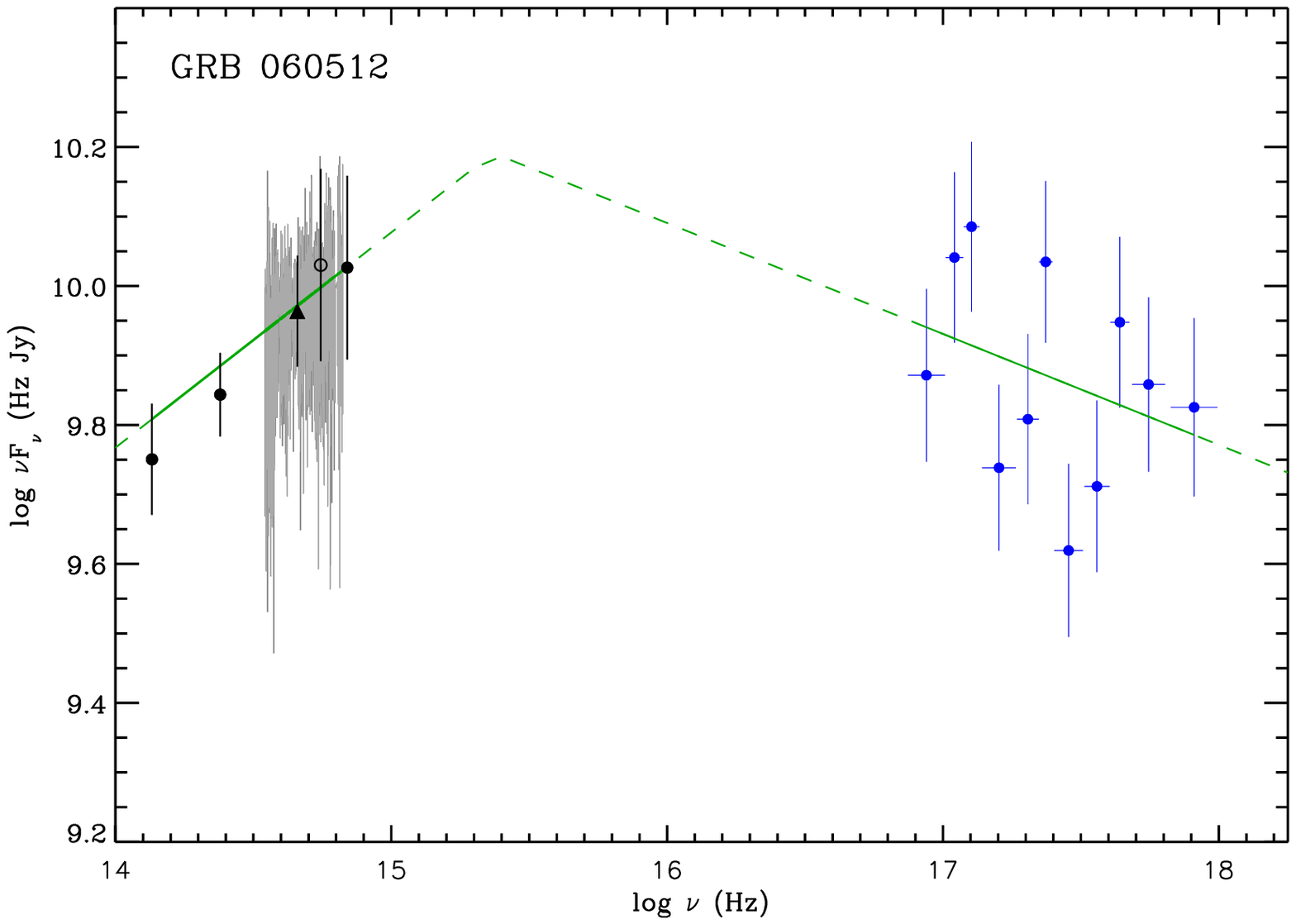}} &
            {\includegraphics[width=0.9\columnwidth,clip=]{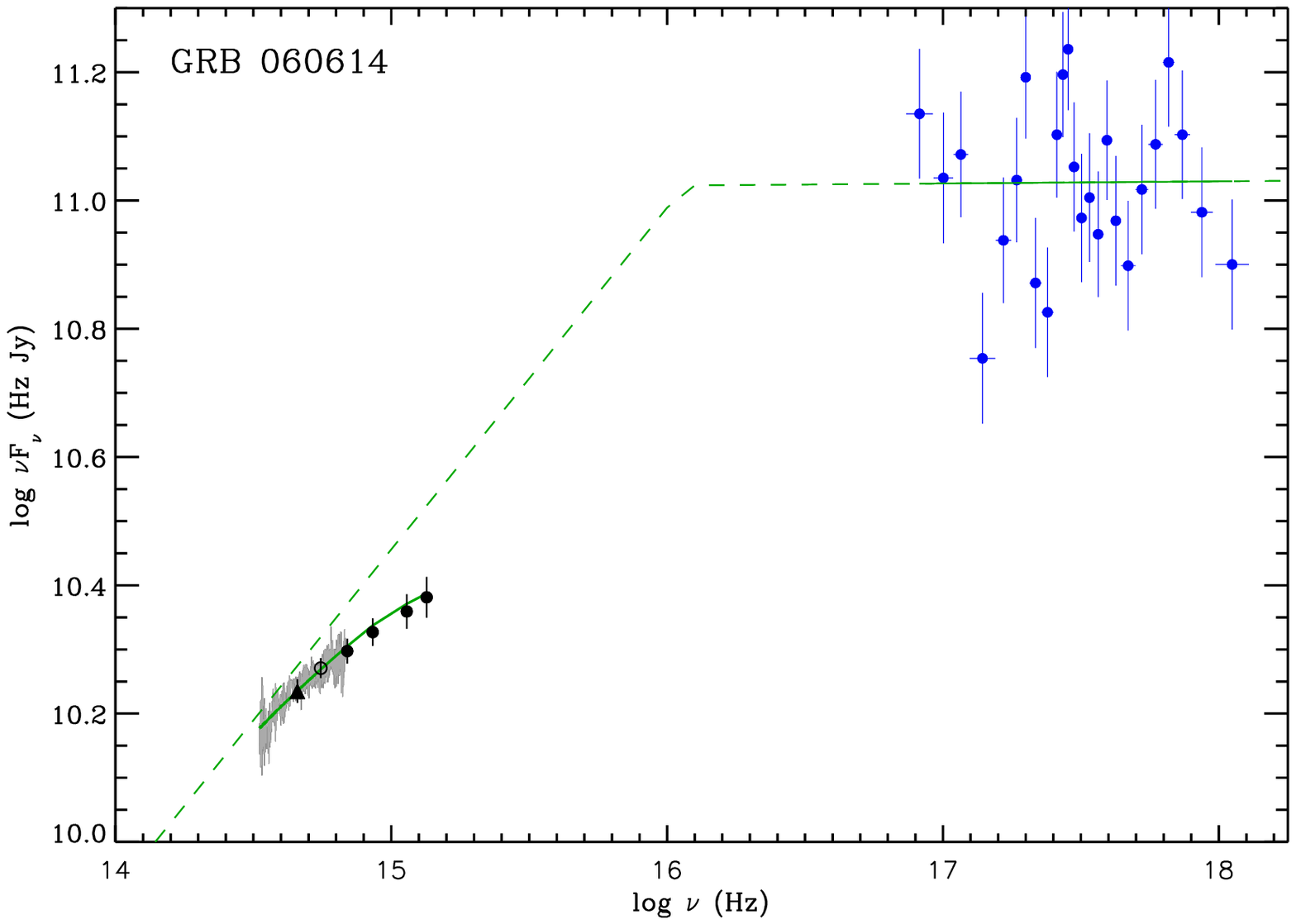}} \\
   {\includegraphics[width=0.9\columnwidth,clip=]{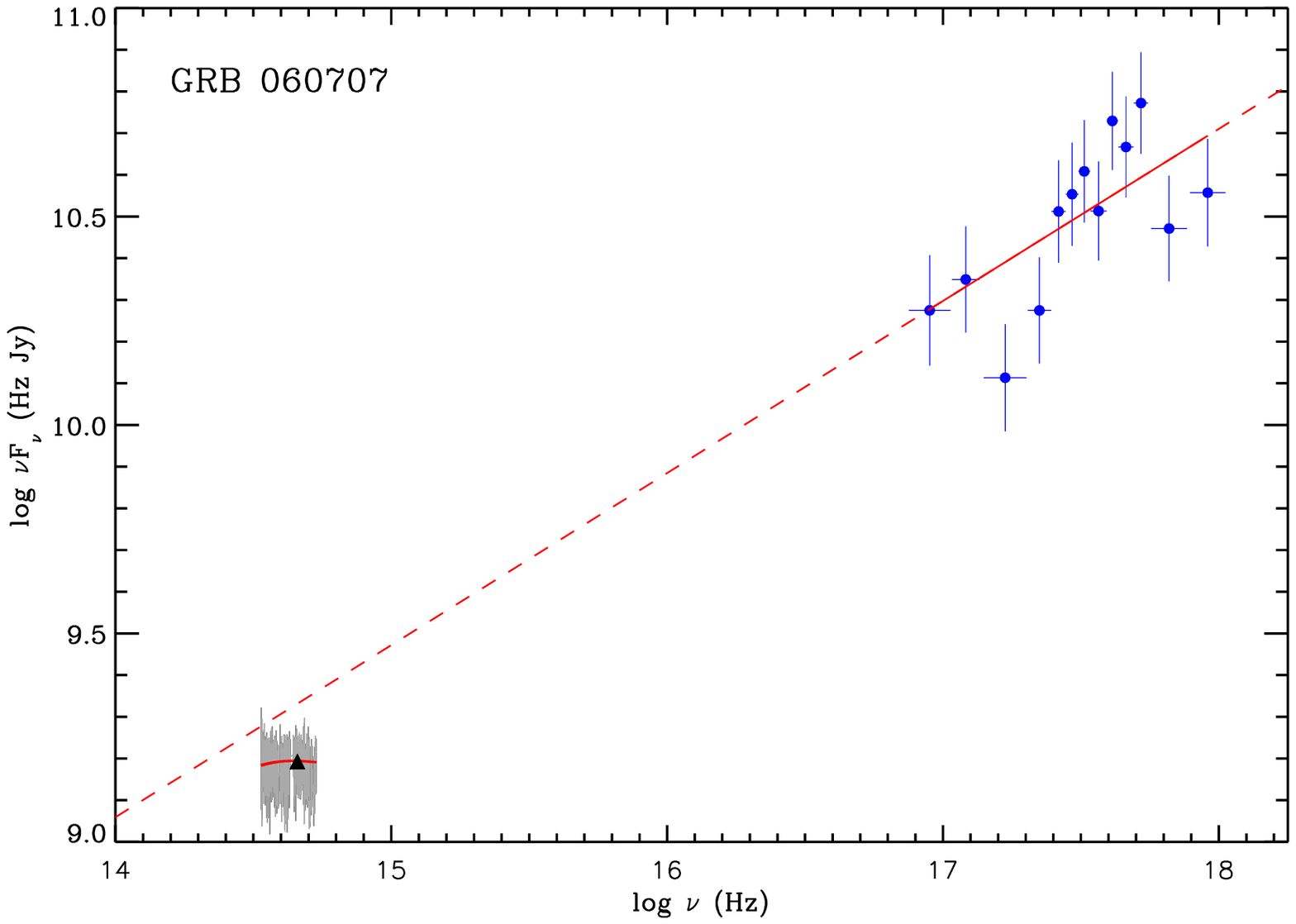}} &
      {\includegraphics[width=0.9\columnwidth,clip=]{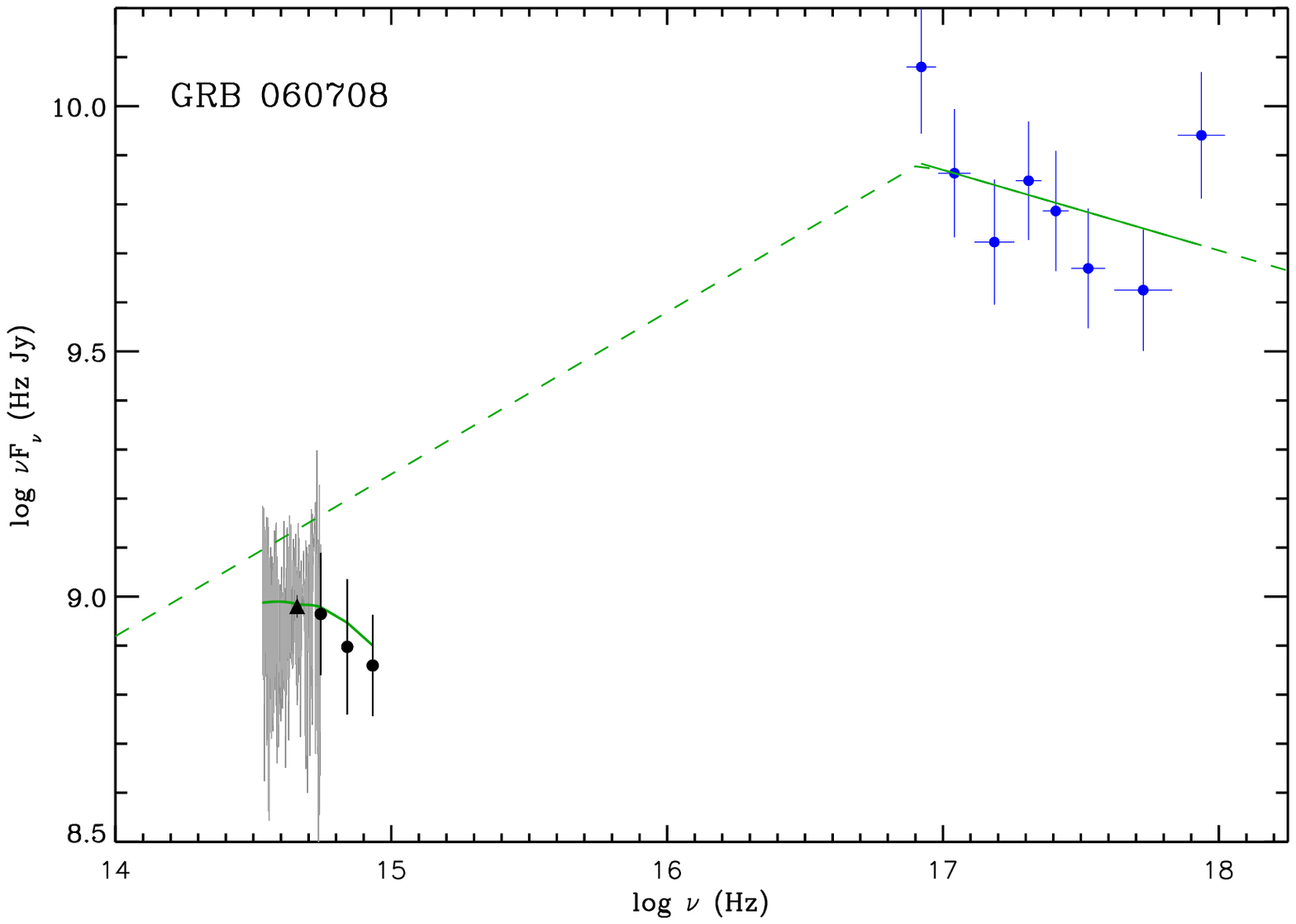}} \\
\end{tabular}
    \caption{Afterglow SEDs of the spectroscopic GRB sample in $\nu F_\nu$ and $\nu$ space. In each figure, we plot the best fit extinguished (solid lines) and extinction corrected spectral model (dashed lines). The grey curve represents the optical spectrum. The X-ray spectra are indicated by blue points. The black triangle corresponds to acquisition camera photometry used for scaling the SED. Black open circles are not included in the spectral fitting because of the optical spectrum wavelength coverage in that region while solid circles represent the data points included in the SED modeling. The red curve shows the best fit single power-law model and the green curve illustrates the broken power-law model.}
\label{sed1}
\end{figure*}
   
\clearpage
\addtocounter{figure}{-1}
  \begin{figure*}
  \begin{tabular}{c c}
      {\includegraphics[width=0.9\columnwidth,clip=]{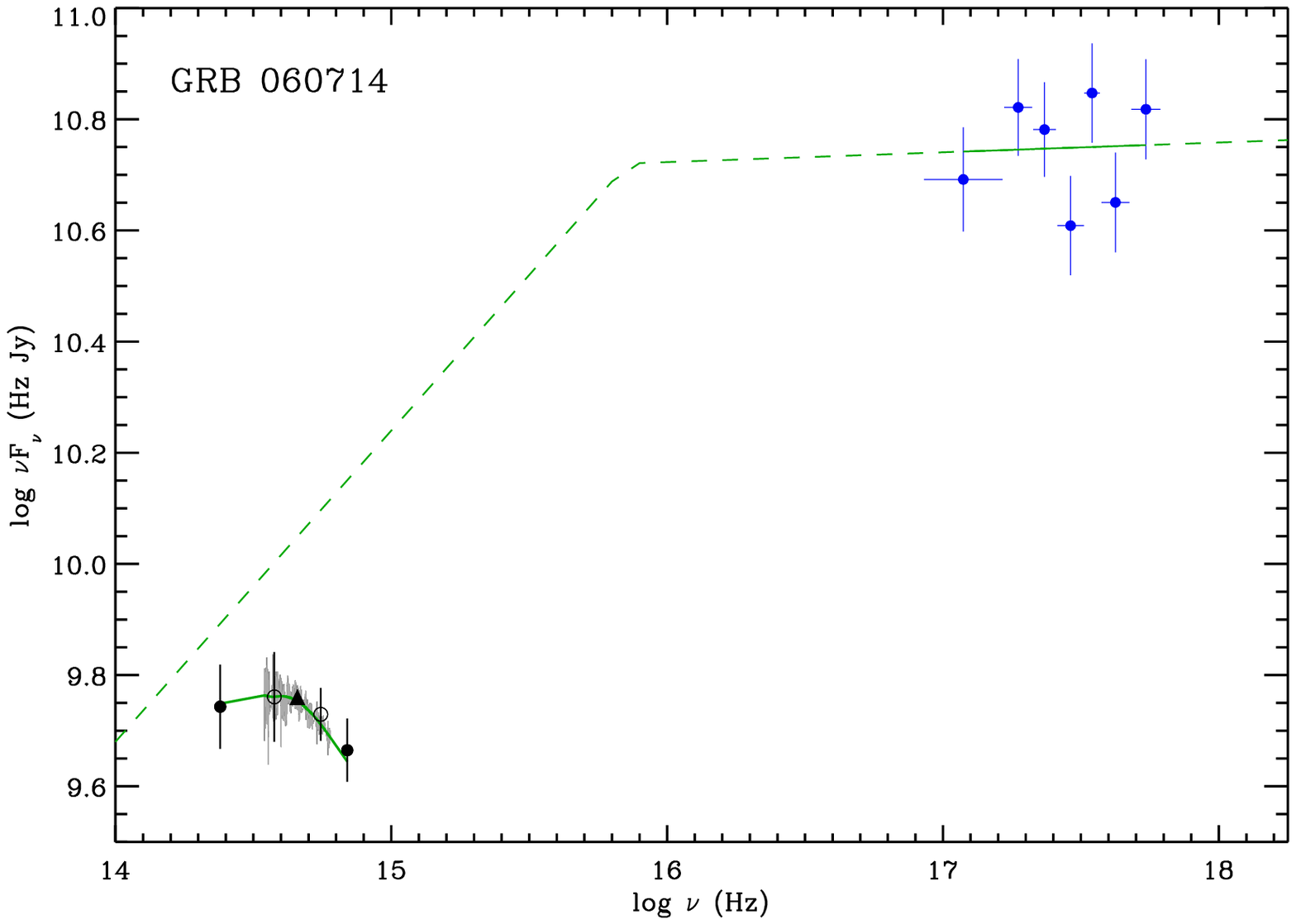}} &
         {\includegraphics[width=0.9\columnwidth,clip=]{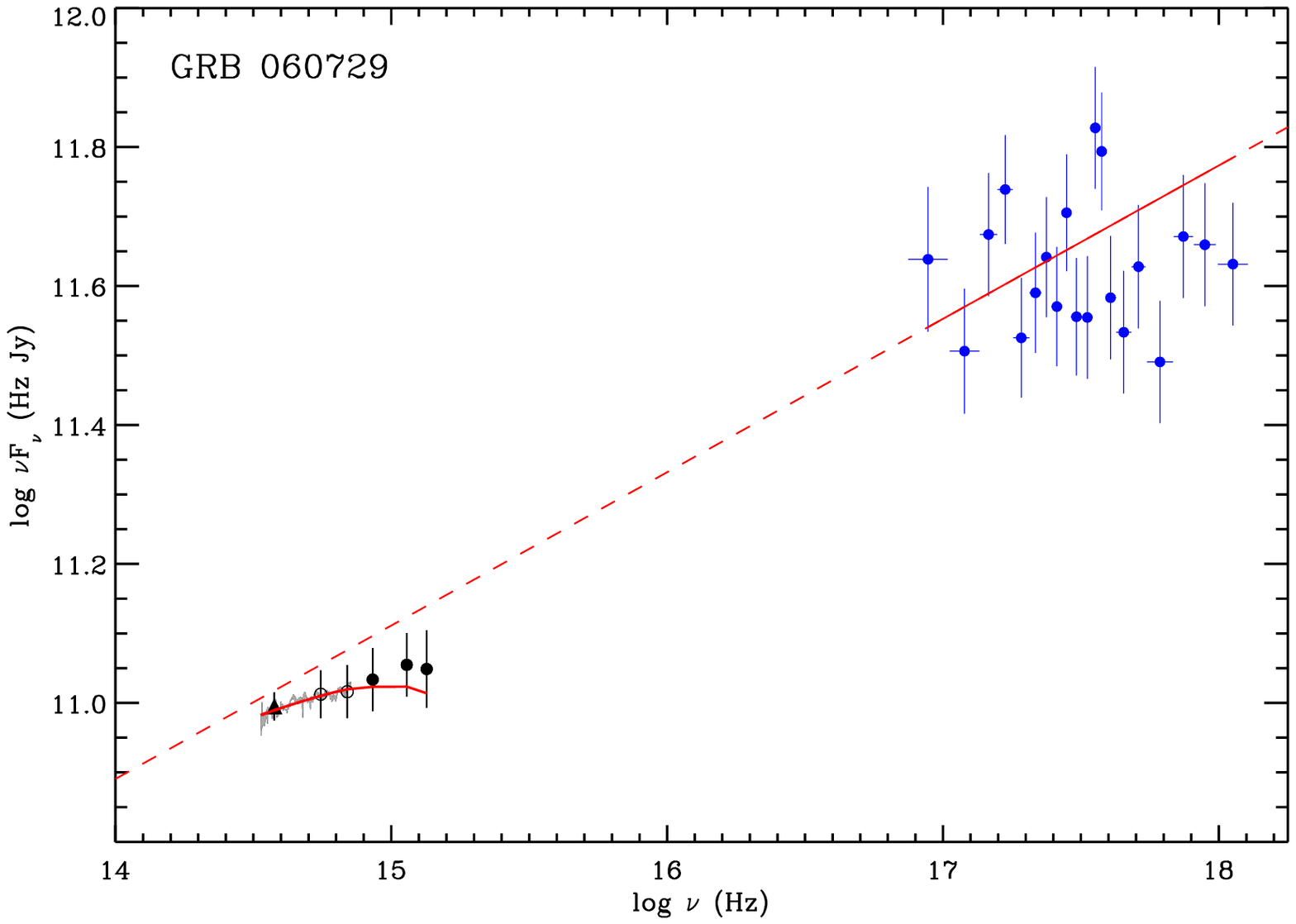}} \\
          {\includegraphics[width=0.9\columnwidth,clip=]{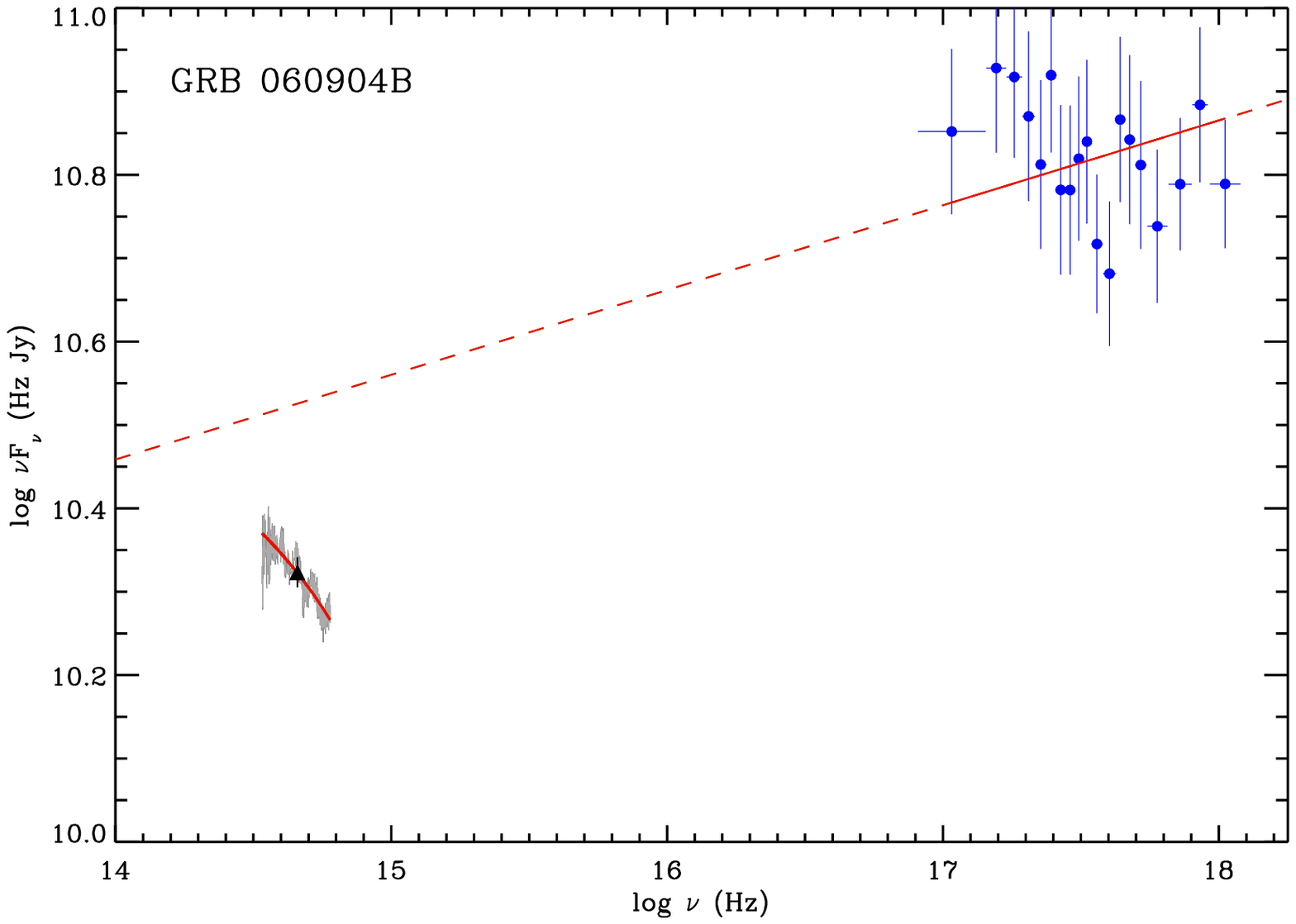}} &
               {\includegraphics[width=0.9\columnwidth,clip=]{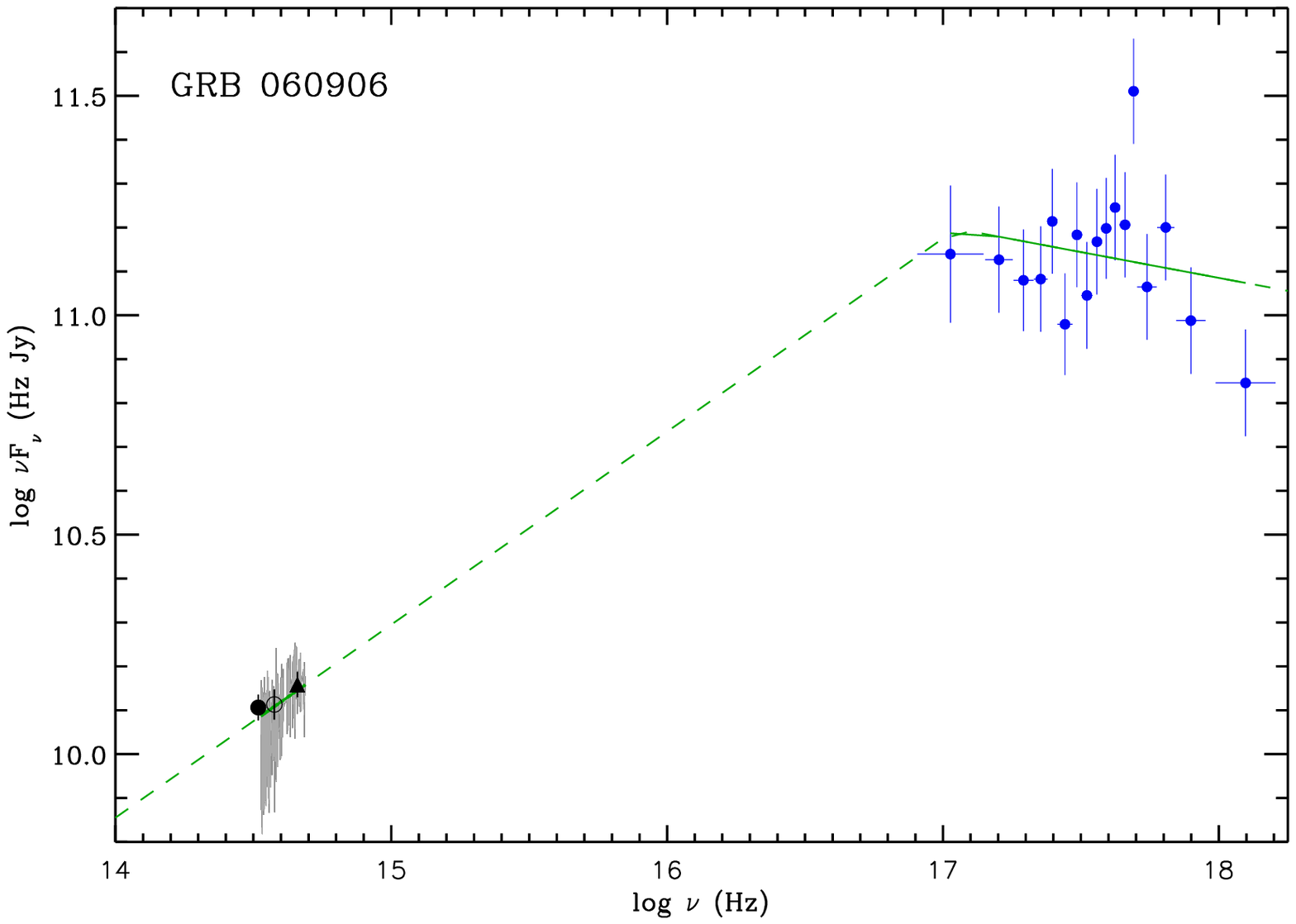}} \\
             {\includegraphics[width=0.9\columnwidth,clip=]{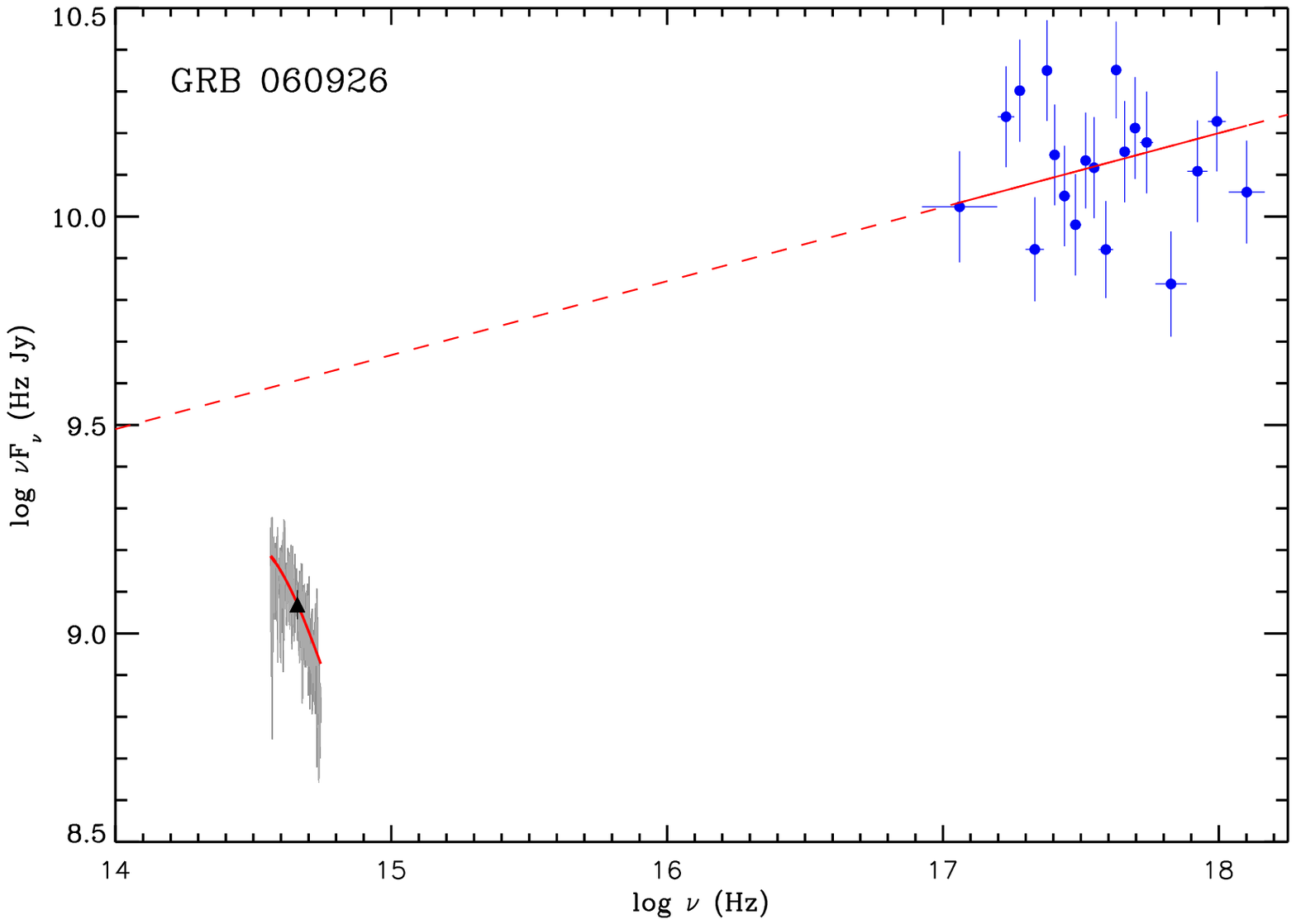}} &
             {\includegraphics[width=0.9\columnwidth,clip=]{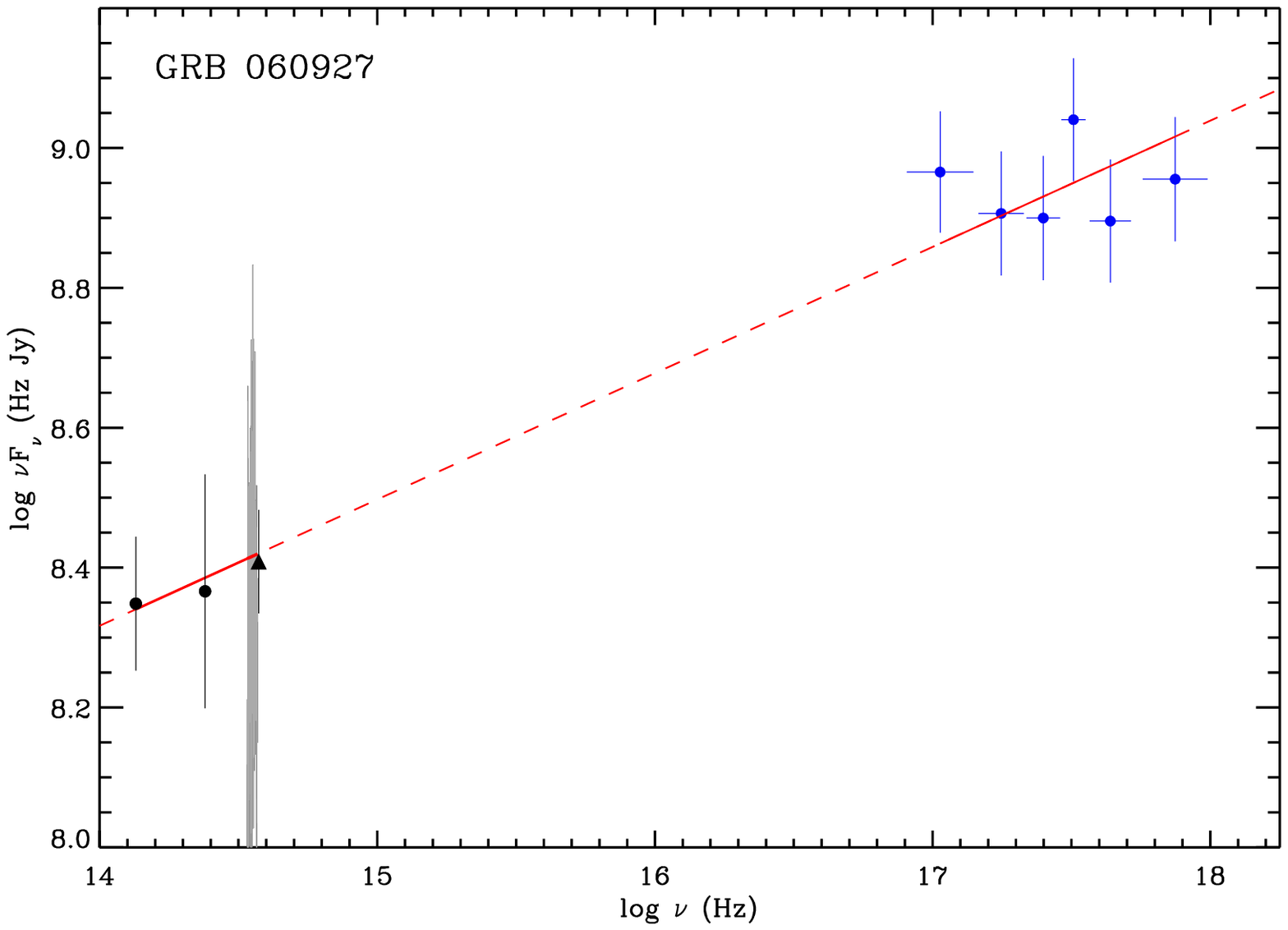}} \\
                  {\includegraphics[width=0.9\columnwidth,clip=]{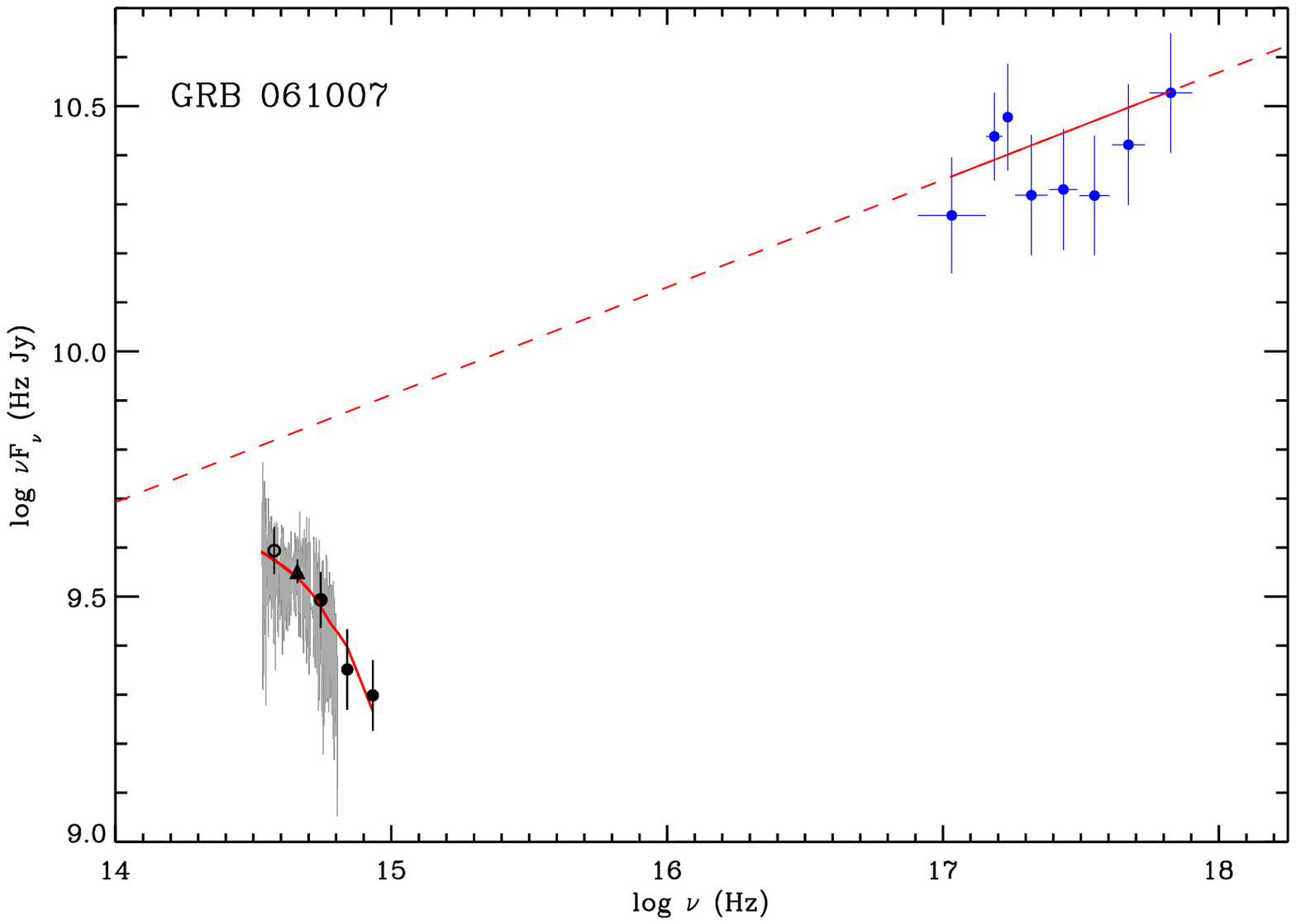}} & 
                    {\includegraphics[width=0.9\columnwidth,clip=]{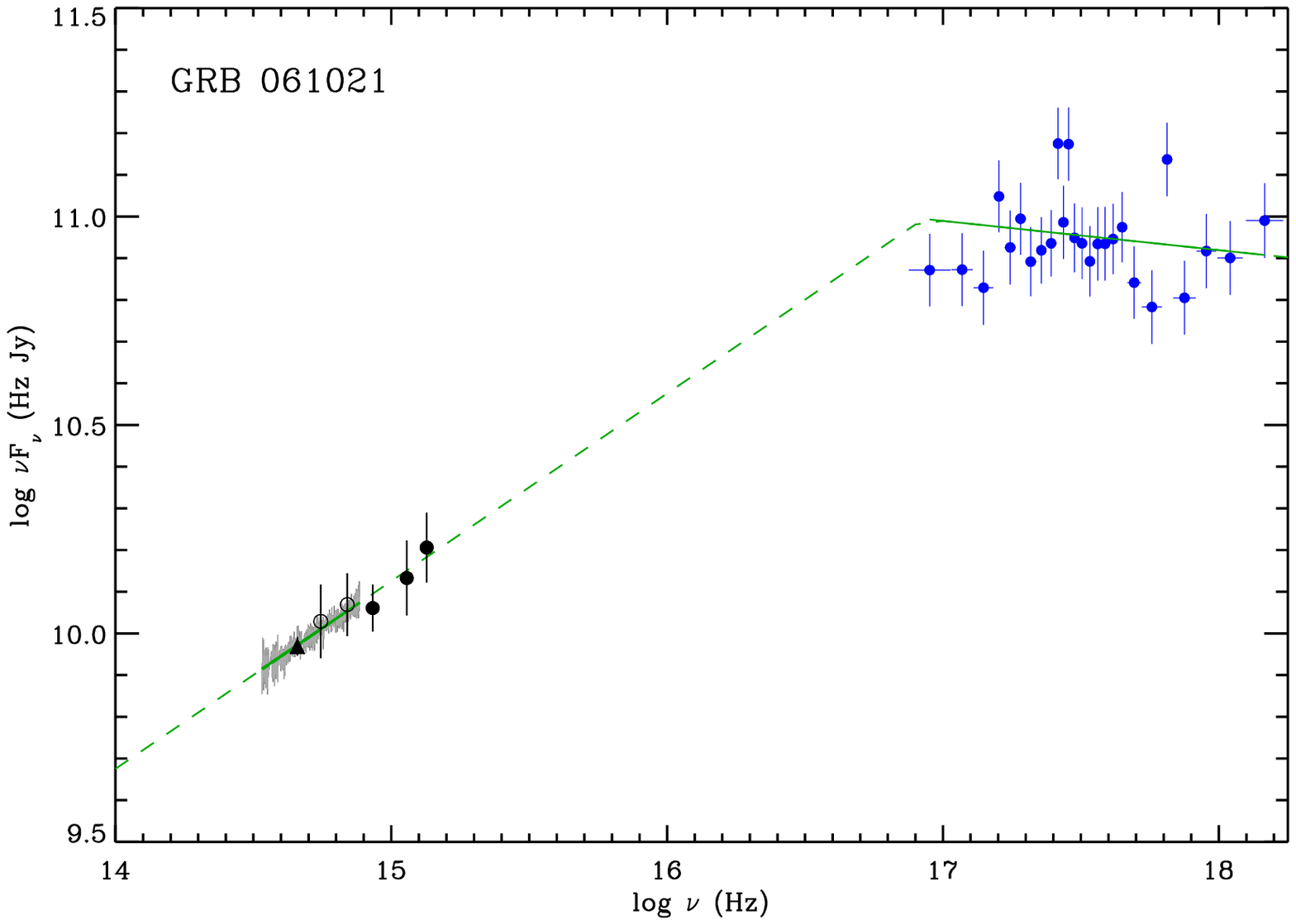}} \\
\end{tabular}
\caption{(continued)}
   \end{figure*}
 \clearpage

\addtocounter{figure}{-1}
  \begin{figure*}
  \begin{tabular}{c c}
        {\includegraphics[width=0.9\columnwidth,clip=]{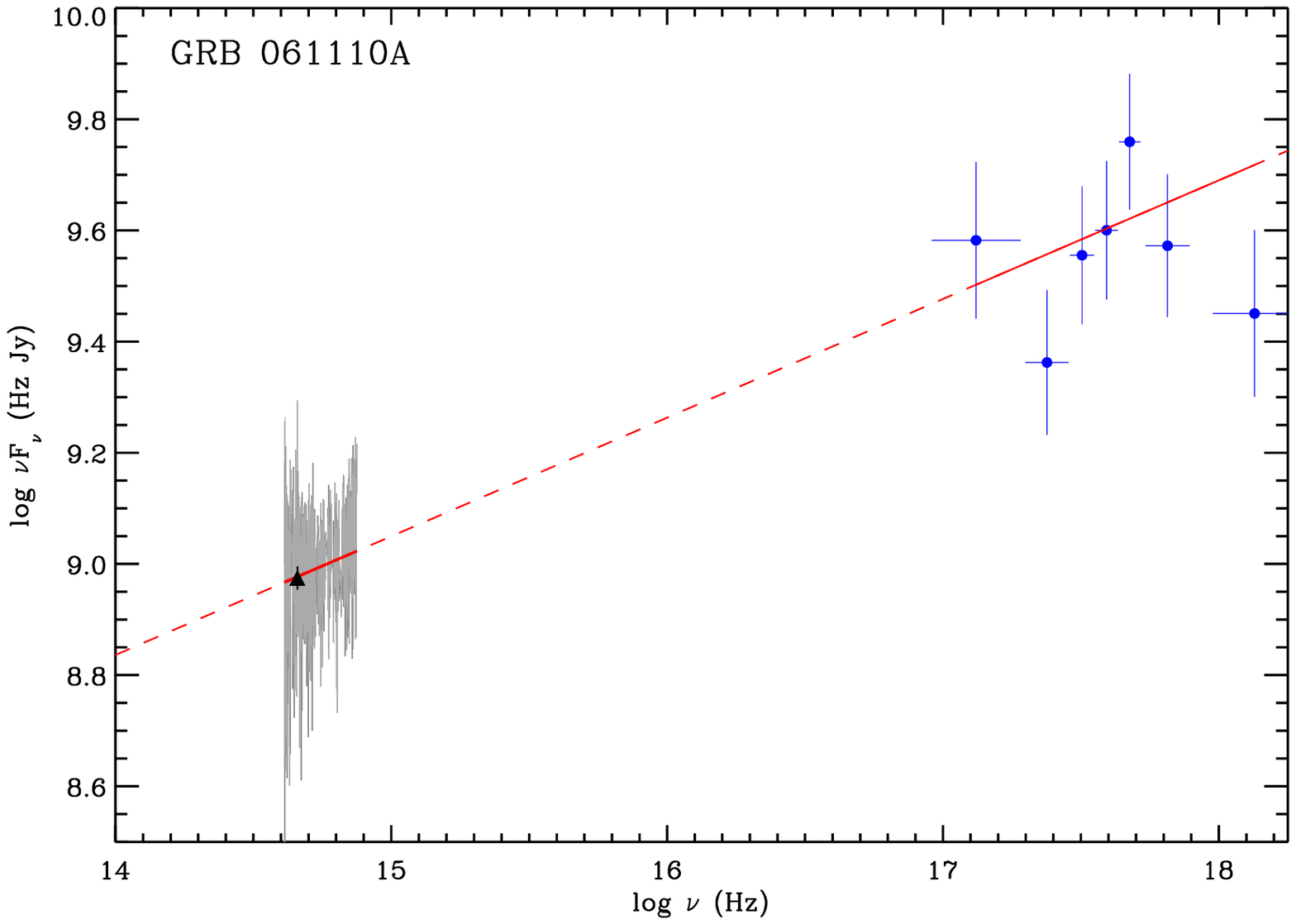}} &
       {\includegraphics[width=0.9\columnwidth,clip=]{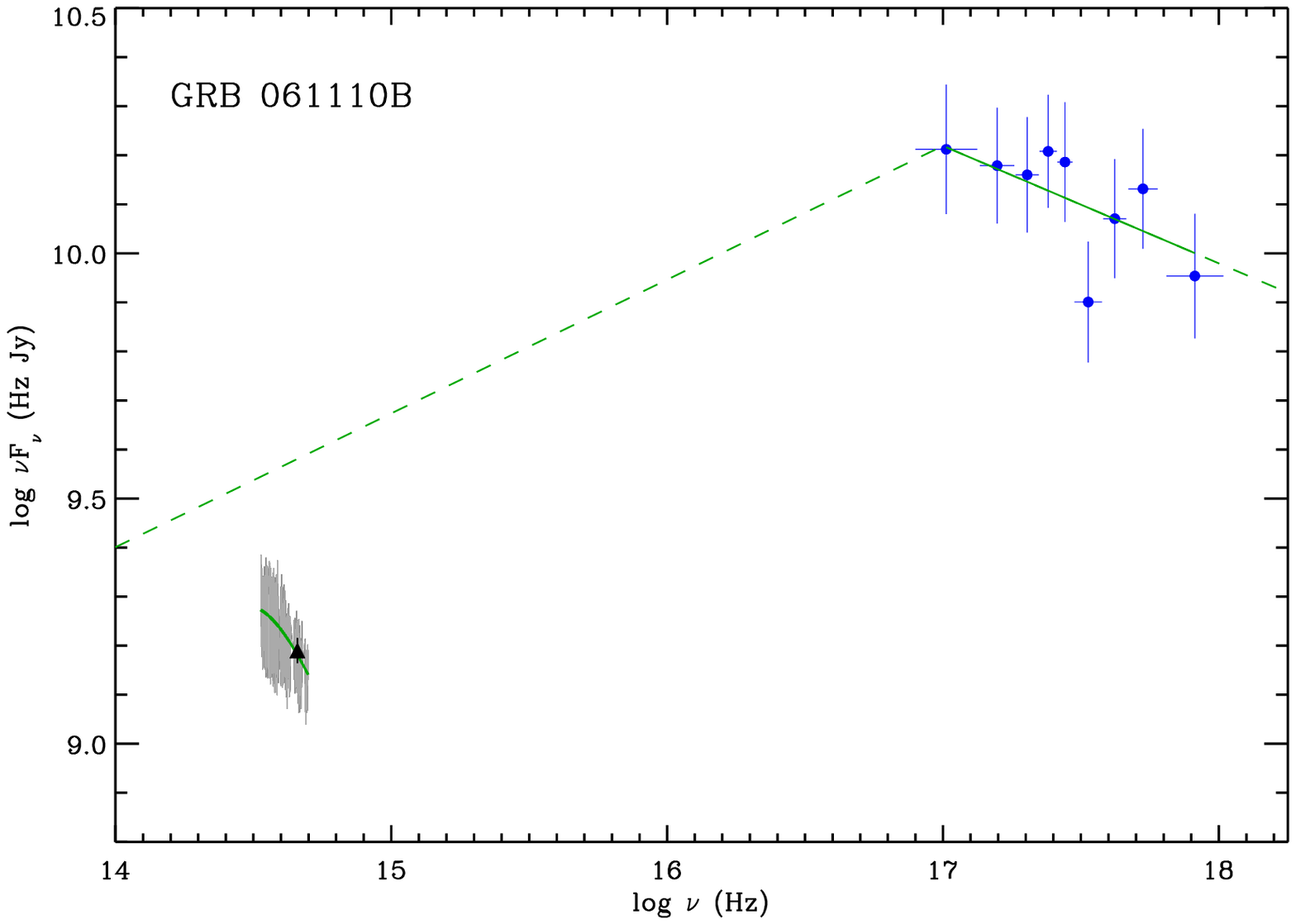}} \\ 
       {\includegraphics[width=0.9\columnwidth,clip=]{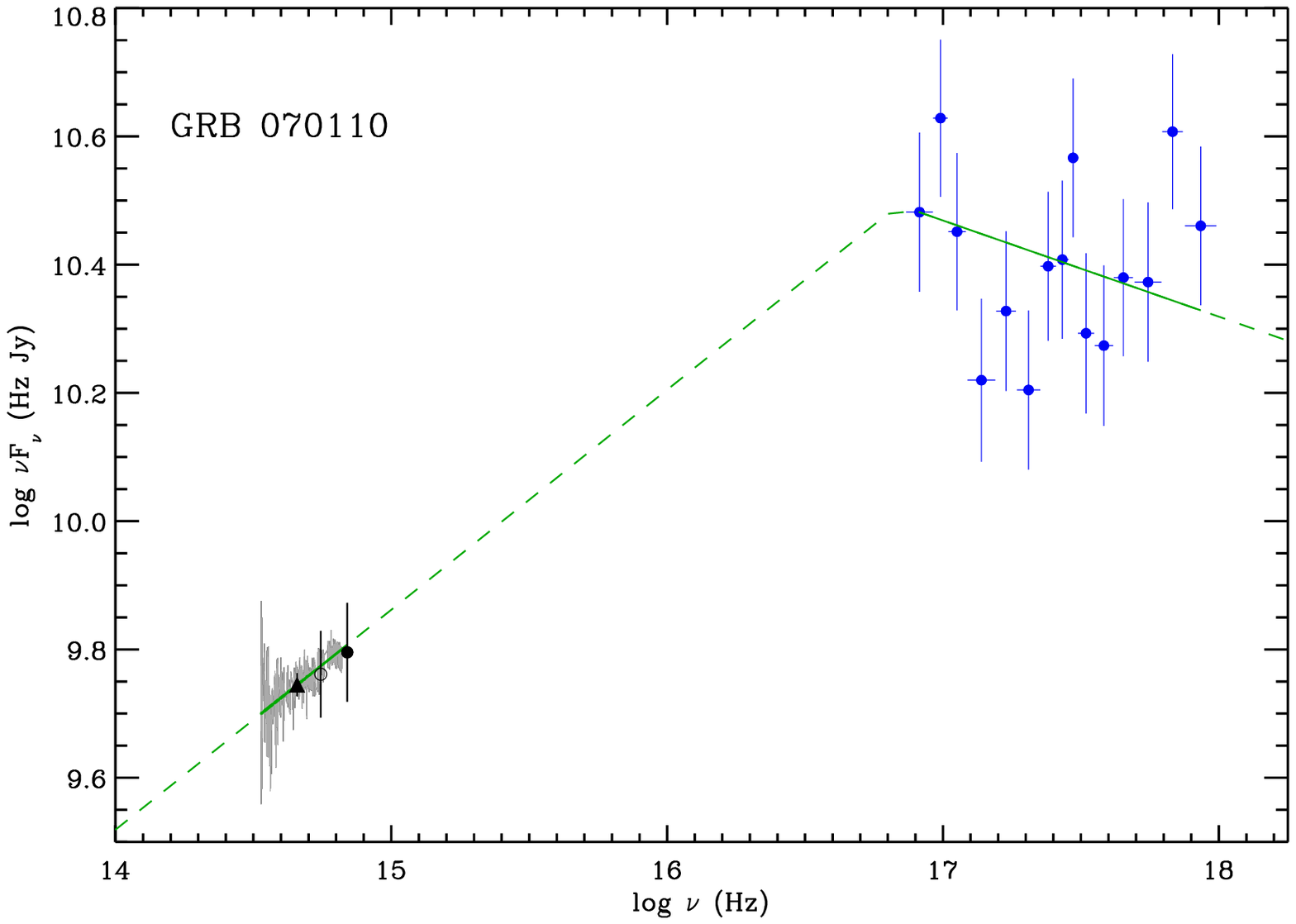}} &
   {\includegraphics[width=0.9\columnwidth,clip=]{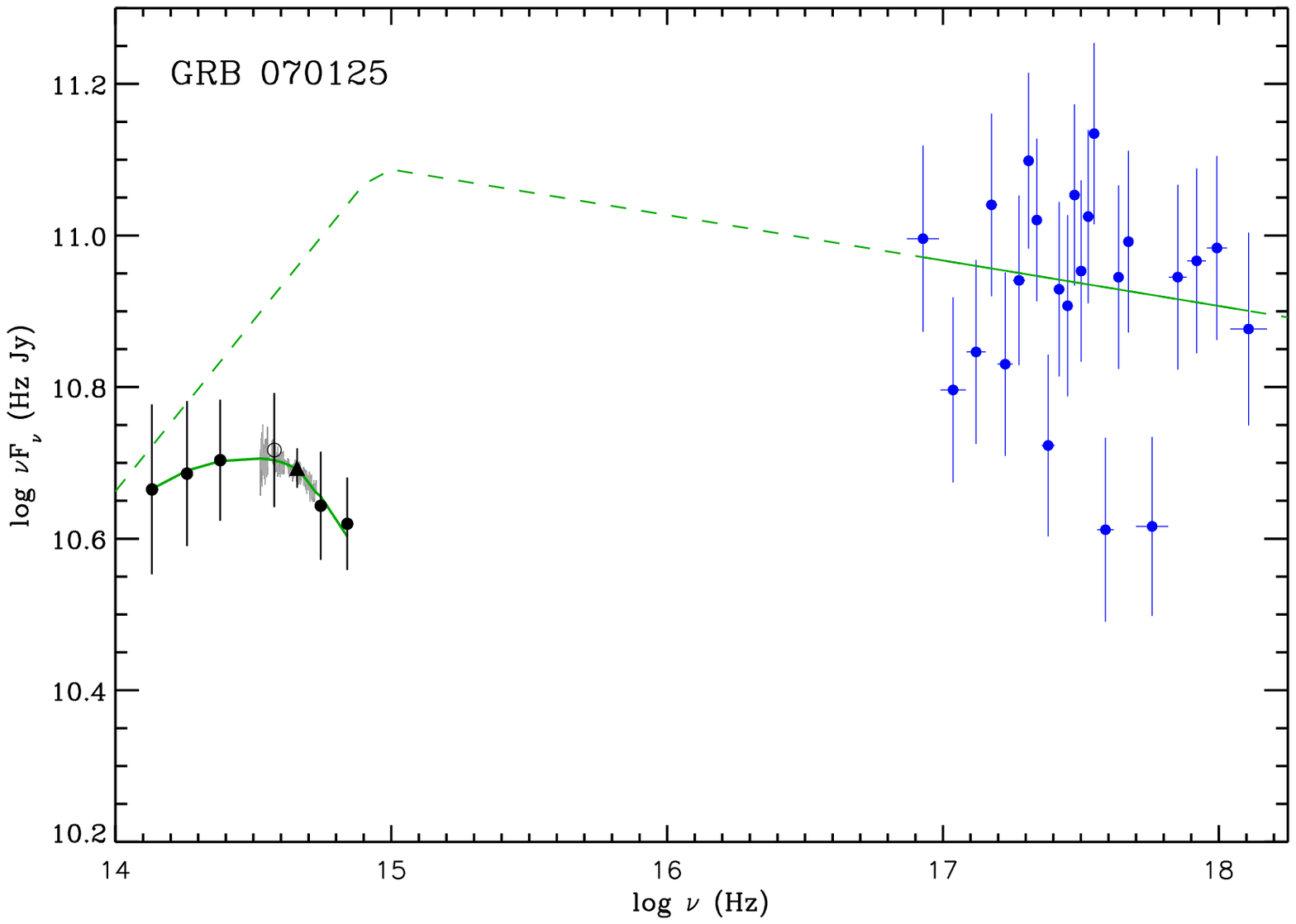}} \\
      {\includegraphics[width=0.9\columnwidth,clip=]{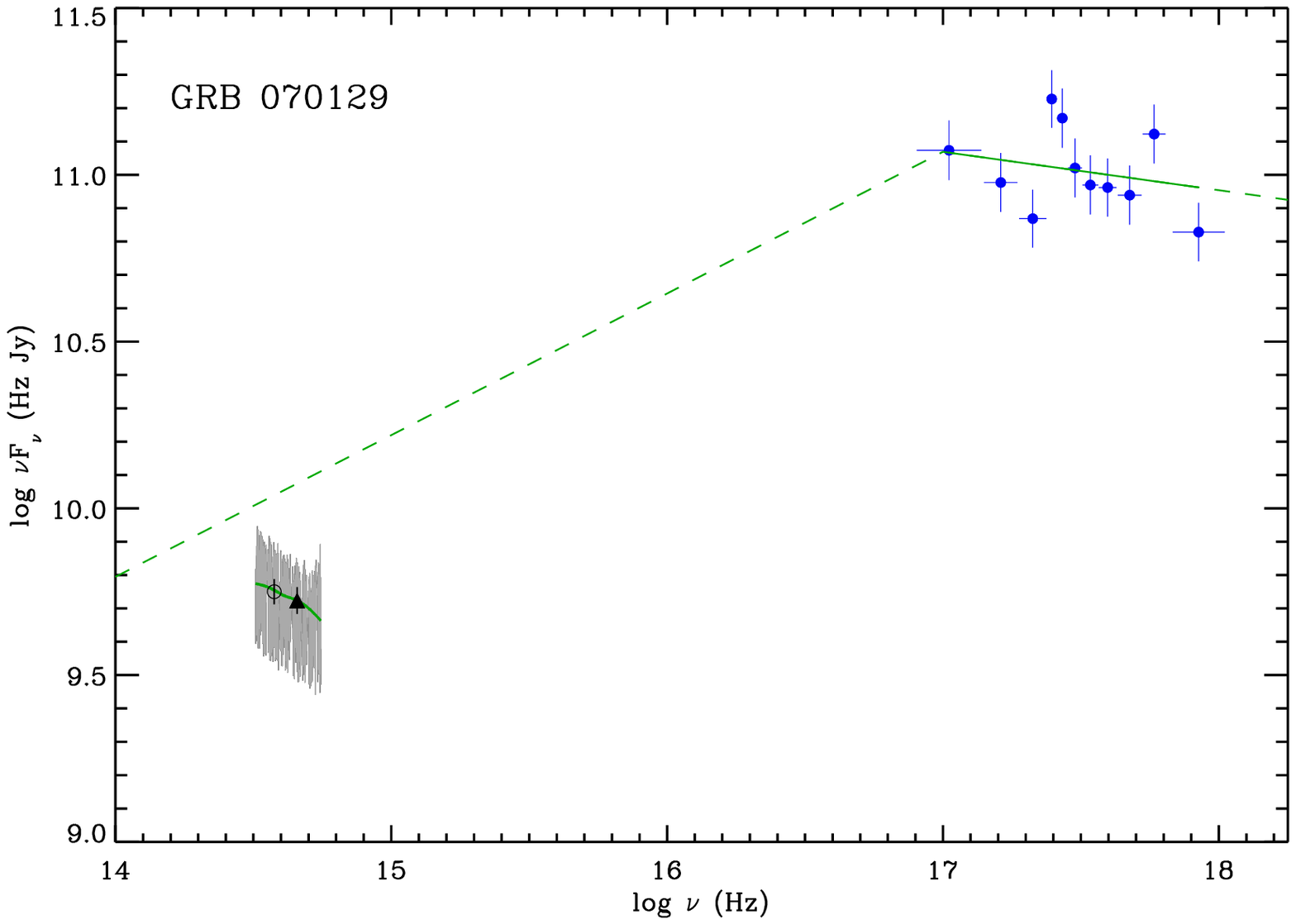}} &
     {\includegraphics[width=0.9\columnwidth,clip=]{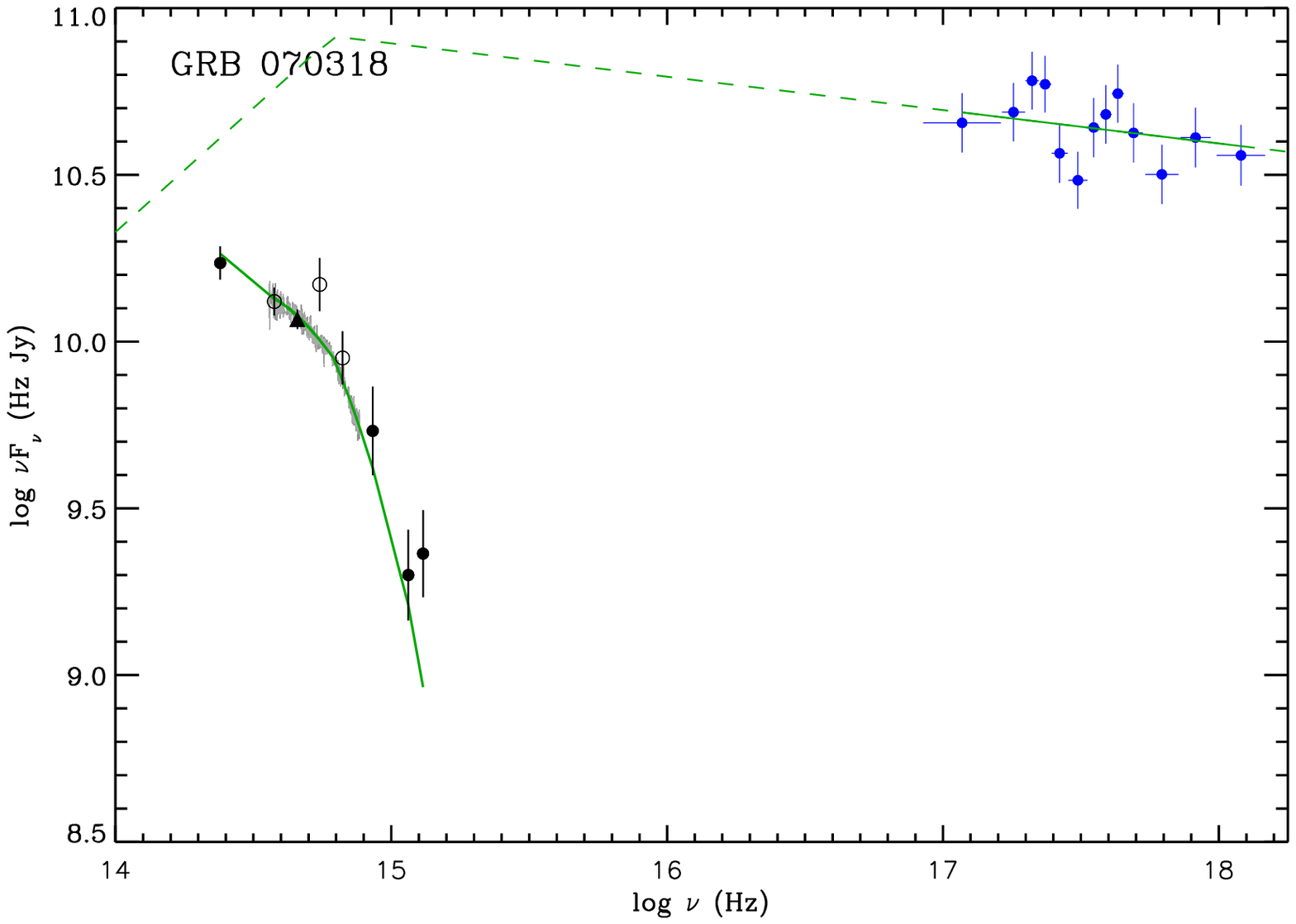}}  \\
          {\includegraphics[width=0.9\columnwidth,clip=]{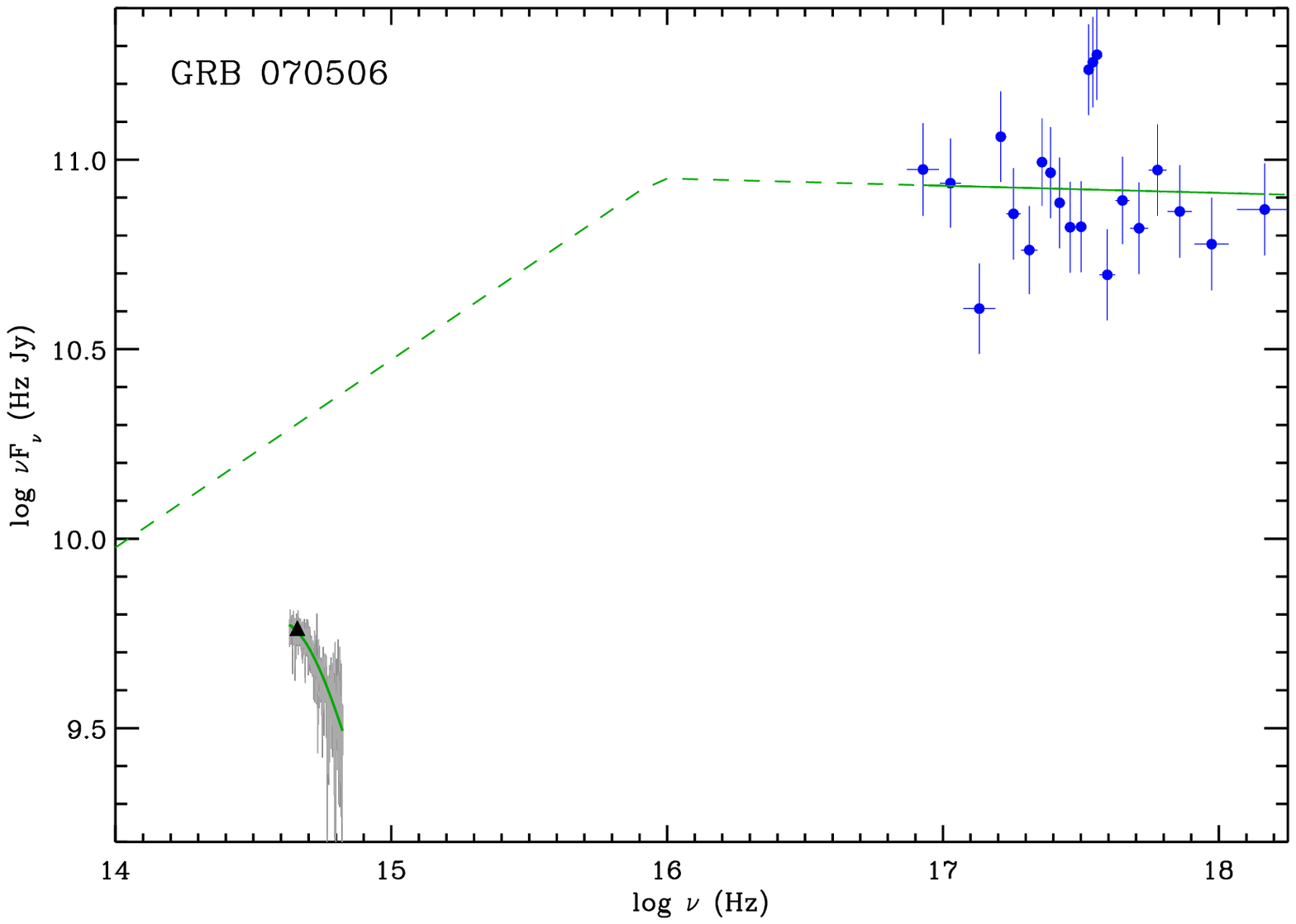}} &
          {\includegraphics[width=0.9\columnwidth,clip=]{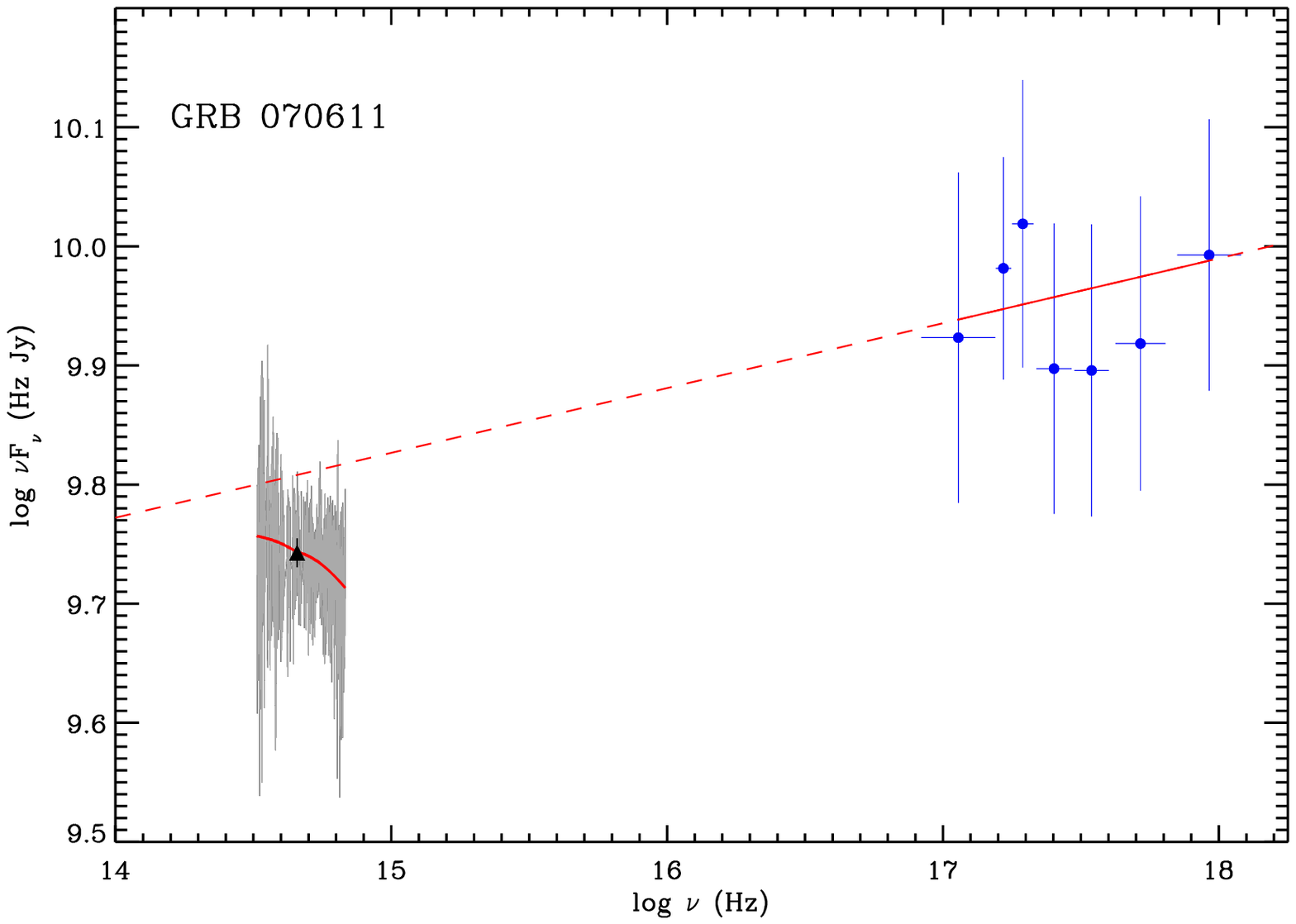}} \\
\end{tabular}
\caption{(continued)}
   \end{figure*}
   \clearpage
  
\addtocounter{figure}{-1}
  \begin{figure*}
  \begin{tabular}{c c}
   {\includegraphics[width=0.9\columnwidth,clip=]{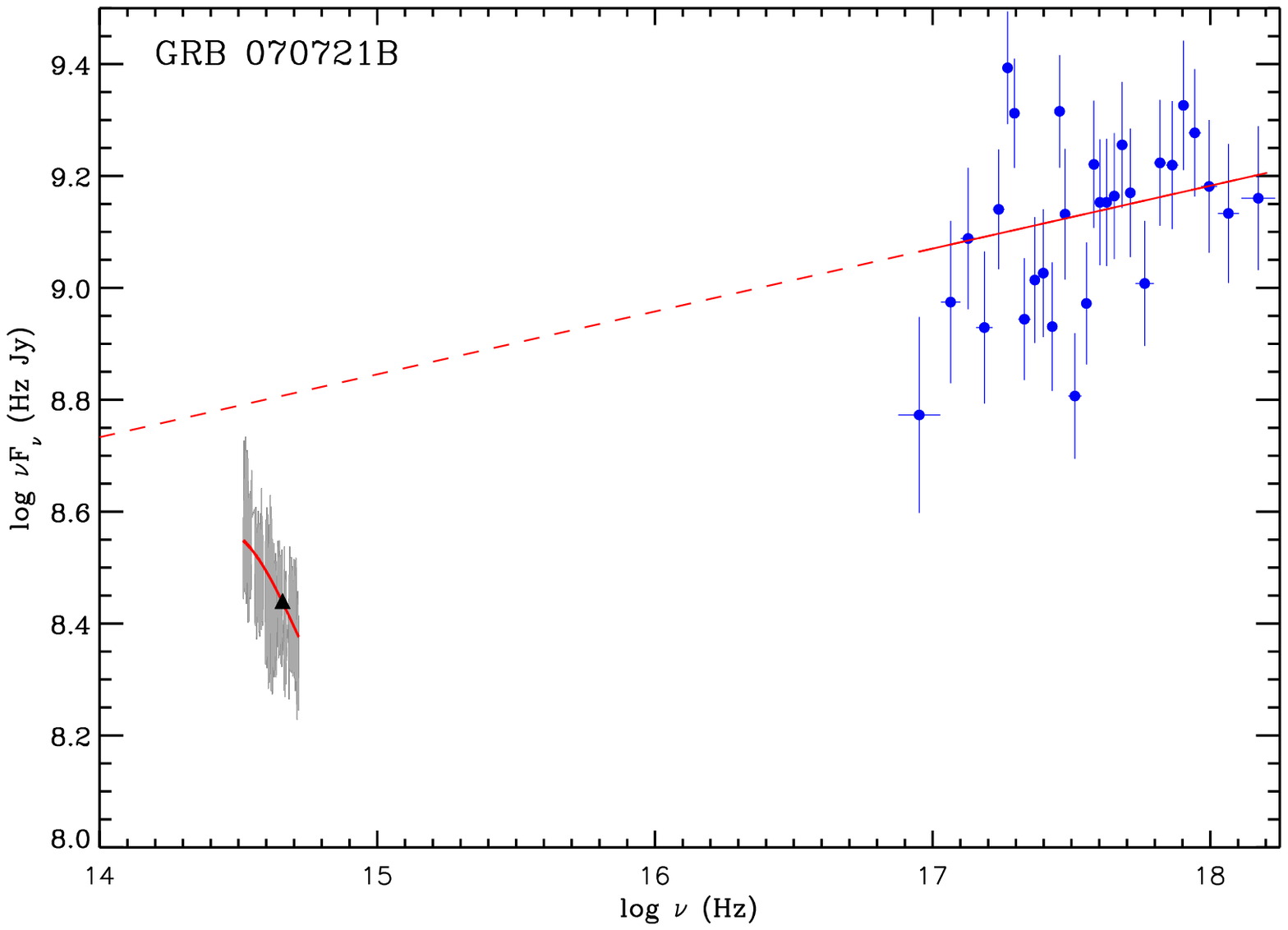}} &
   {\includegraphics[width=0.9\columnwidth,clip=]{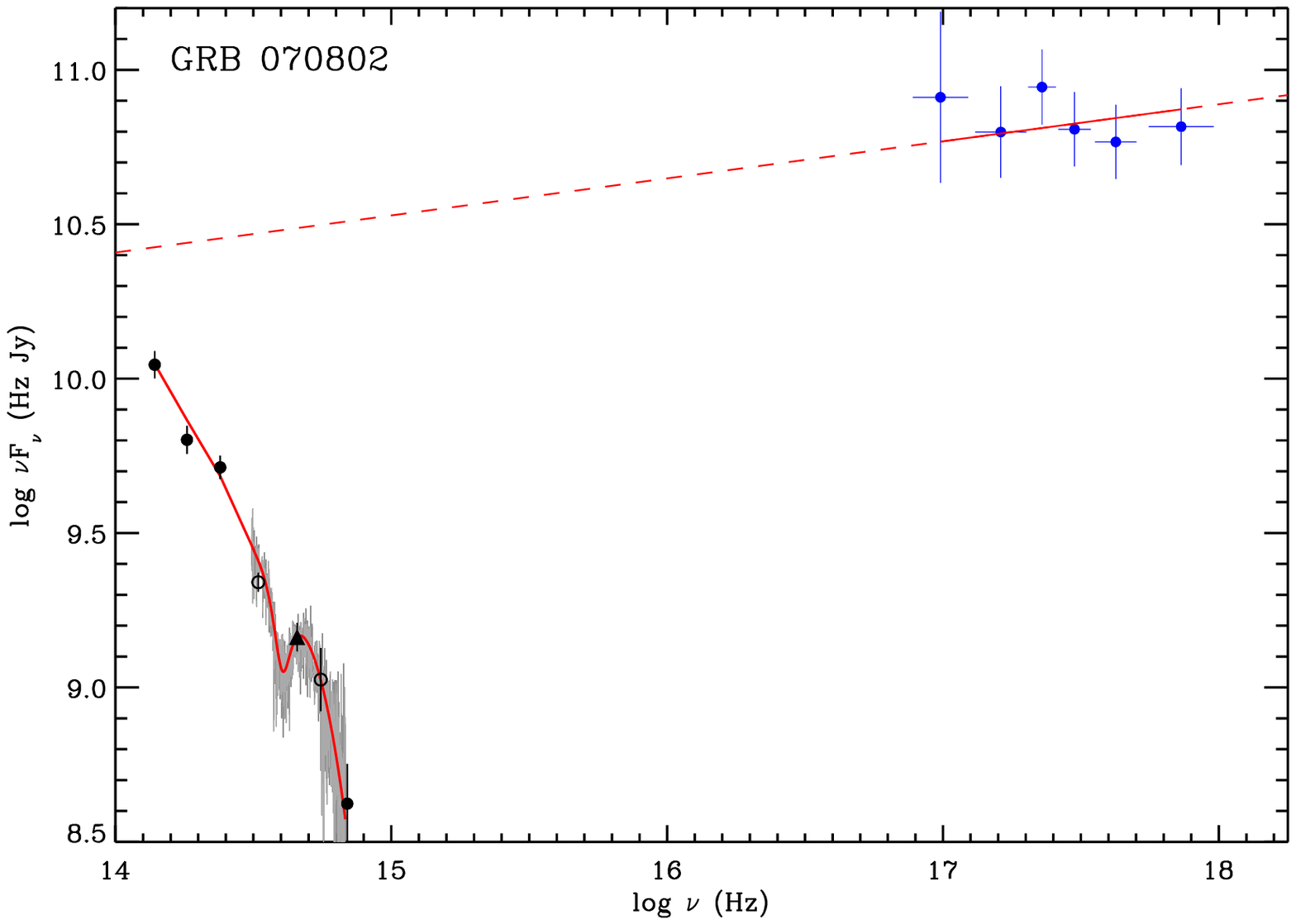}}  \\
   {\includegraphics[width=0.9\columnwidth,clip=]{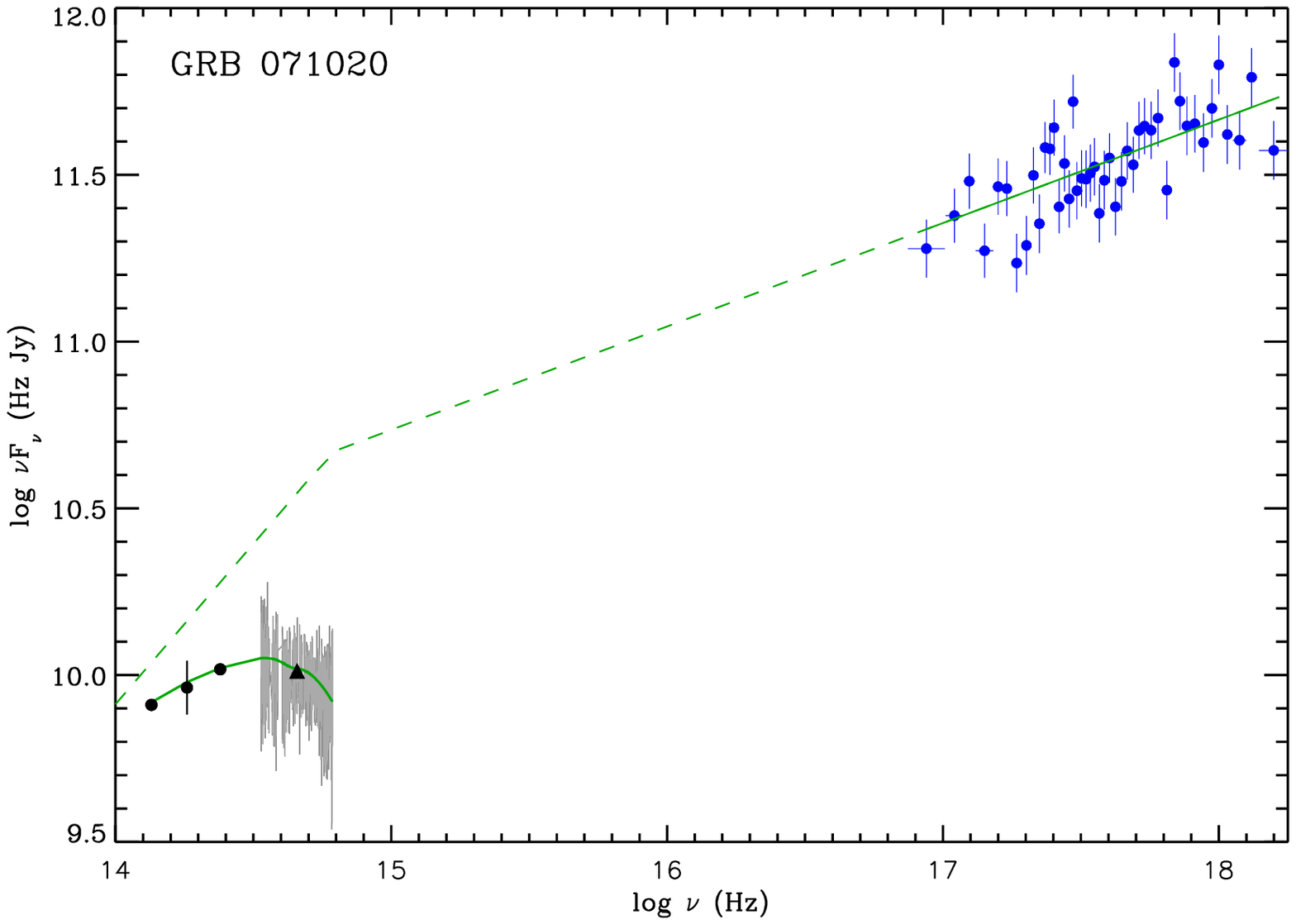}} &
   {\includegraphics[width=0.9\columnwidth,clip=]{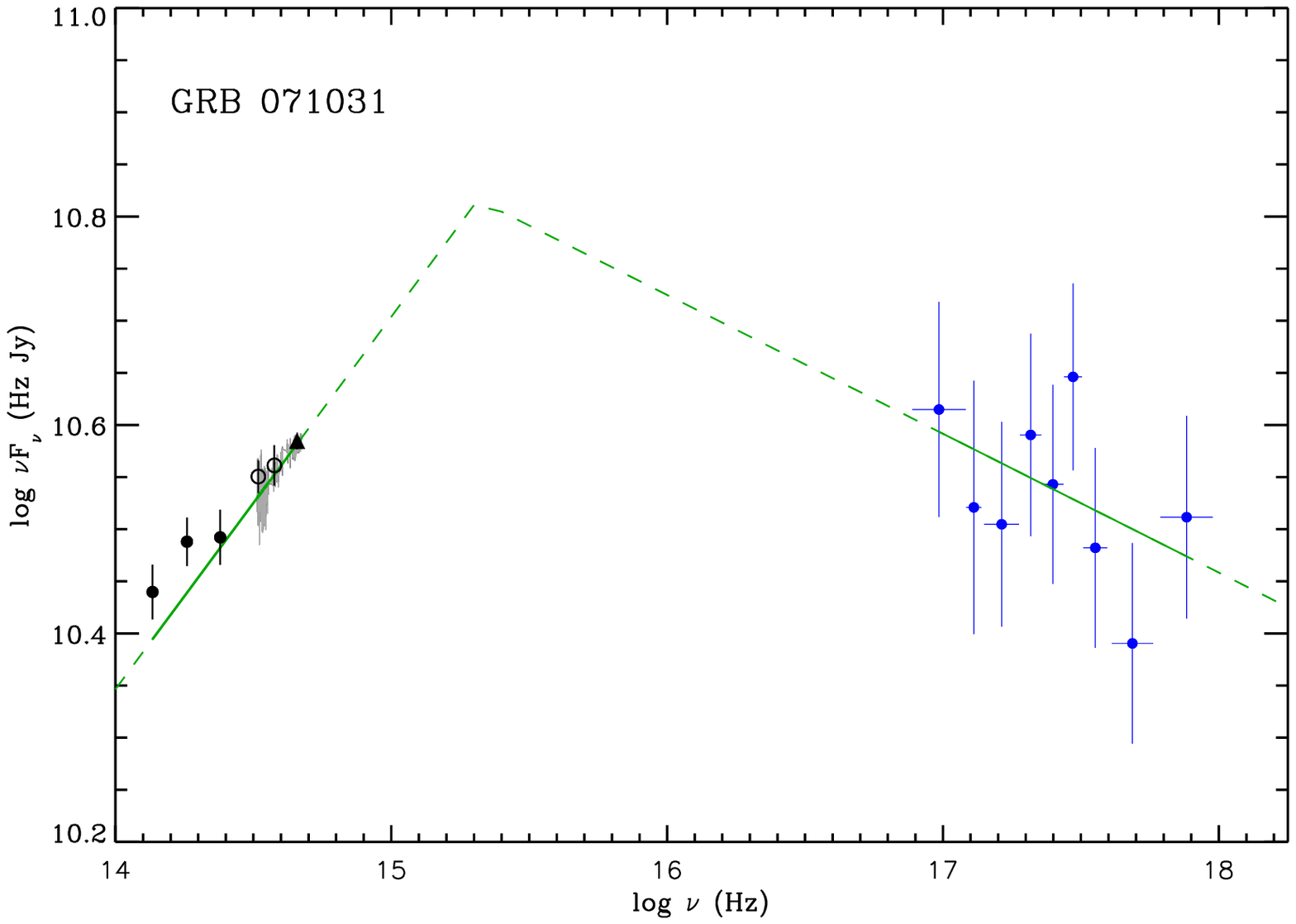}} \\
   {\includegraphics[width=0.9\columnwidth,clip=]{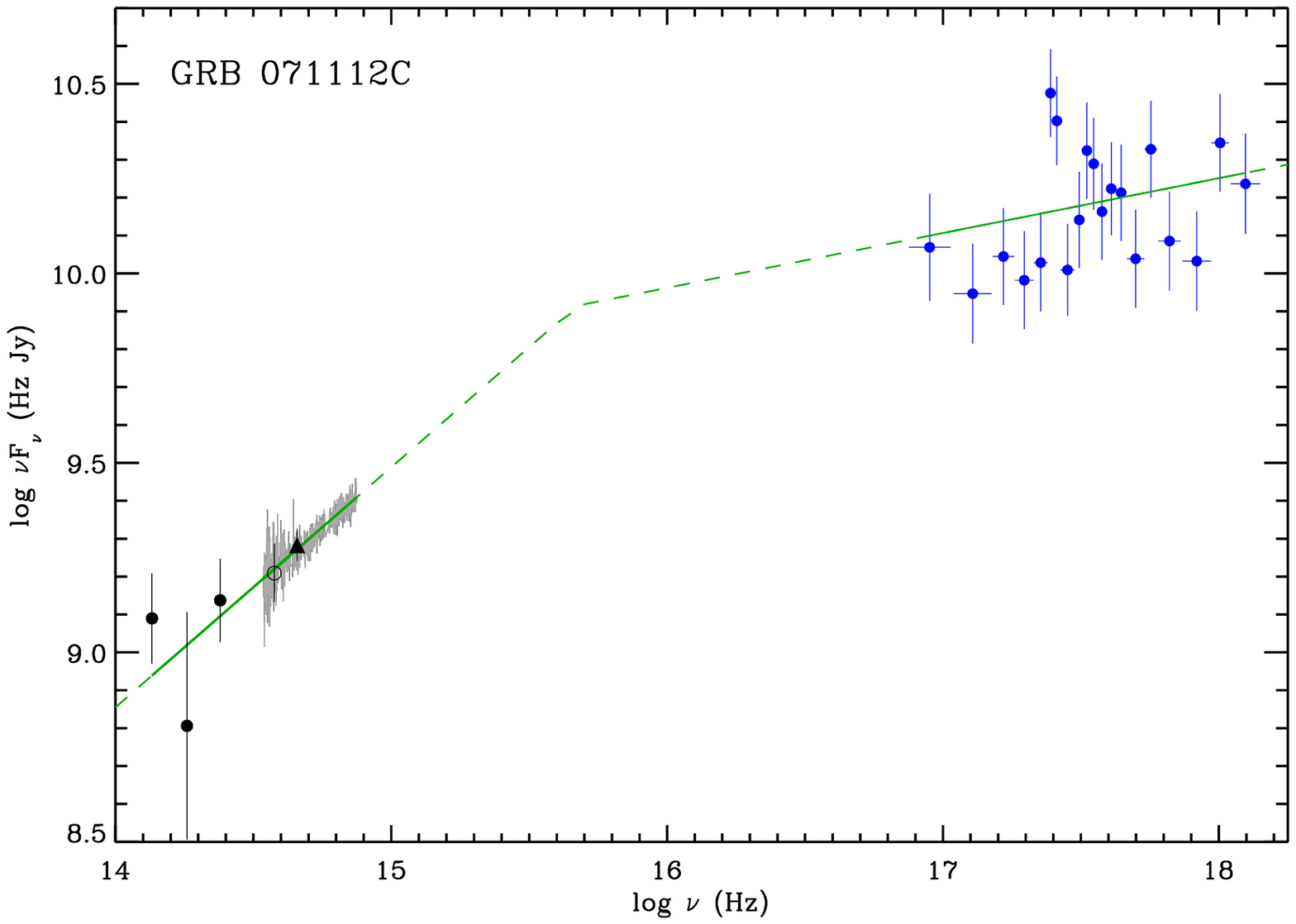}} &
   {\includegraphics[width=0.9\columnwidth,clip=]{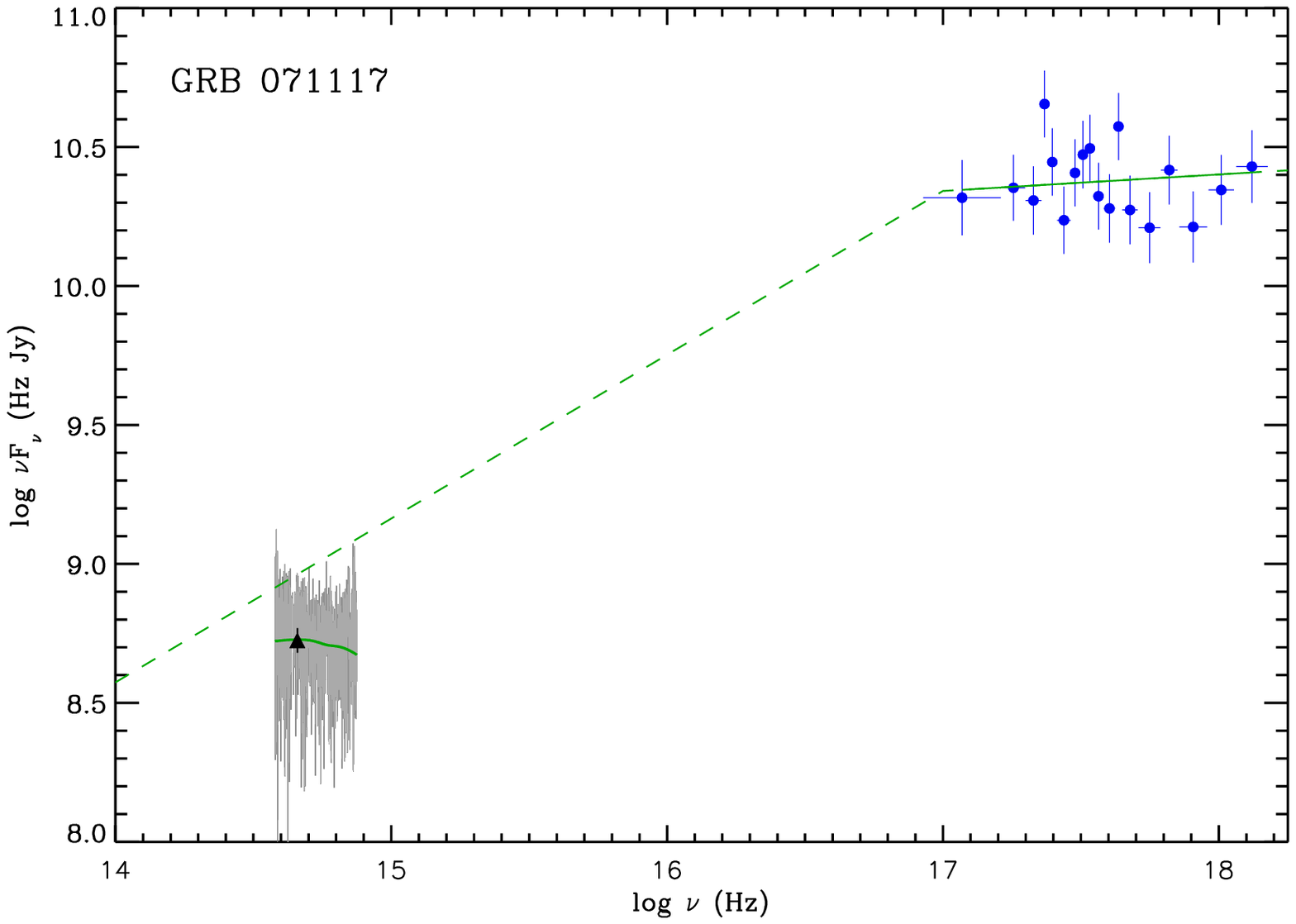}}  \\
      {\includegraphics[width=0.9\columnwidth,clip=]{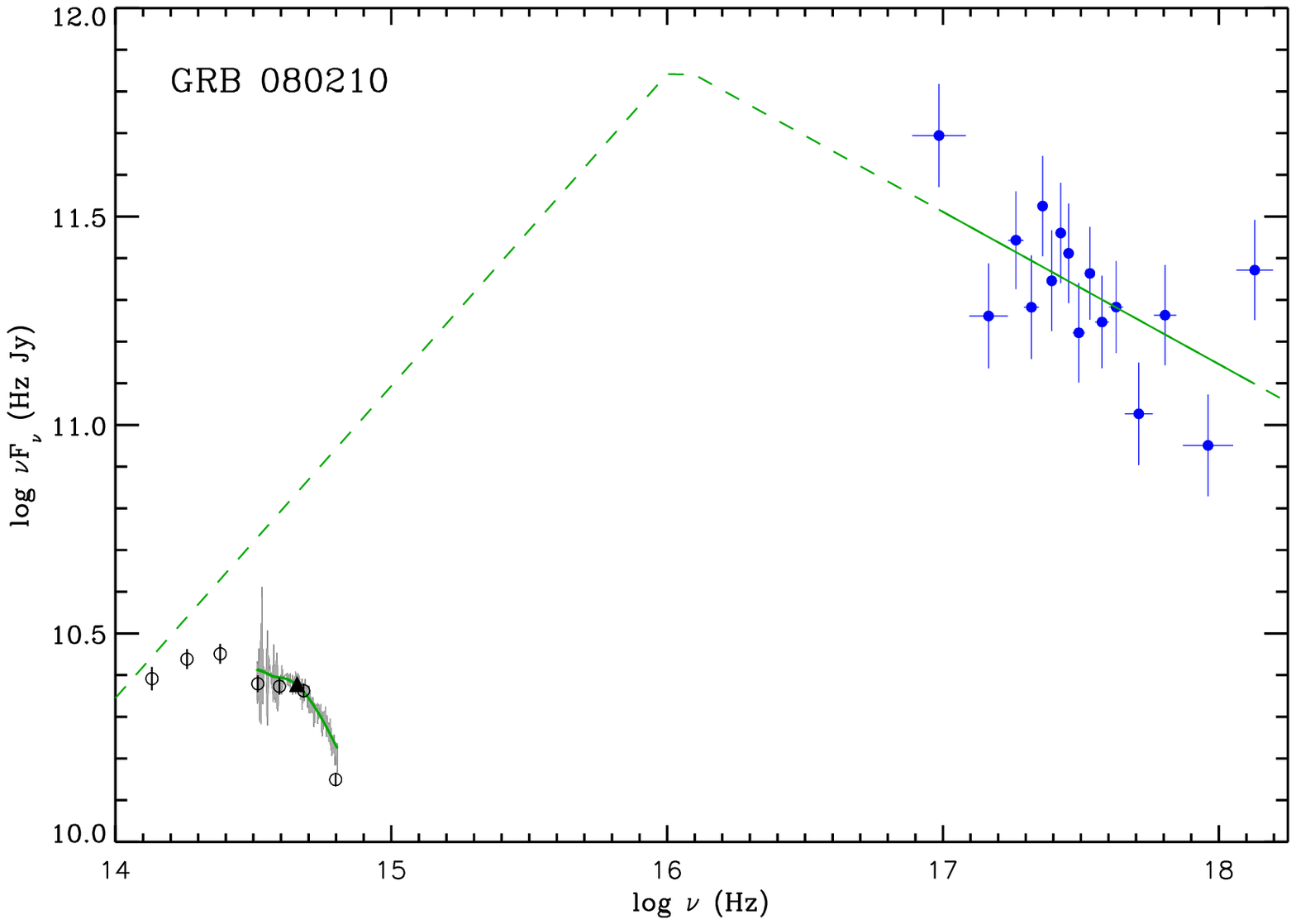}} &
            {\includegraphics[width=0.9\columnwidth,clip=]{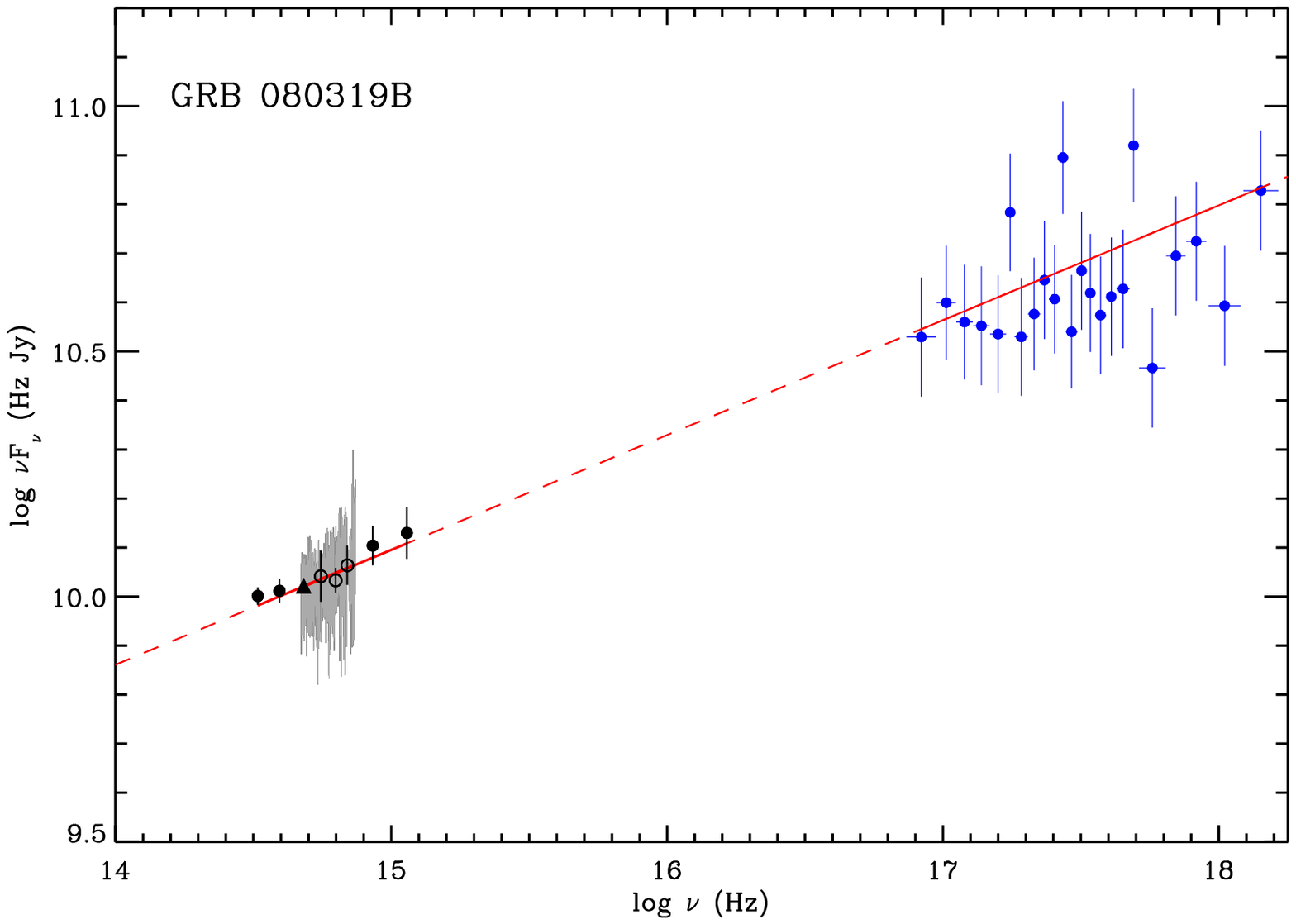}} \\
\end{tabular}
\caption{(continued)}
   \end{figure*}
   \clearpage
      
\addtocounter{figure}{-1}
  \begin{figure*}
  \begin{tabular}{c c}
     {\includegraphics[width=0.9\columnwidth,clip=]{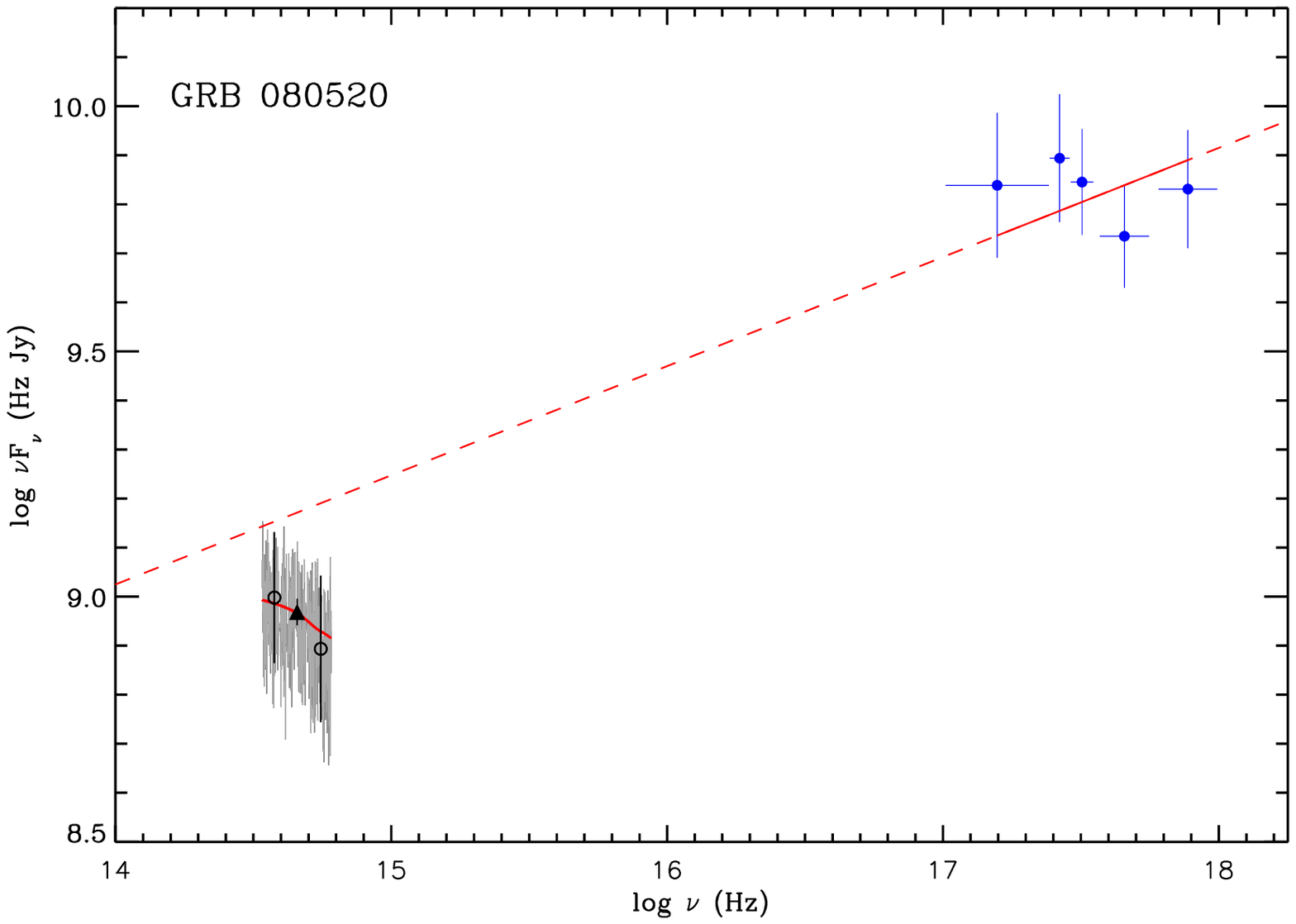}}  &
          {\includegraphics[width=0.9\columnwidth,clip=]{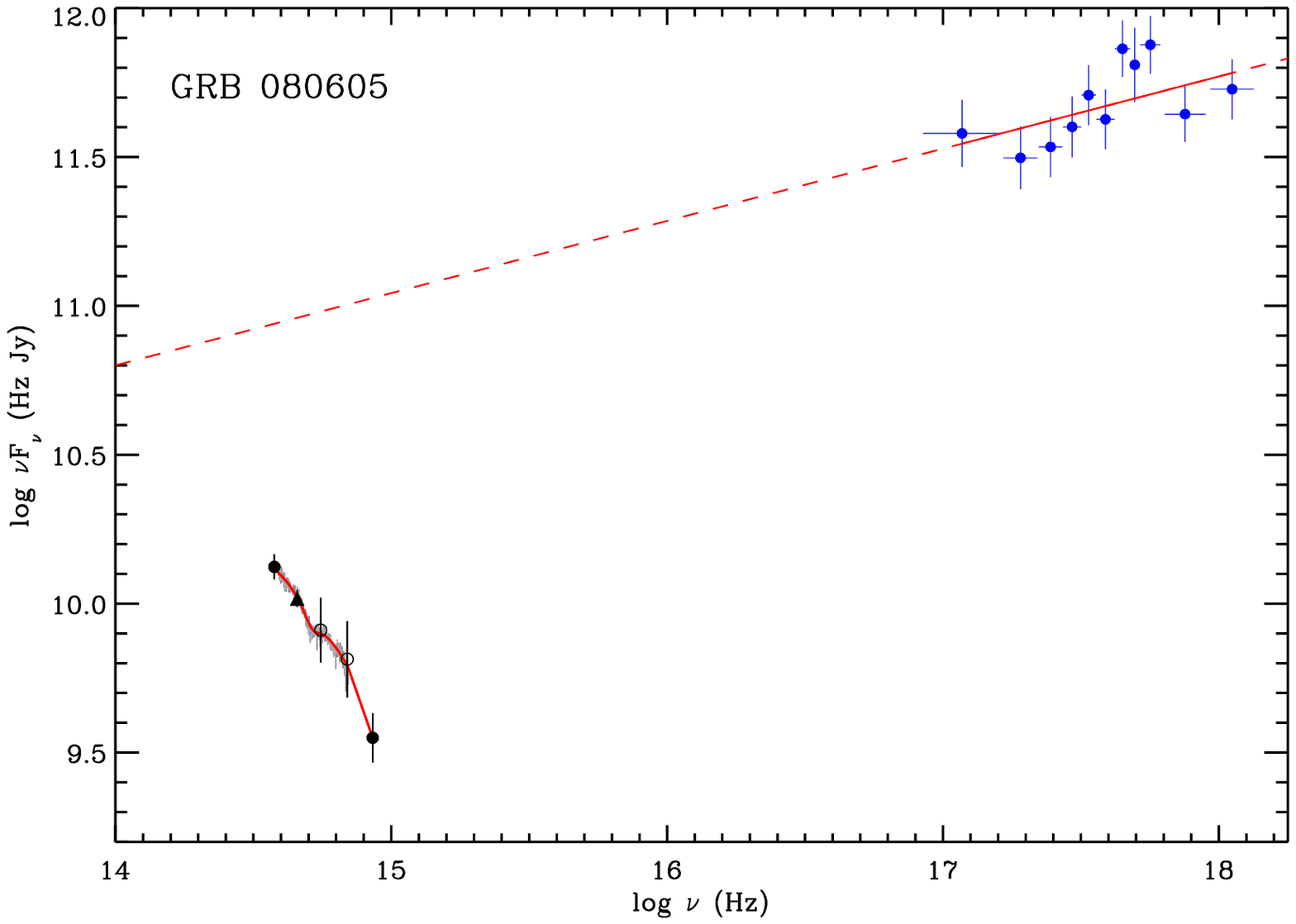}}   \\
   {\includegraphics[width=0.9\columnwidth,clip=]{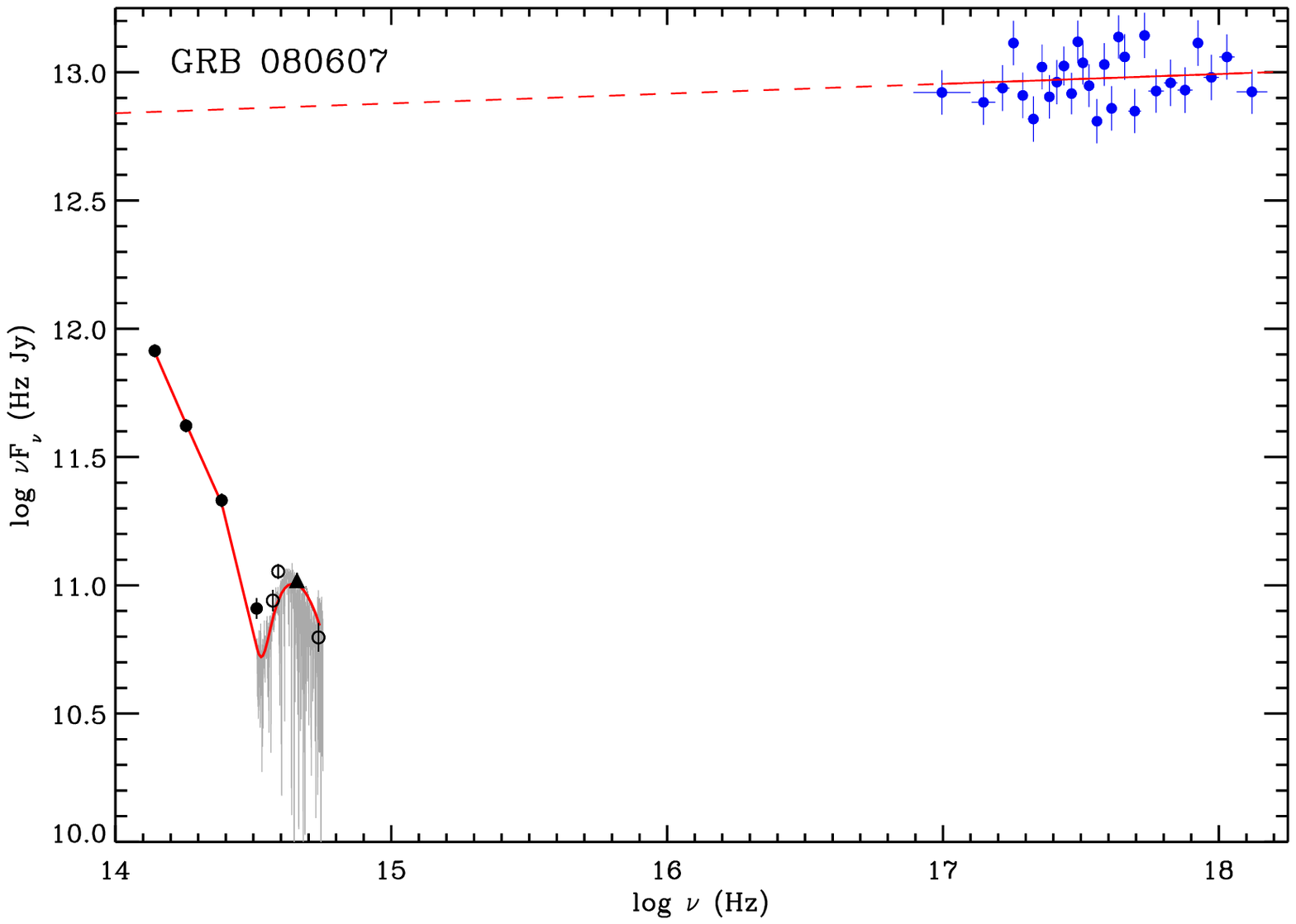}} &
        {\includegraphics[width=0.9\columnwidth,clip=]{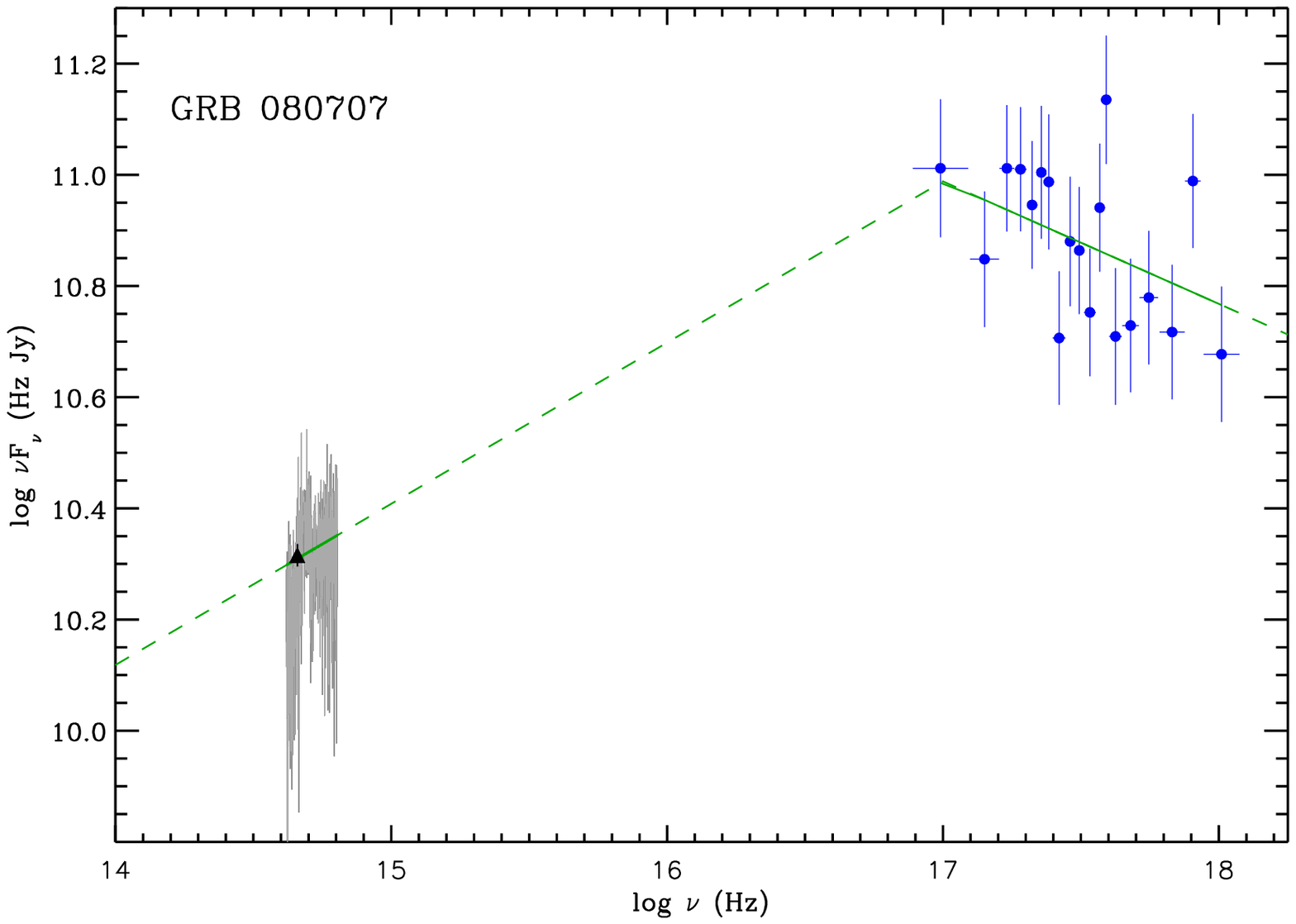}} \\
   {\includegraphics[width=0.9\columnwidth,clip=]{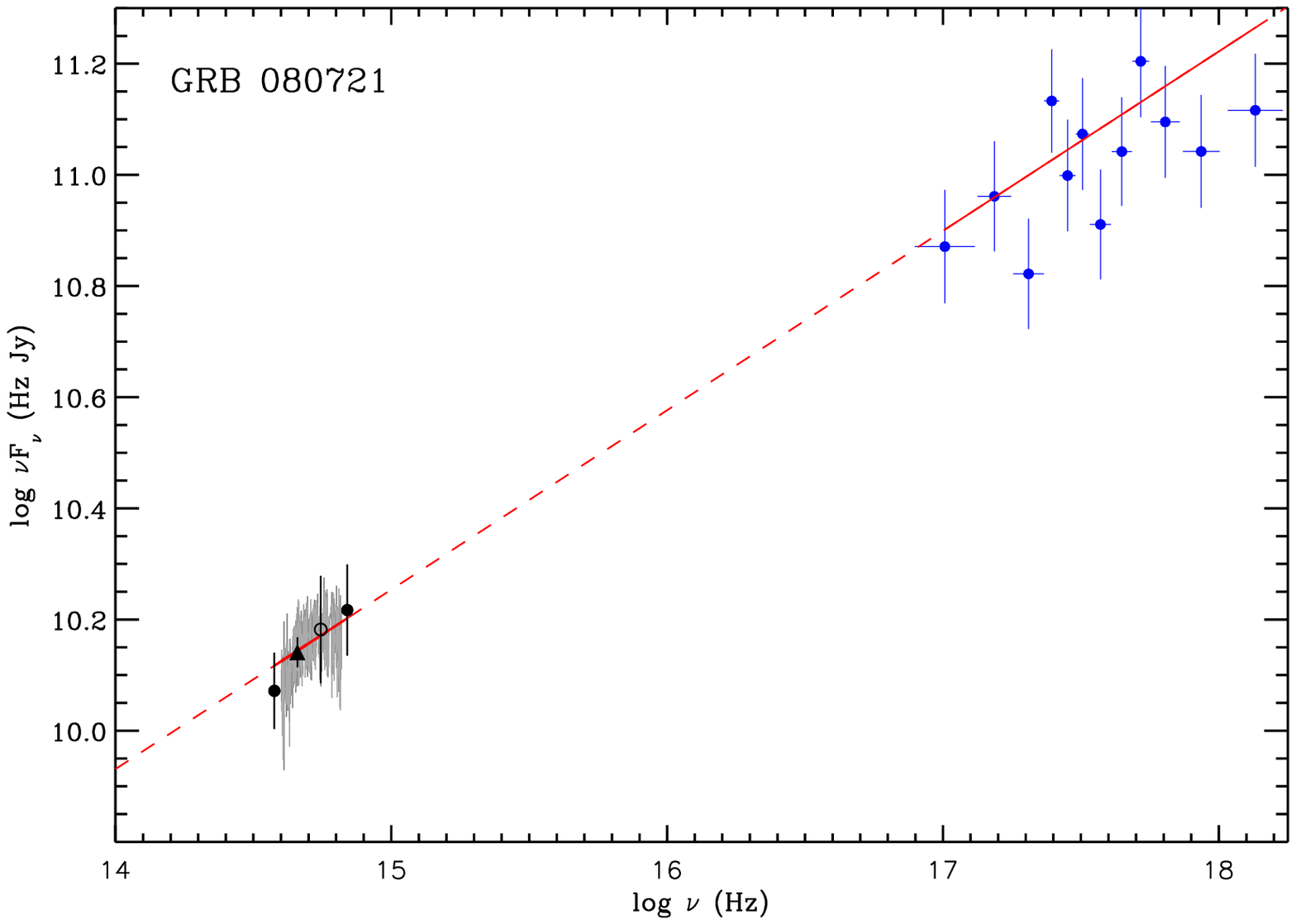}} &
      {\includegraphics[width=0.9\columnwidth,clip=]{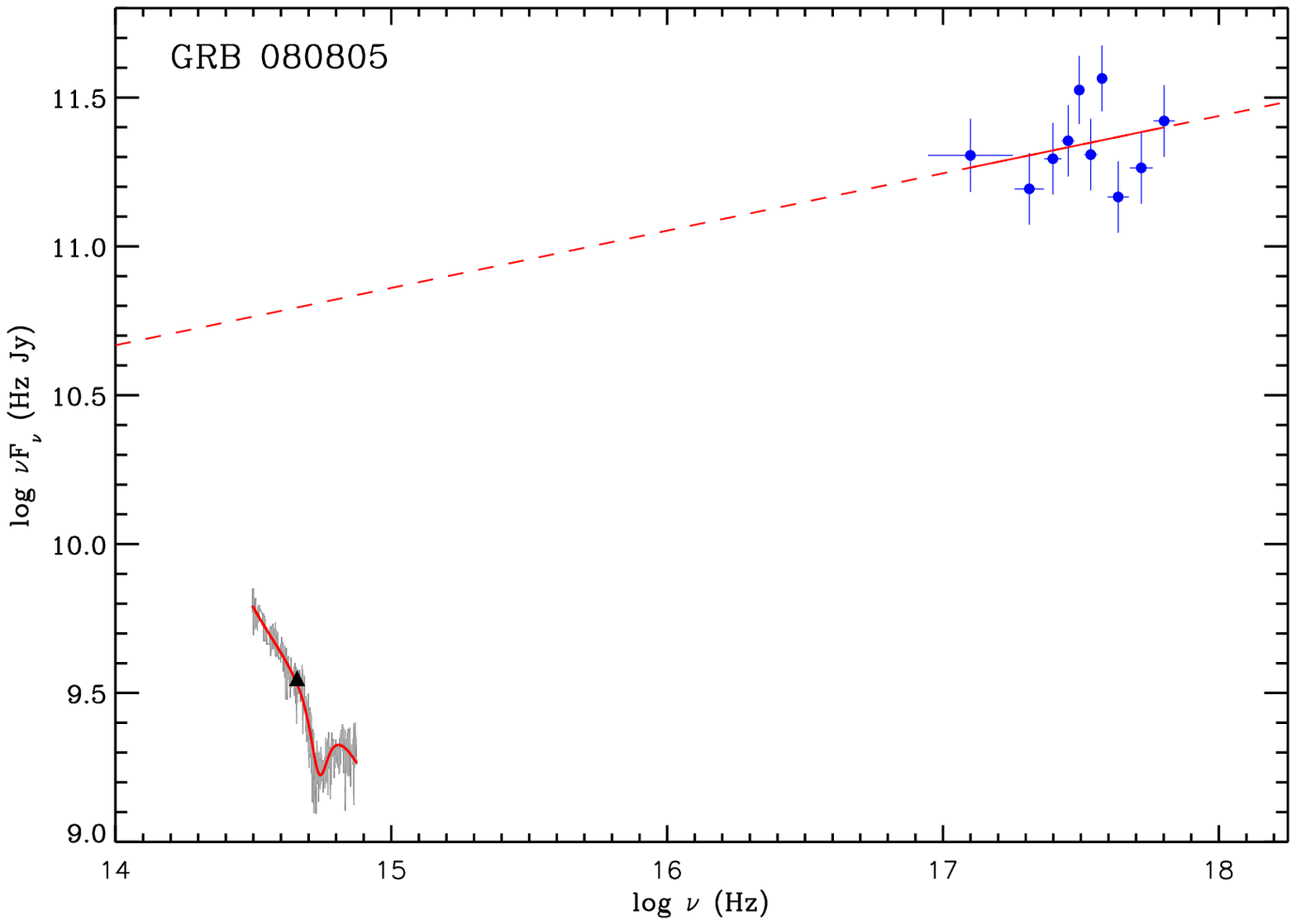}} \\
         {\includegraphics[width=0.9\columnwidth,clip=]{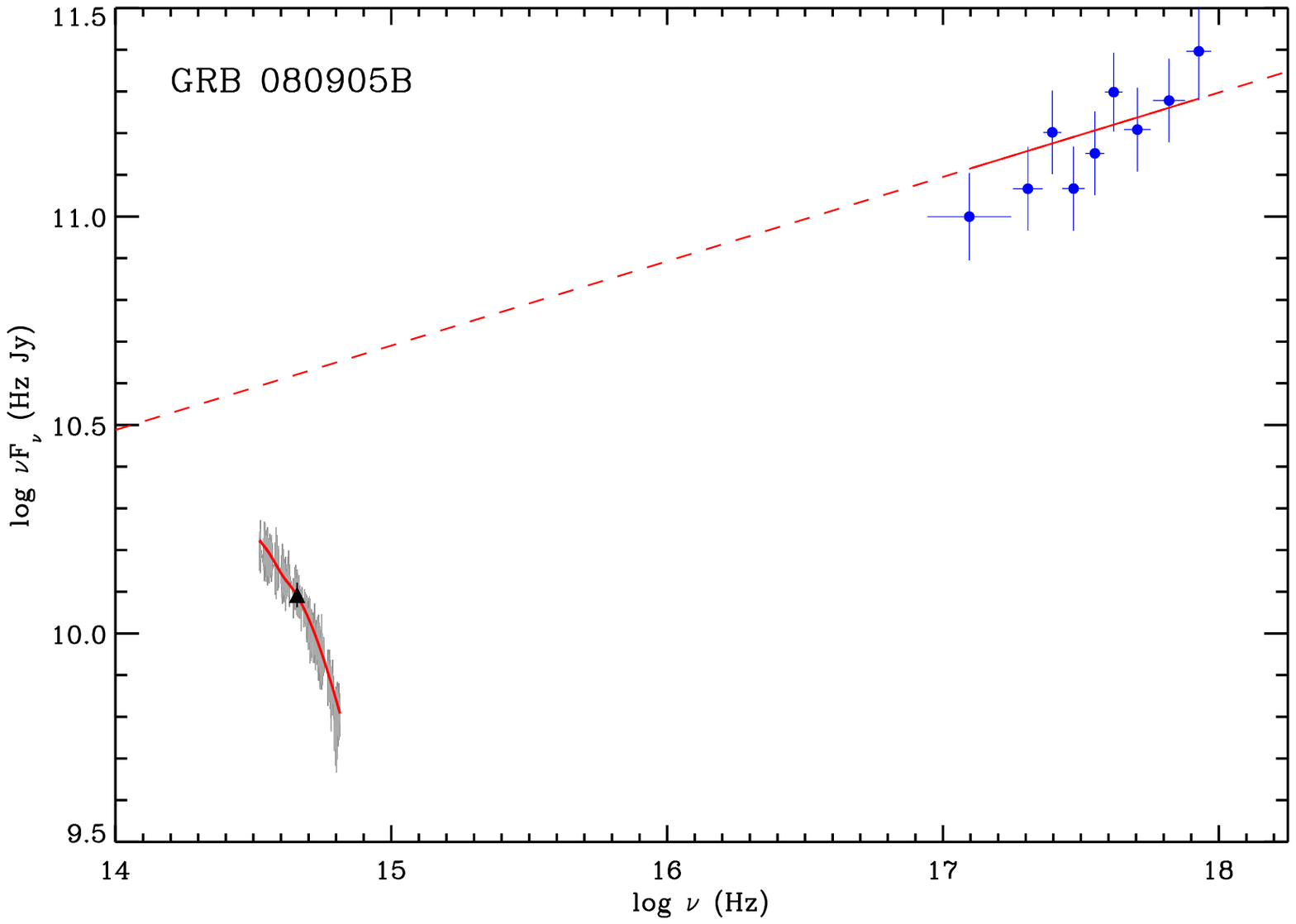}} &
        {\includegraphics[width=0.9\columnwidth,clip=]{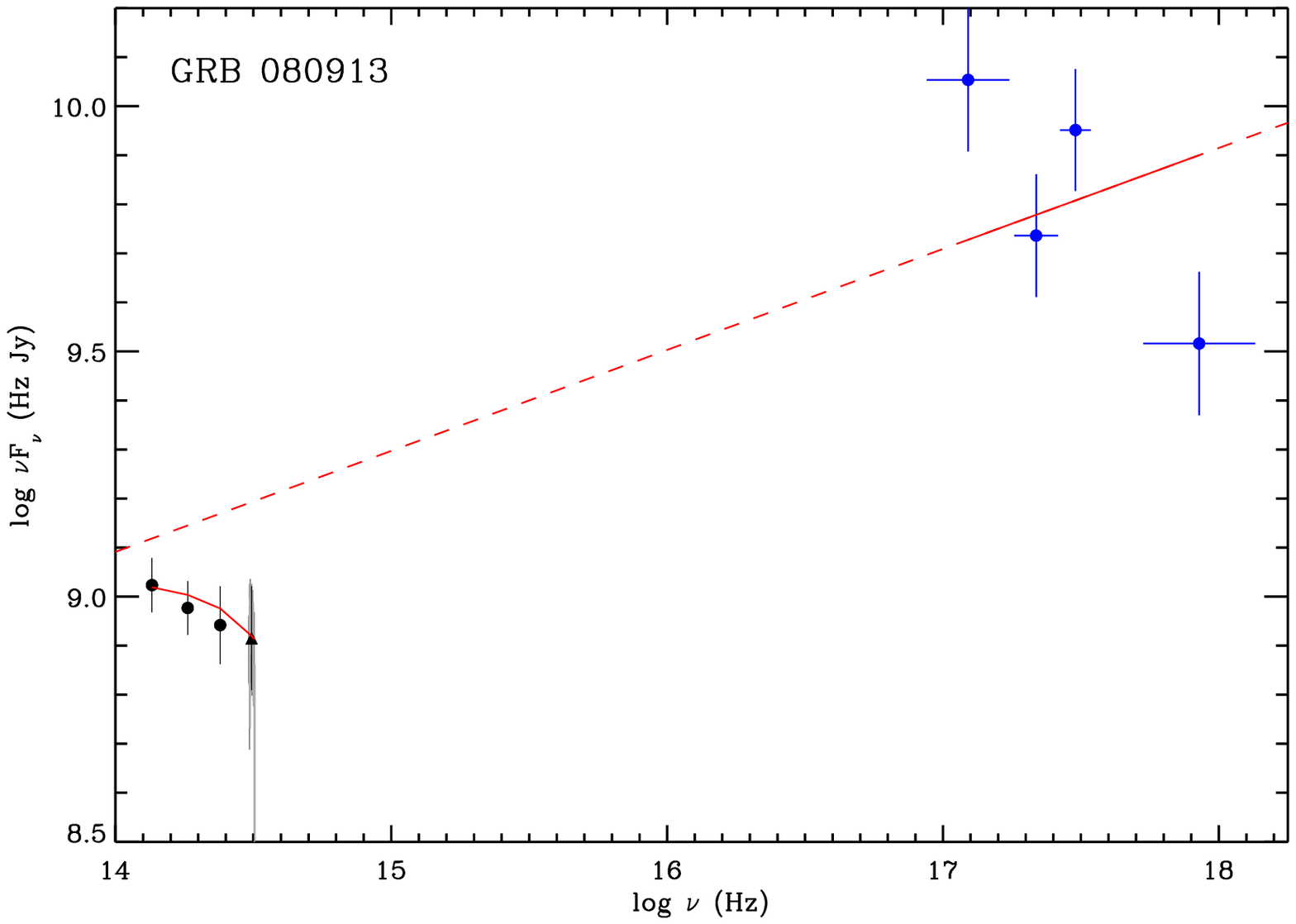}} \\
\end{tabular}
\caption{(continued)}
   \end{figure*}
   \clearpage

\addtocounter{figure}{-1}
  \begin{figure*}
  \begin{tabular}{c c}
          {\includegraphics[width=0.9\columnwidth,clip=]{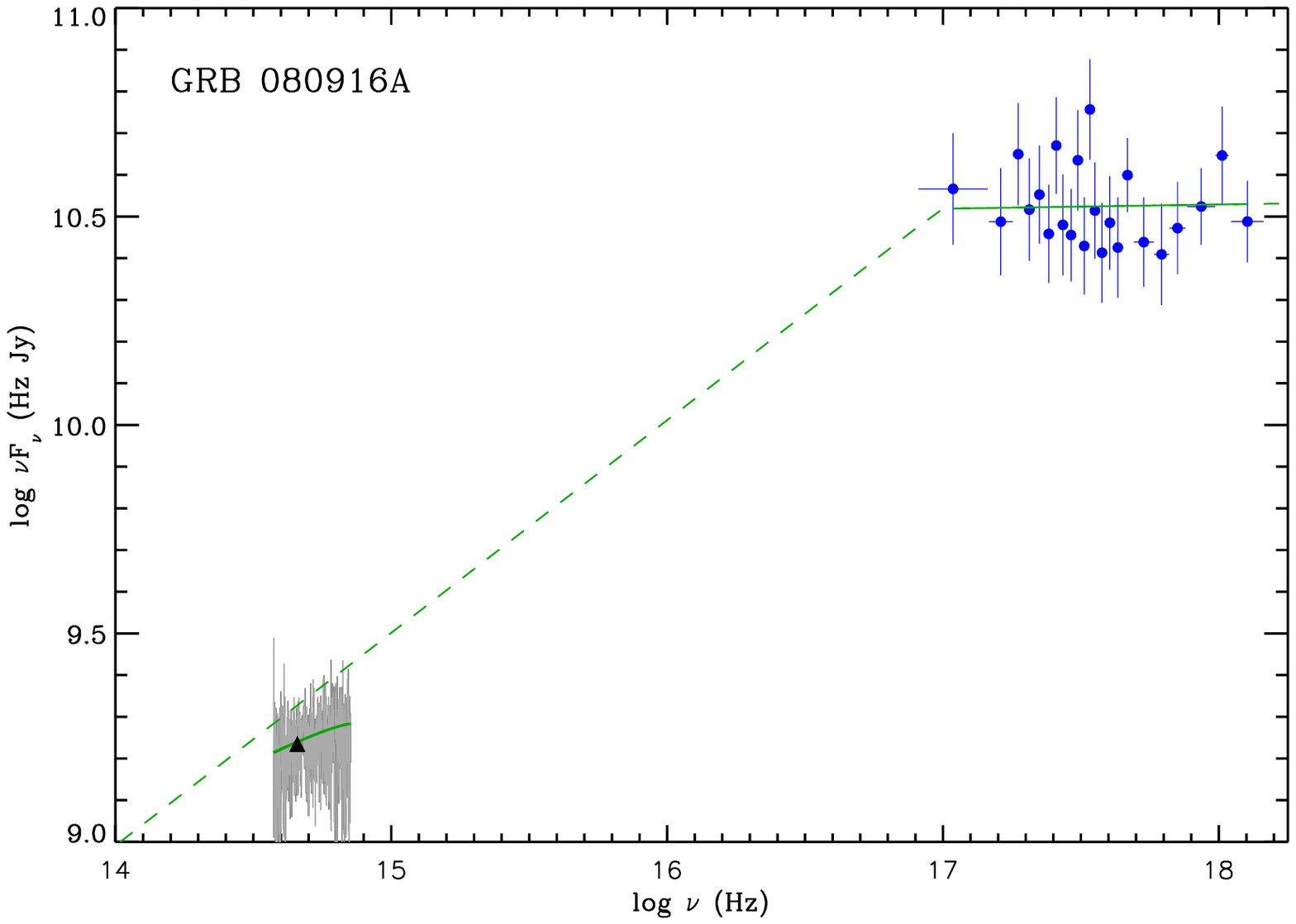}} &
   {\includegraphics[width=0.9\columnwidth,clip=]{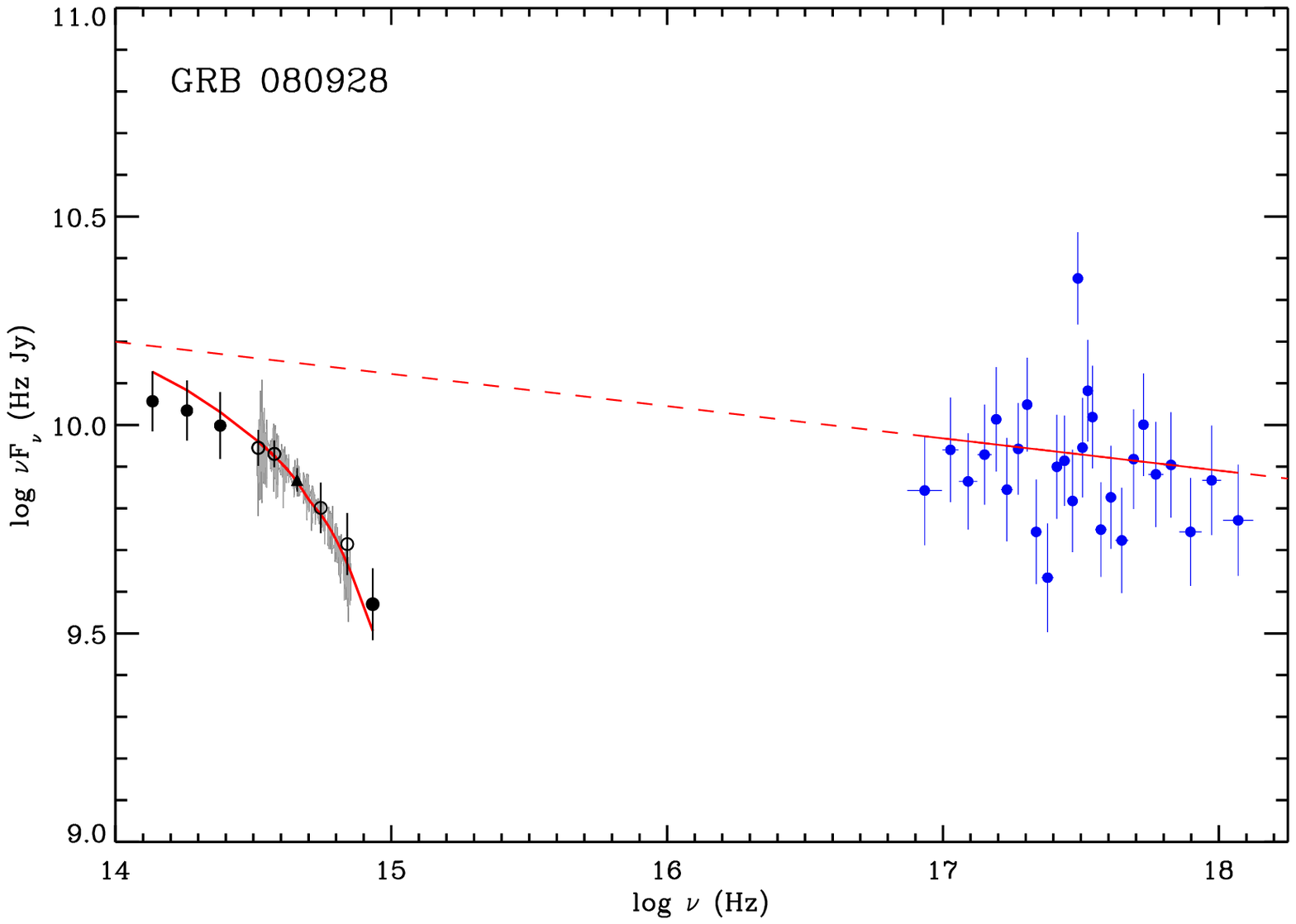}}  \\
   \end{tabular}
\caption{(continued)}
   \end{figure*}
\clearpage
      \begin{figure*}
        \begin{tabular}{c c}
  {\includegraphics[width=0.9\columnwidth,clip=]{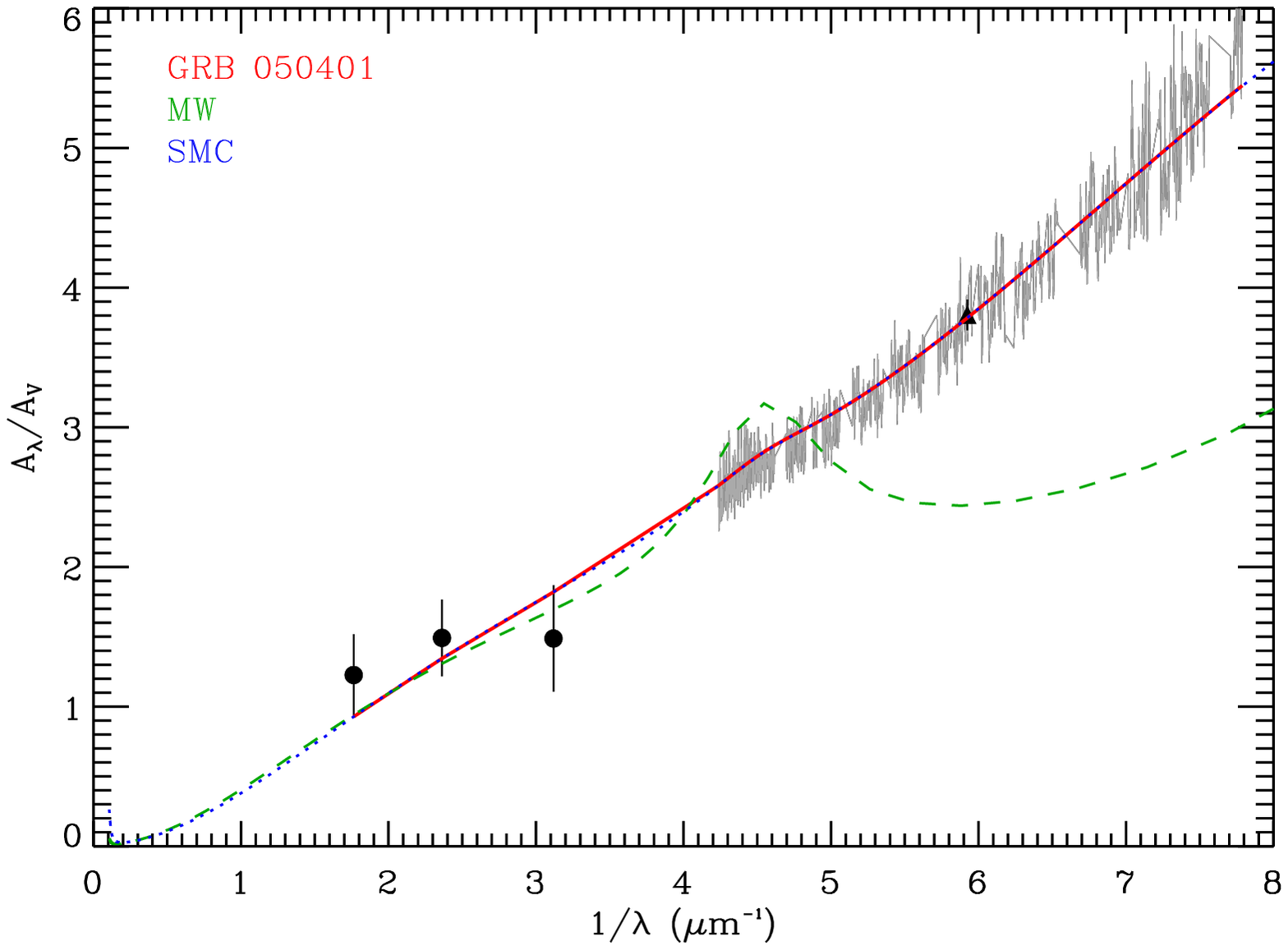}} &
  {\includegraphics[width=0.9\columnwidth,clip=]{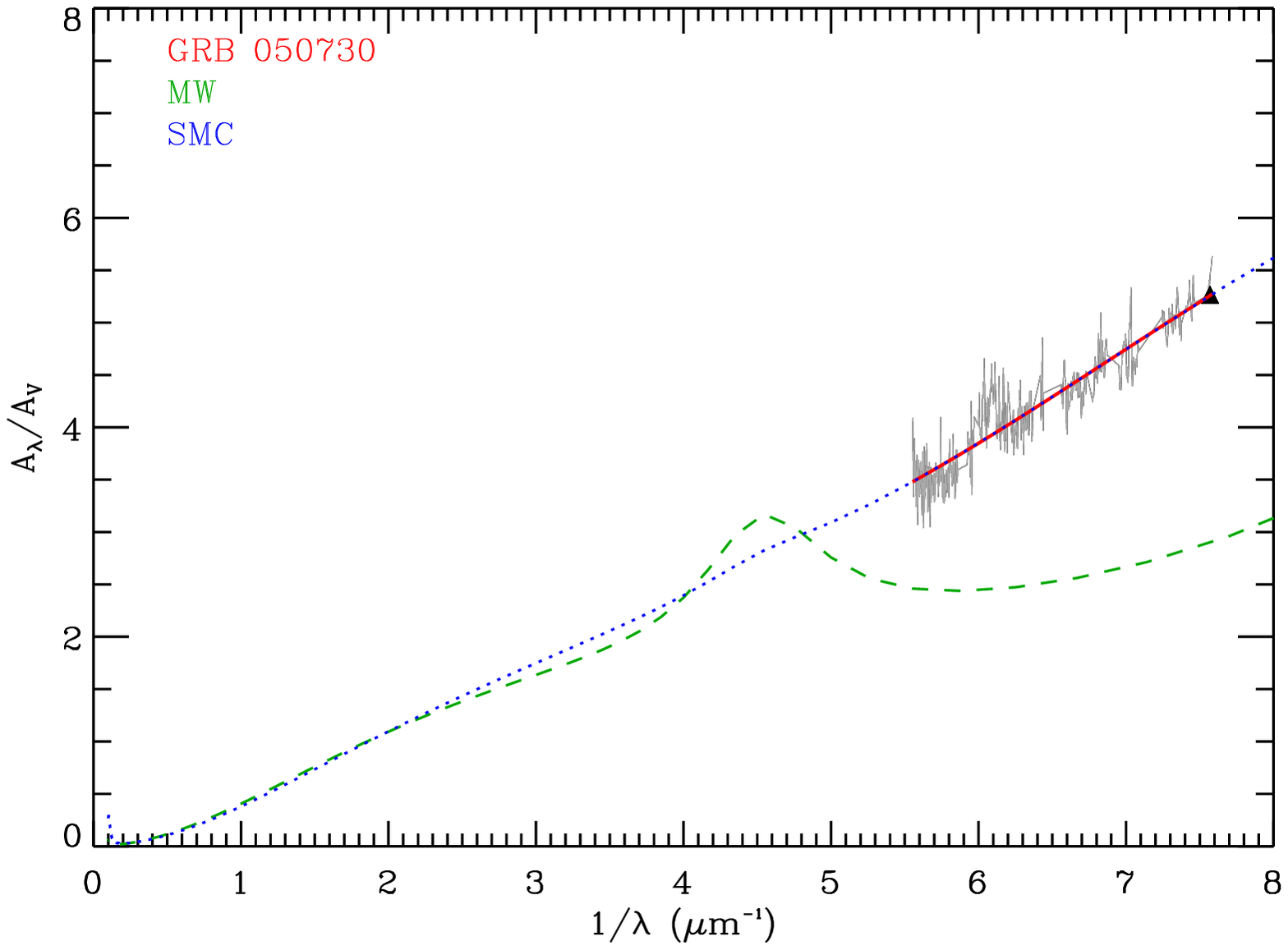}}  \\
   {\includegraphics[width=0.9\columnwidth,clip=]{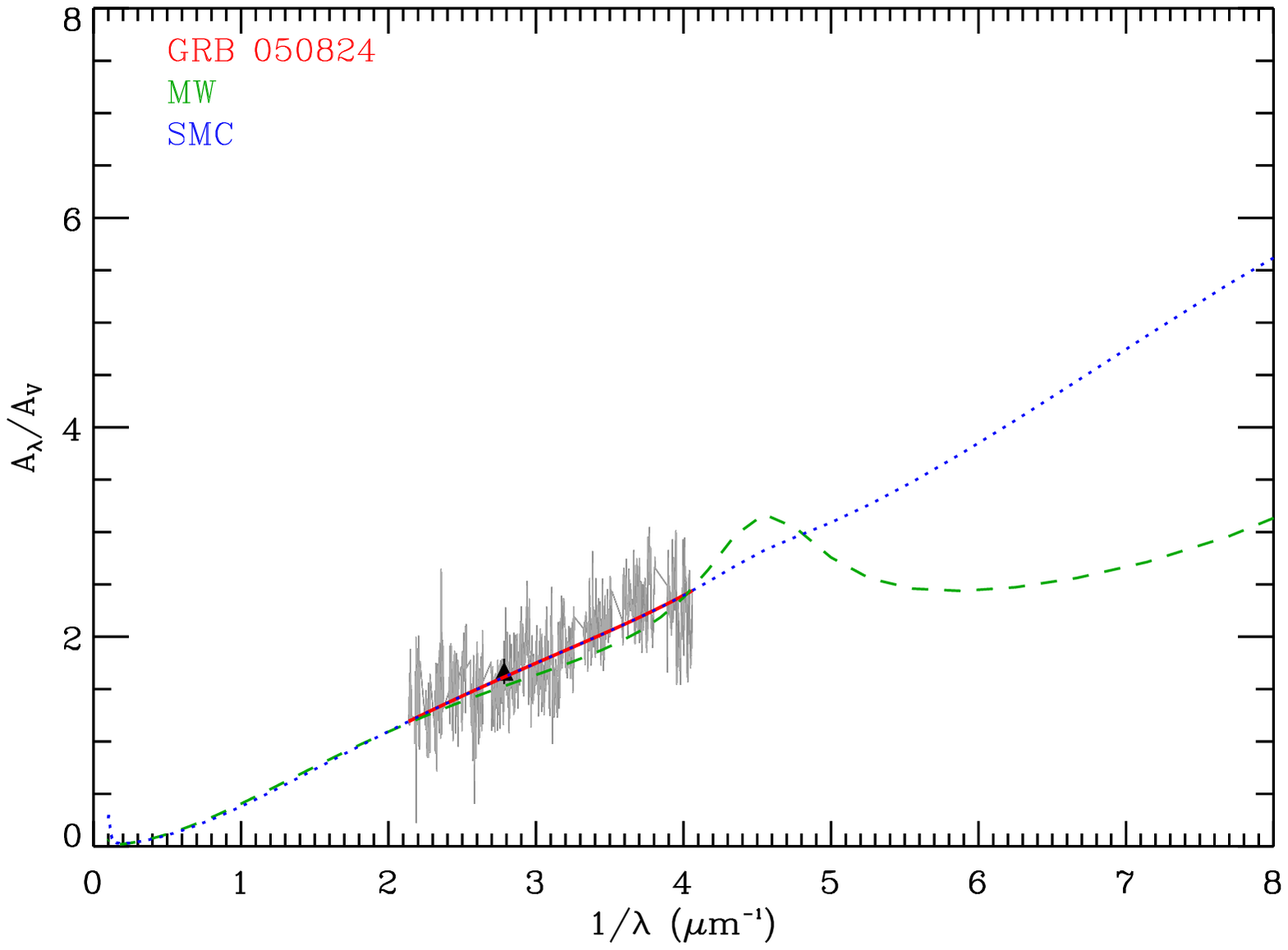}} &
   {\includegraphics[width=0.9\columnwidth,clip=]{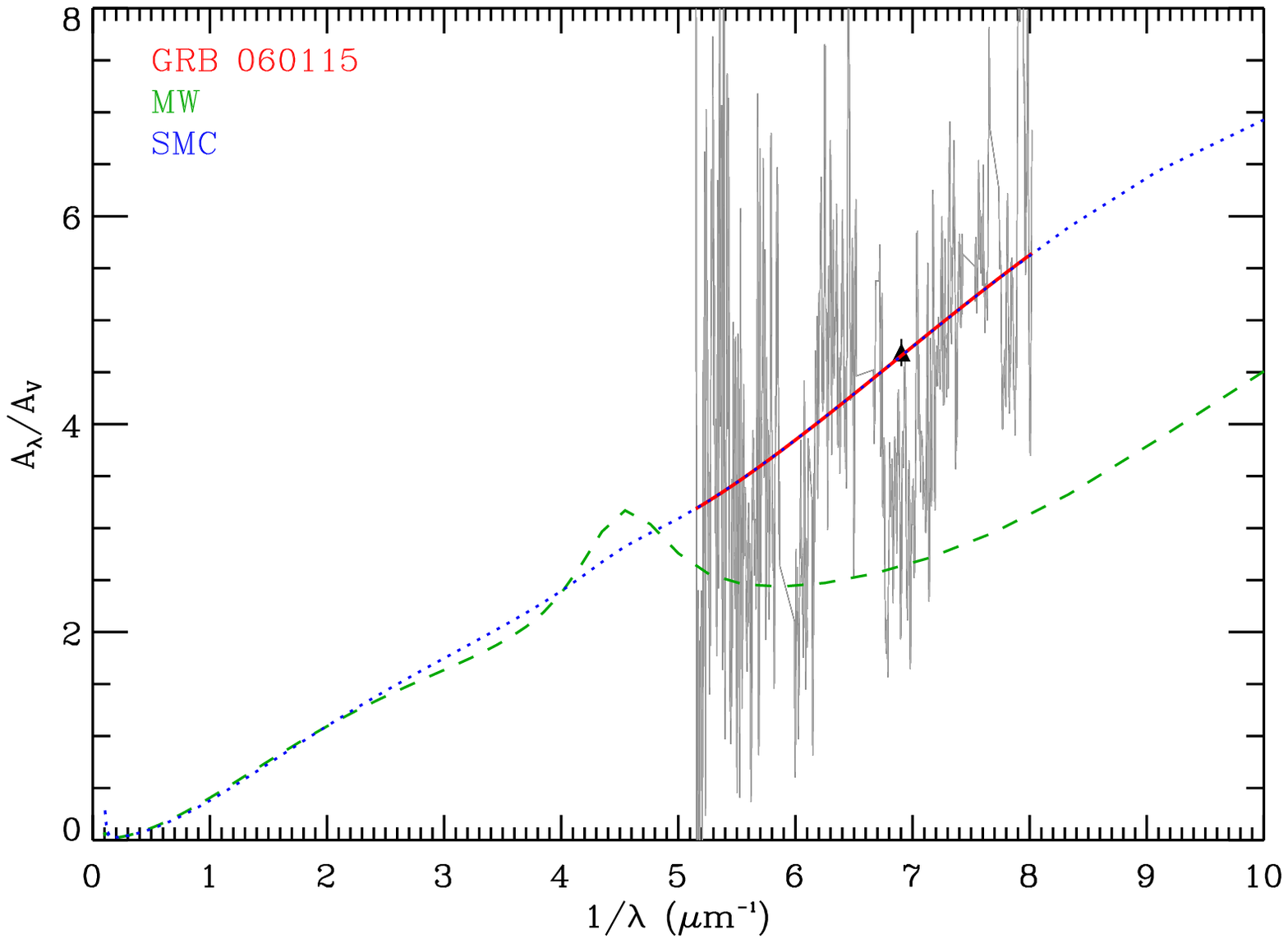}} \\
   {\includegraphics[width=0.9\columnwidth,clip=]{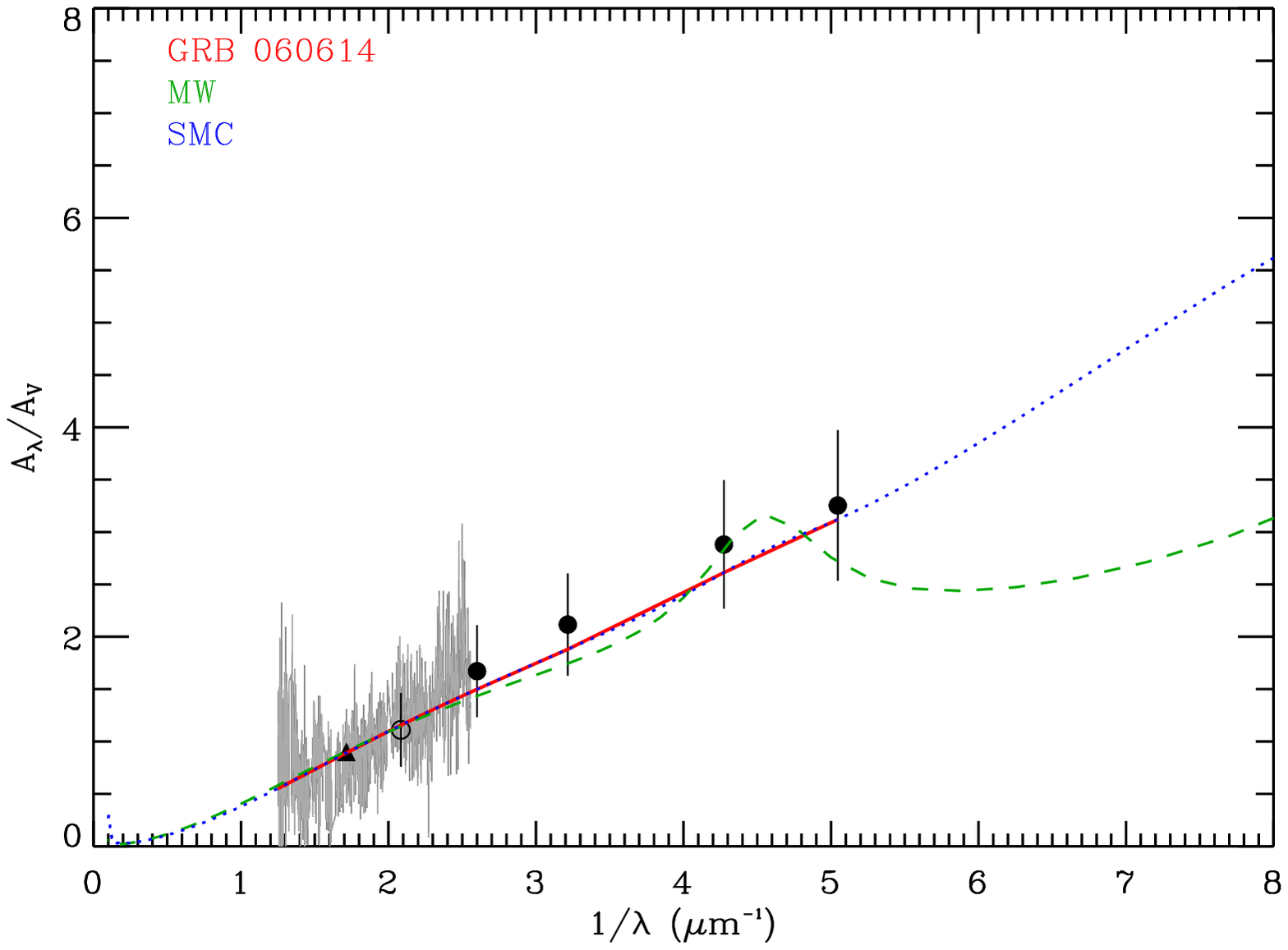}} &
   {\includegraphics[width=0.9\columnwidth,clip=]{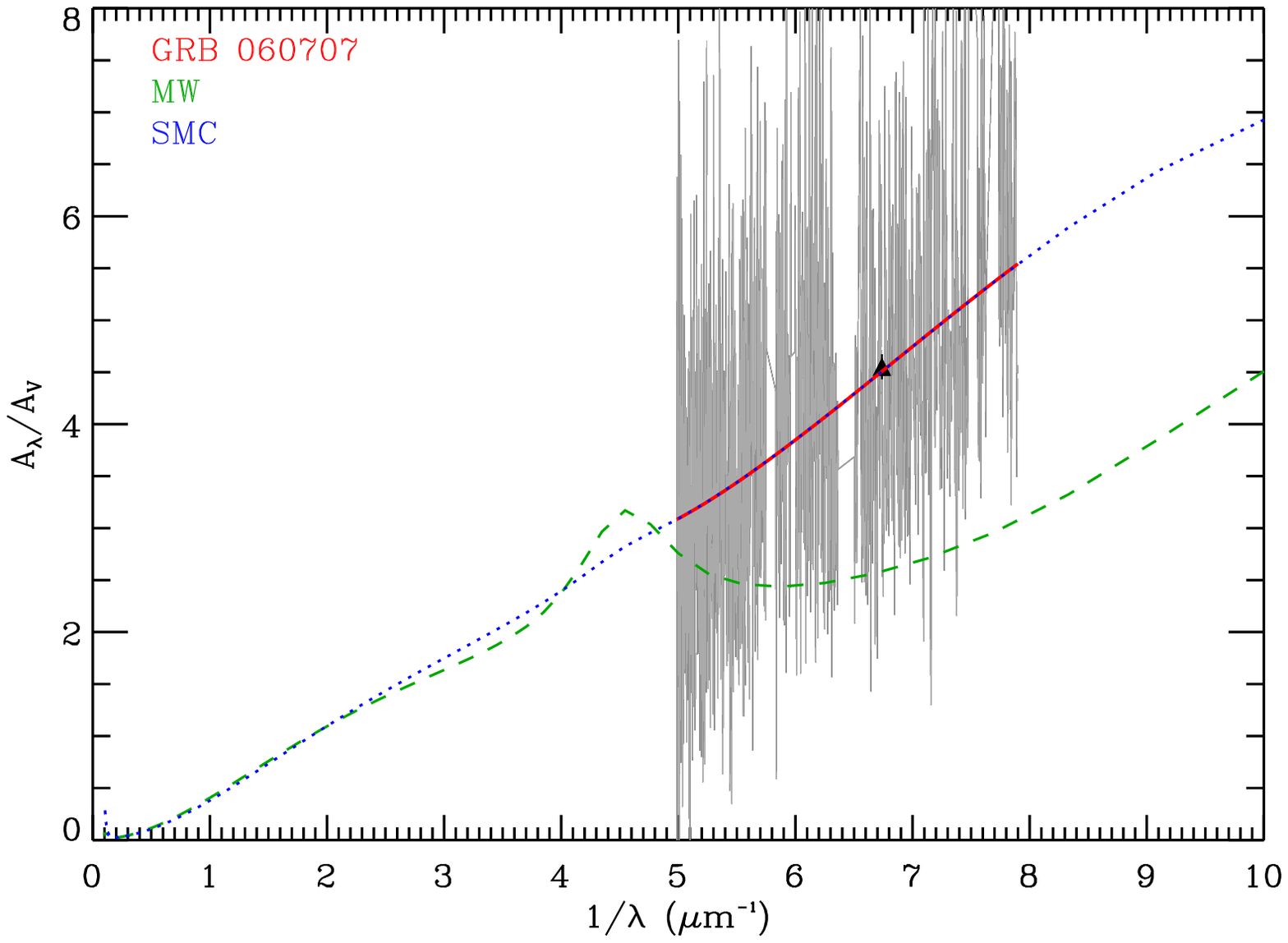}} \\
       {\includegraphics[width=0.9\columnwidth,clip=]{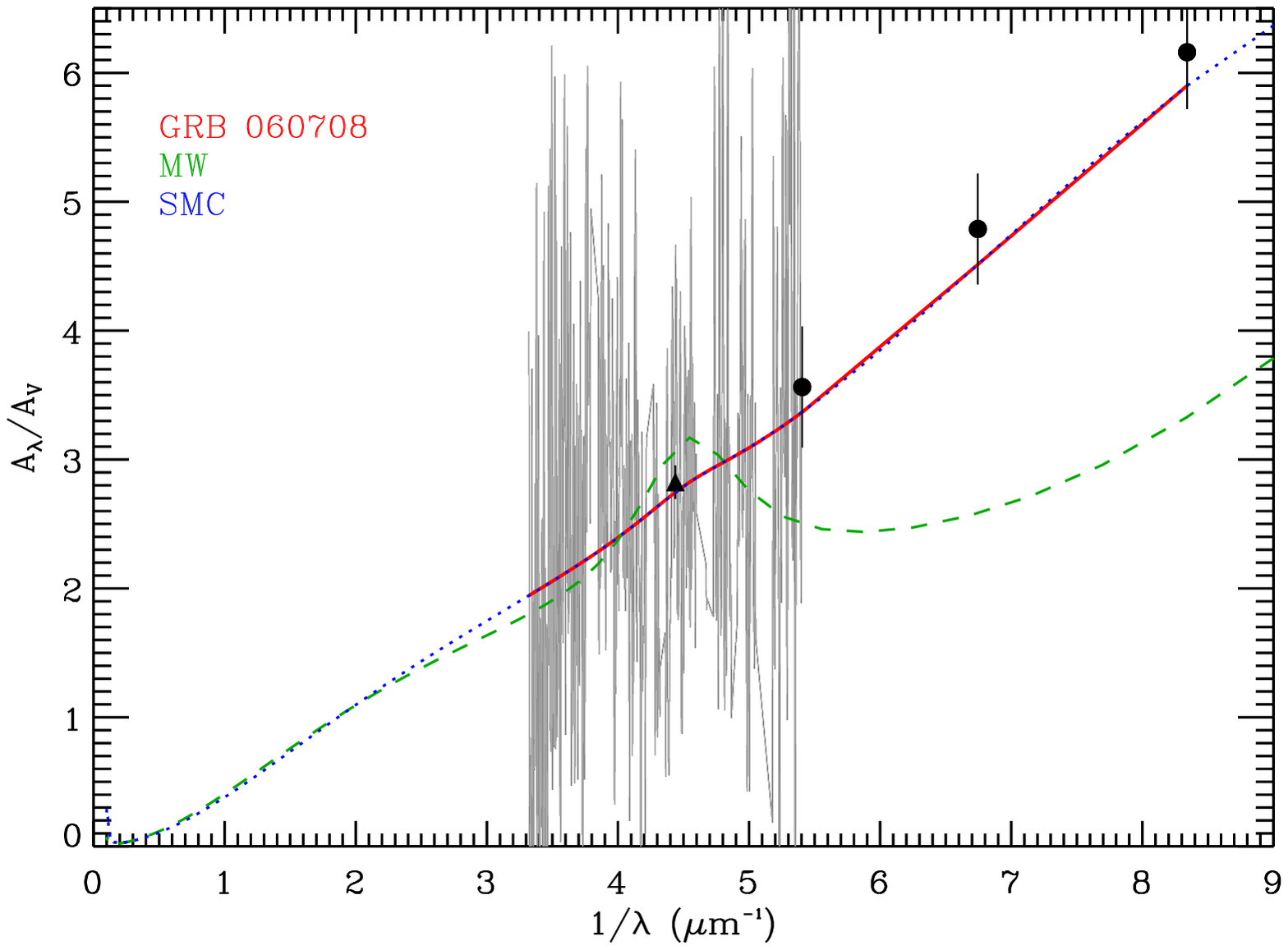}}  &
   {\includegraphics[width=0.9\columnwidth,clip=]{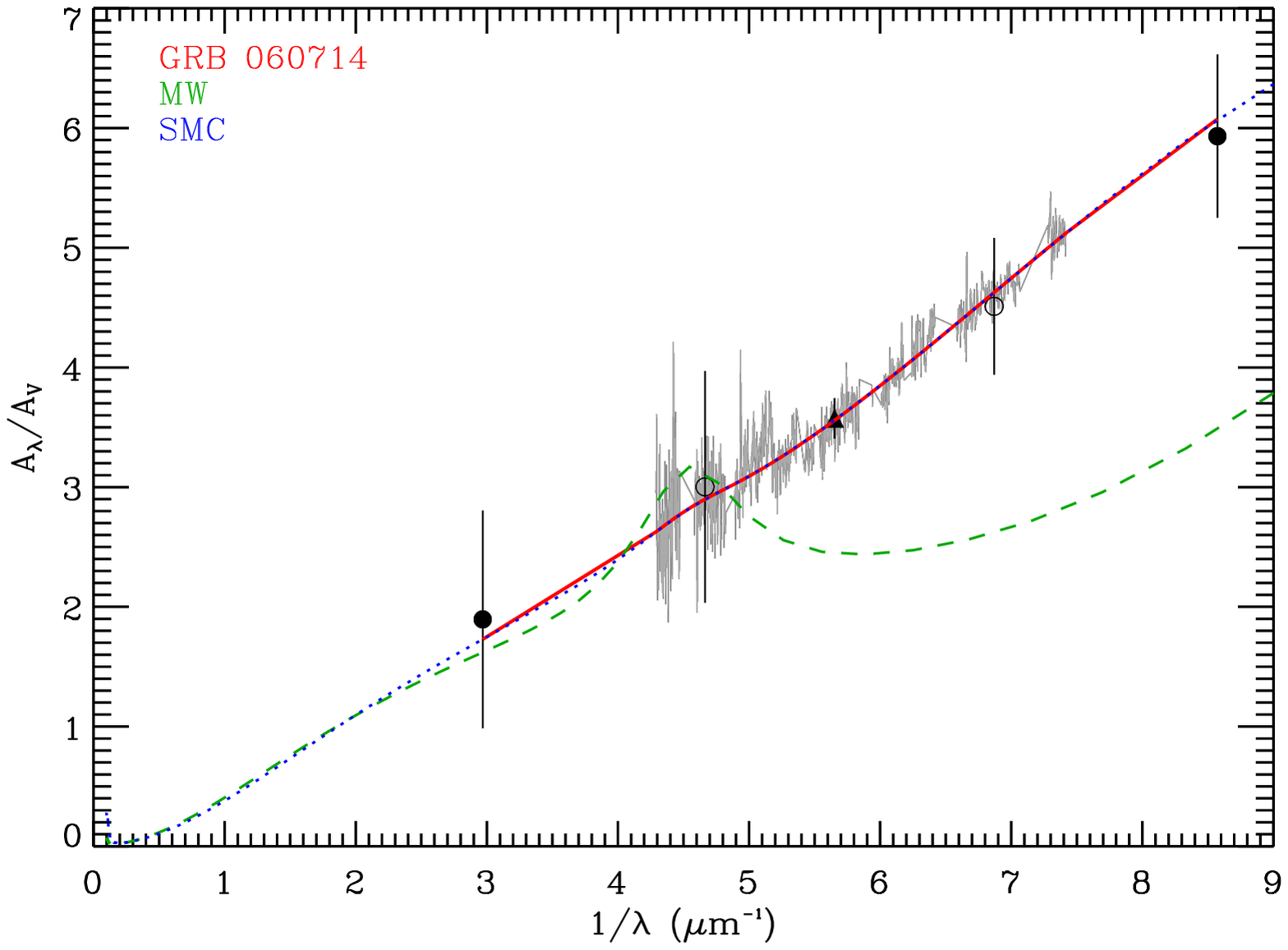}} \\
        \end{tabular}
\caption{Absolute extinction curves of the spectroscopic GRB sample. The extinction curves are based on the best fit models given in Table \ref{fitresult}. The grey curve represents the optical spectrum. The black triangles correspond to acquisition camera photometry used for scaling the afterglow SED. Black open circles are not included in the spectral fitting because of the optical spectrum wavelength coverage in that region while solid circles represents the photometric data points included in the SED modeling. The red solid curve corresponds to the best dust model for each GRB. Also shown are the Milky Way (green dashed line) and SMC (blue dotted line) extinction models taken from \citet{pei}.}
        \label{extinction1}
   \end{figure*}
   \clearpage
   
\addtocounter{figure}{-1}
  \begin{figure*}
  \begin{tabular}{c c}
   {\includegraphics[width=0.9\columnwidth,clip=]{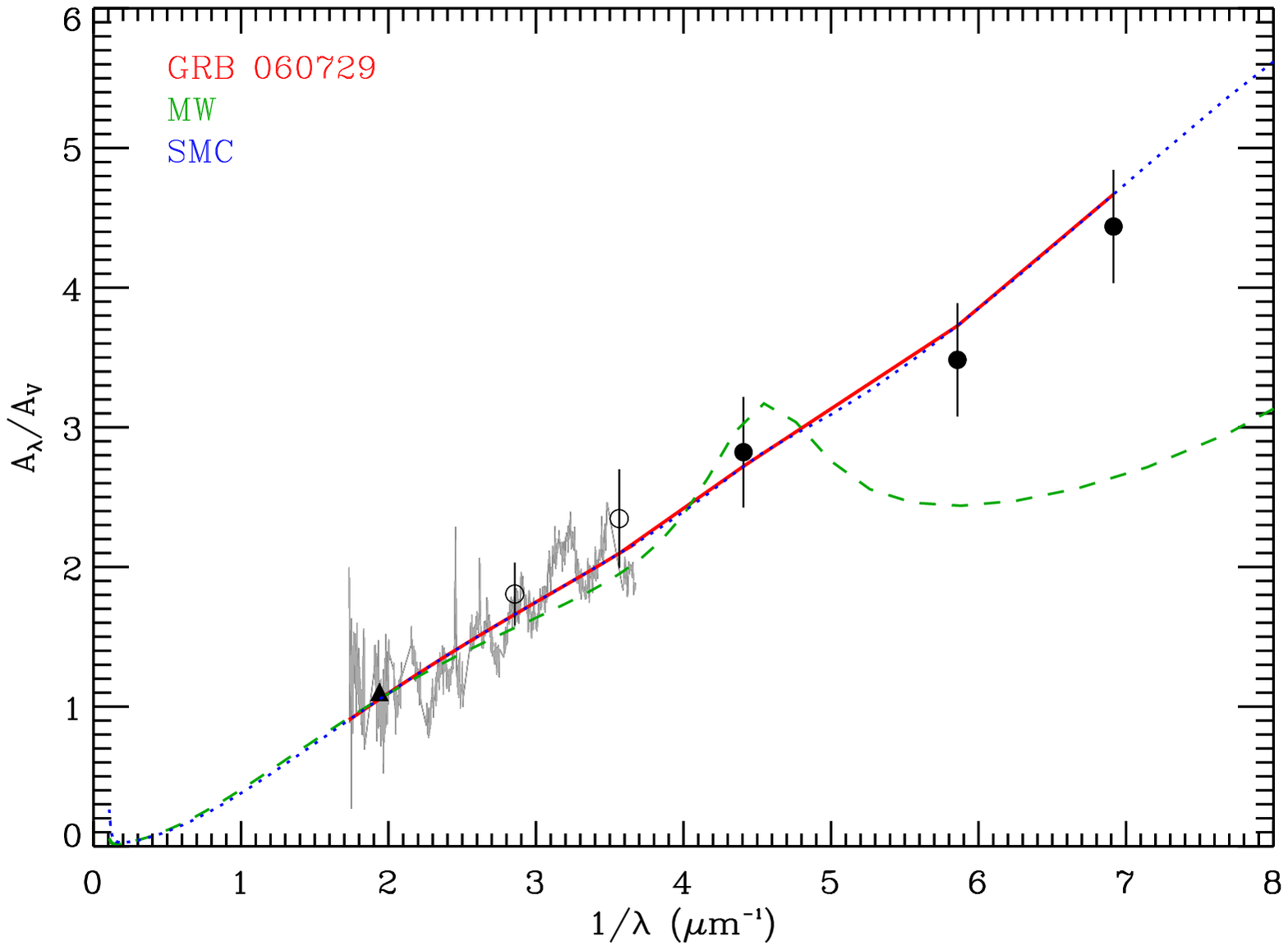}}  &
      {\includegraphics[width=0.9\columnwidth,clip=]{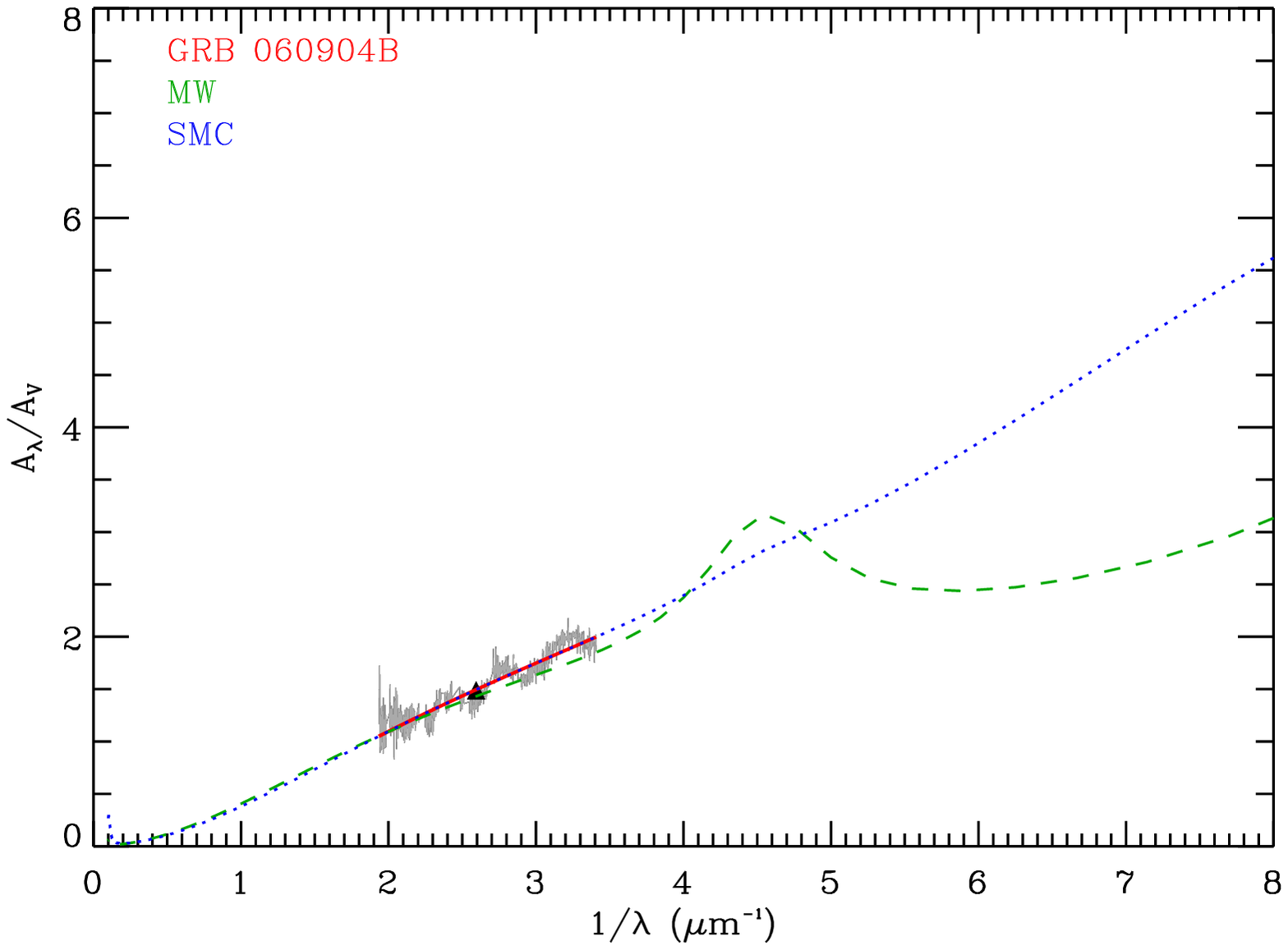}} \\
      {\includegraphics[width=0.9\columnwidth,clip=]{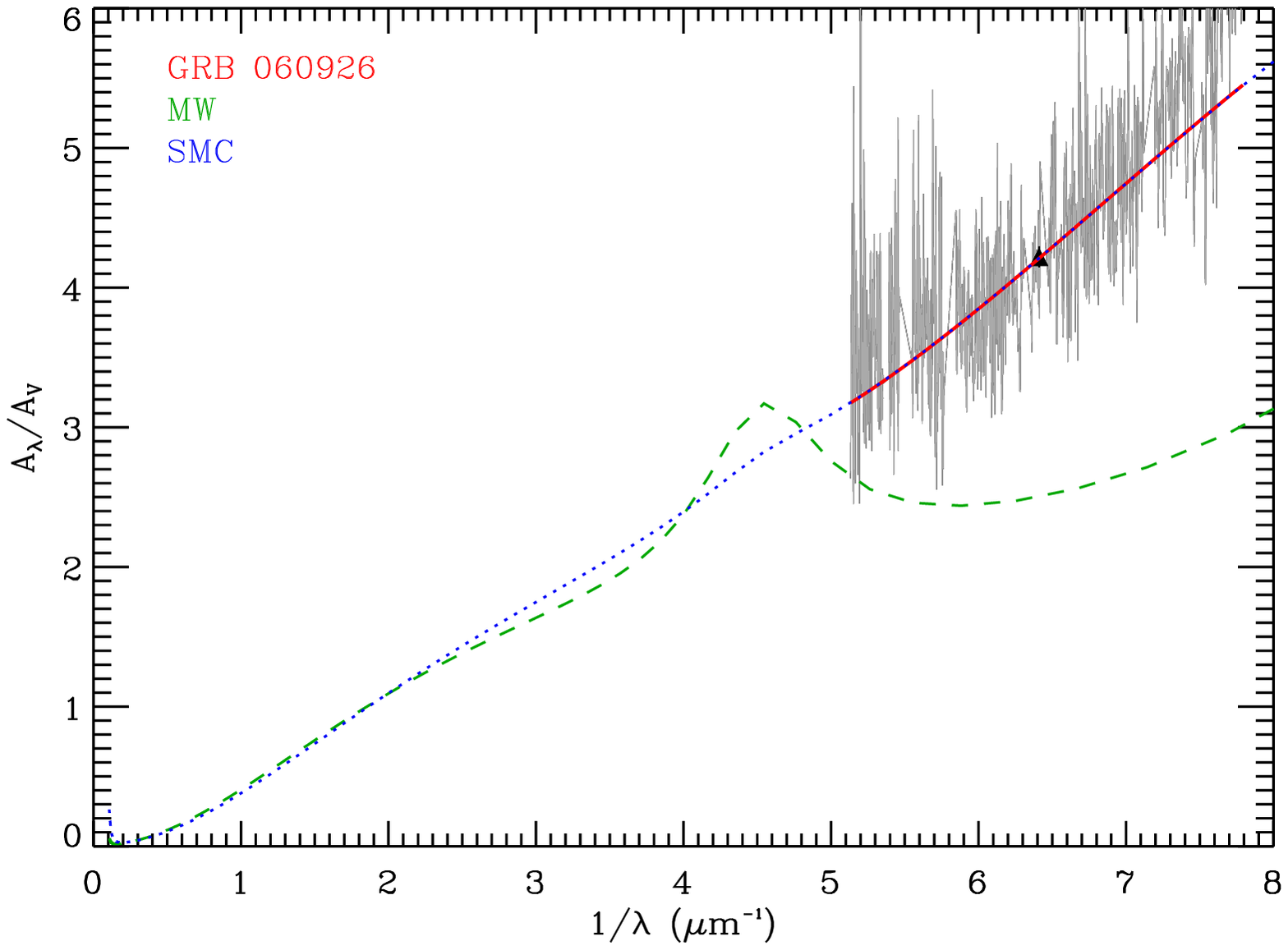}}  &
   {\includegraphics[width=0.9\columnwidth,clip=]{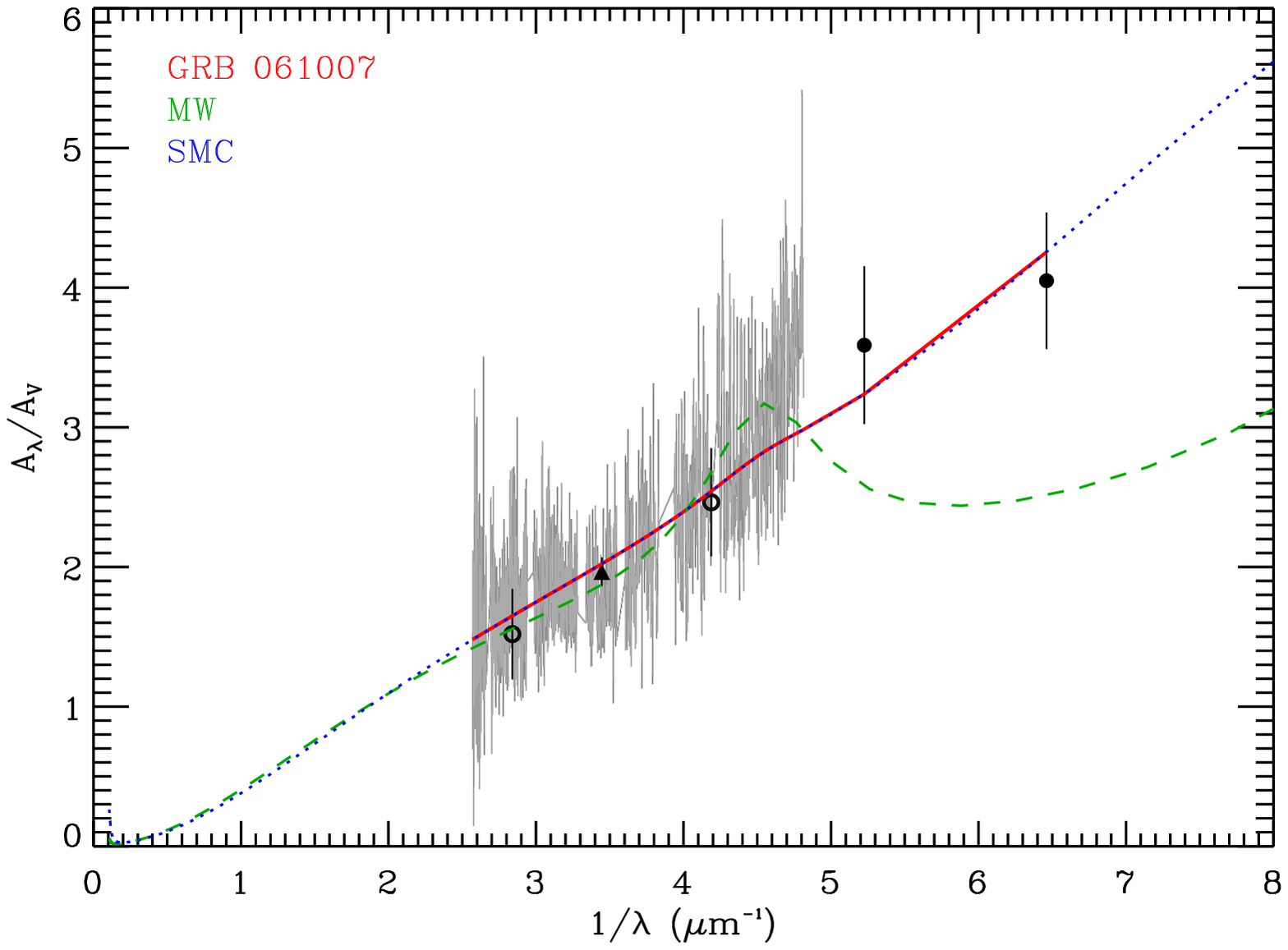}}  \\ 
   {\includegraphics[width=0.9\columnwidth,clip=]{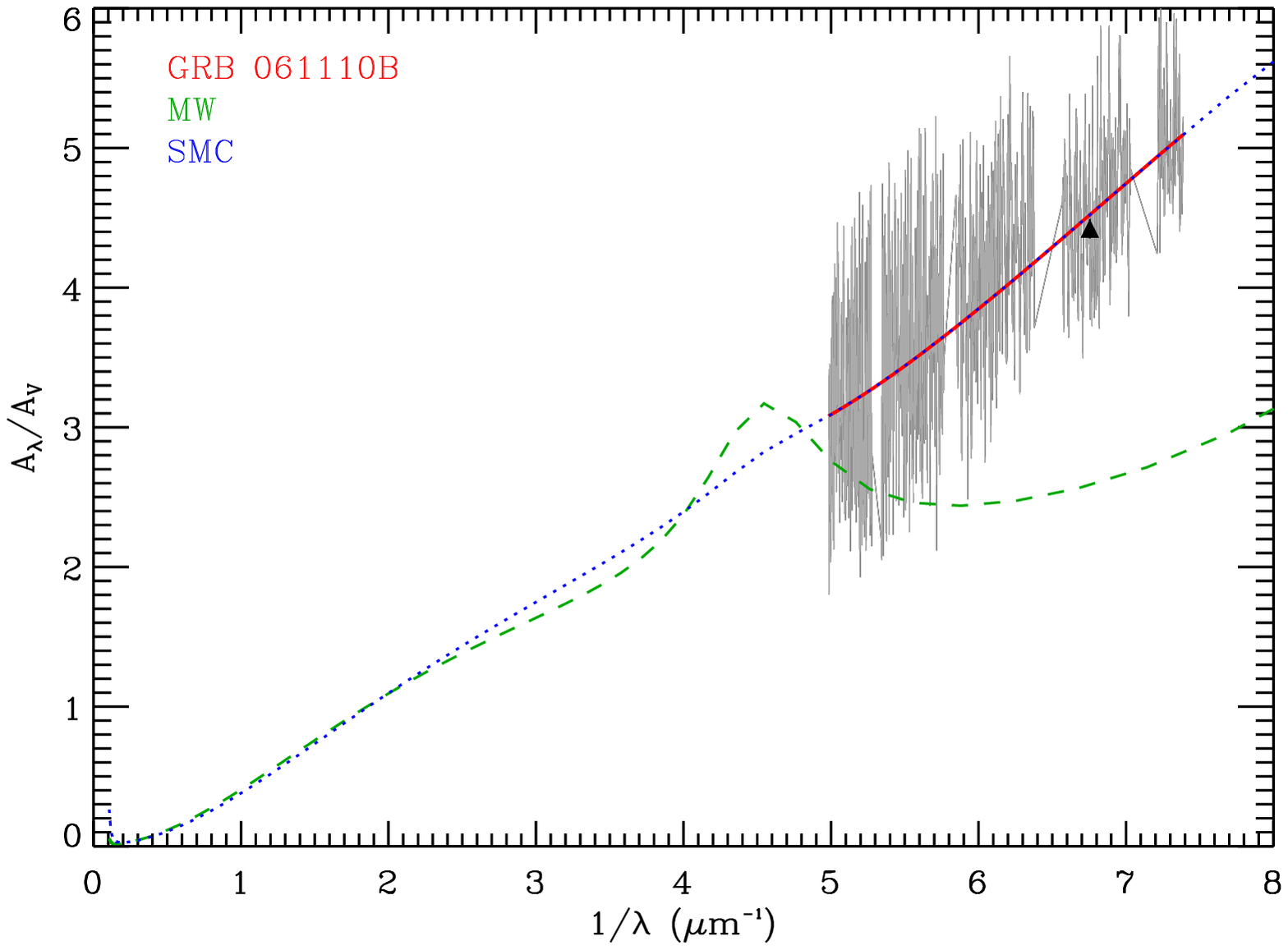}}  &
          {\includegraphics[width=0.9\columnwidth,clip=]{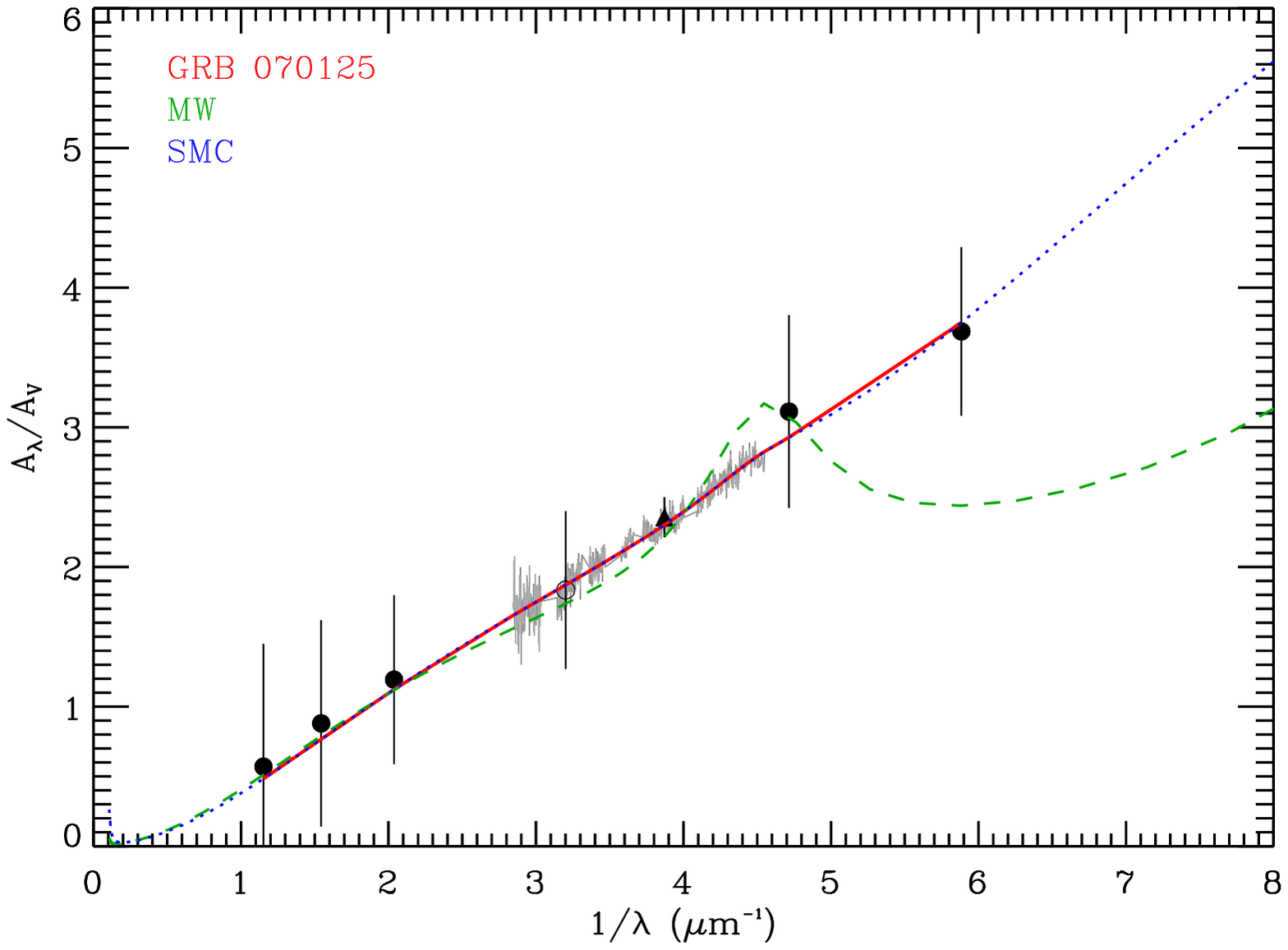}}  \\
             {\includegraphics[width=0.9\columnwidth,clip=]{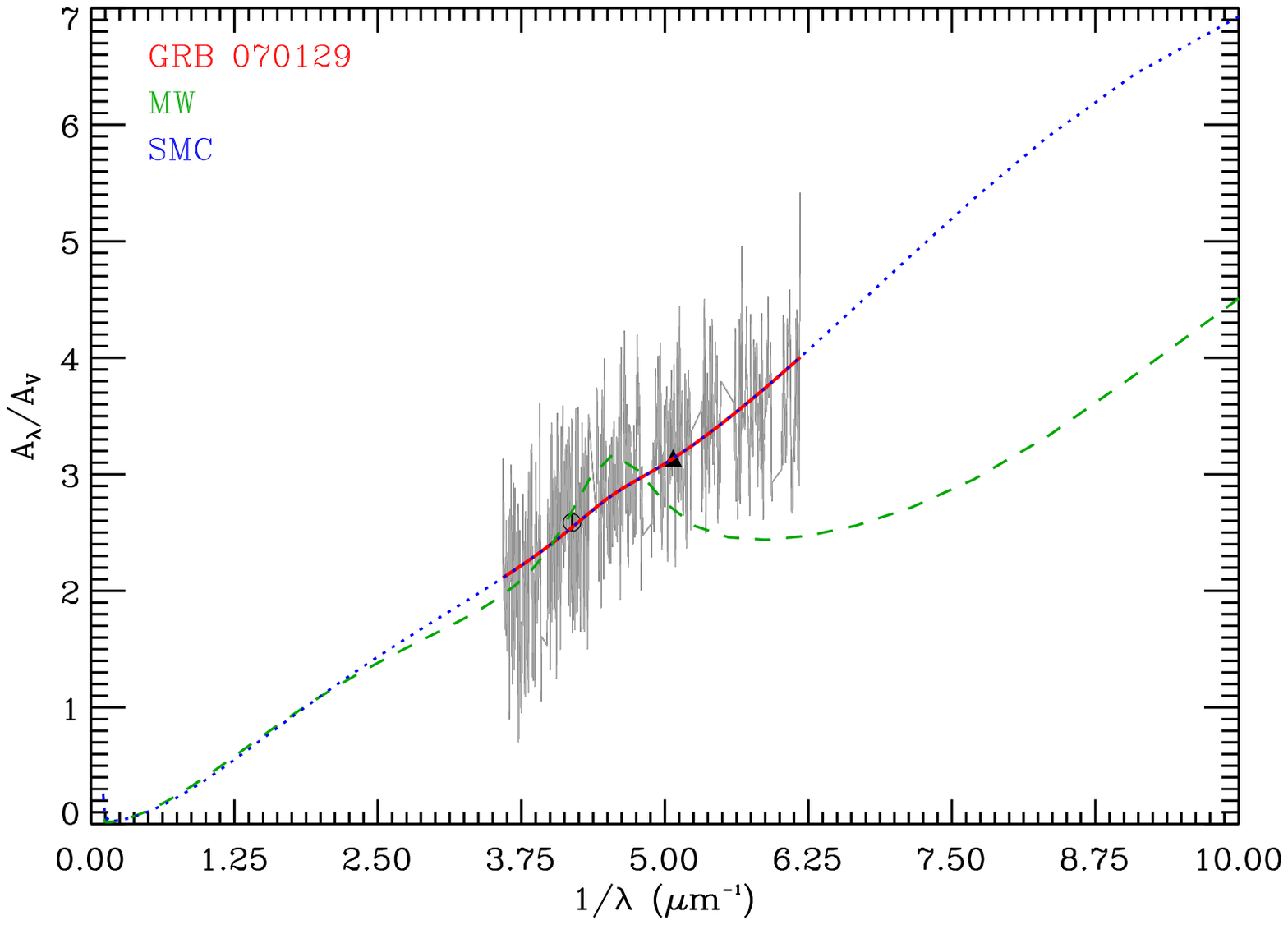}}  &
      {\includegraphics[width=0.9\columnwidth,clip=]{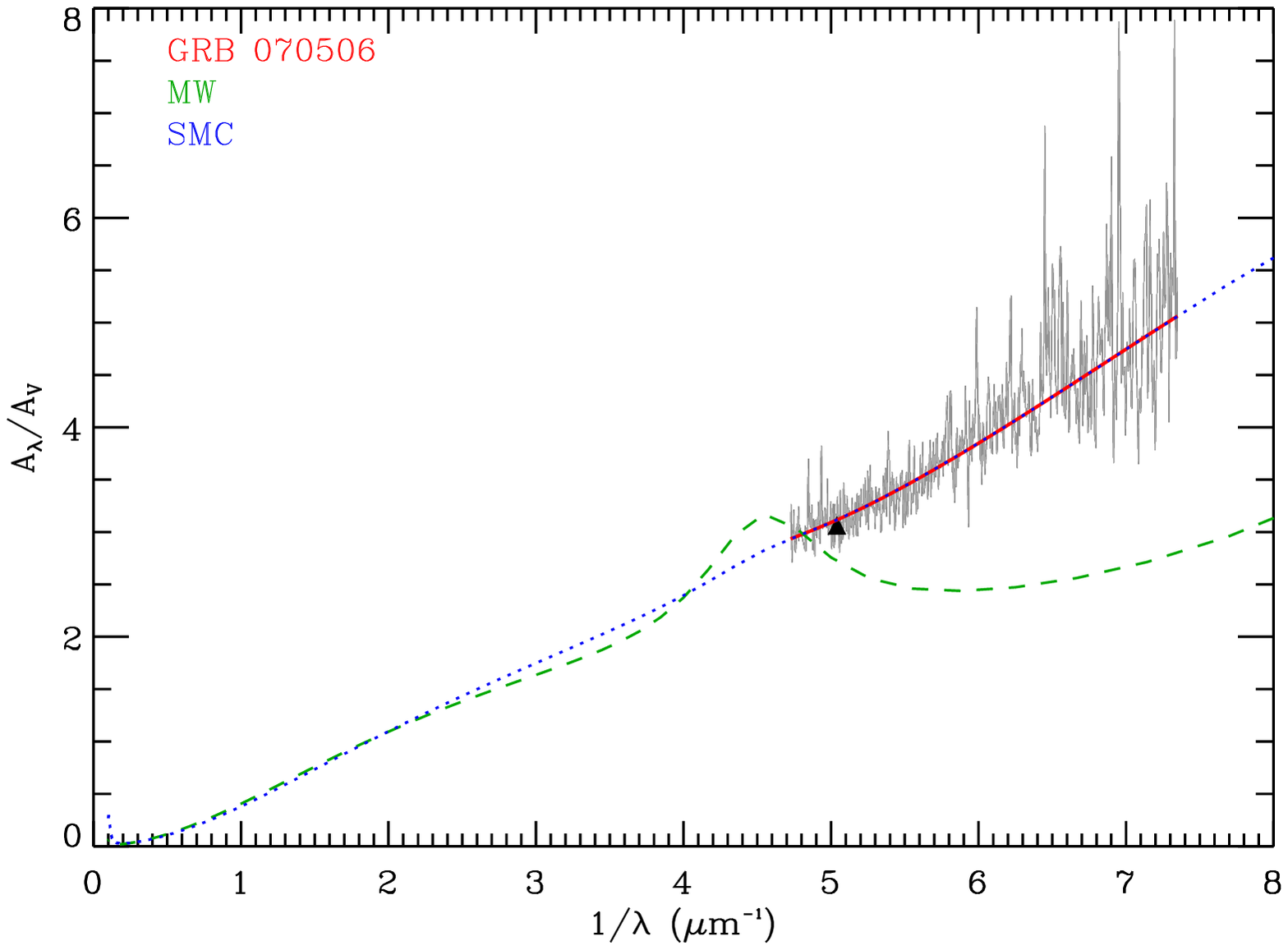}} \\
\end{tabular}
\caption{(continued)}
   \end{figure*}
   \clearpage
   
\addtocounter{figure}{-1}
  \begin{figure*}
  \begin{tabular}{c c}
   {\includegraphics[width=0.9\columnwidth,clip=]{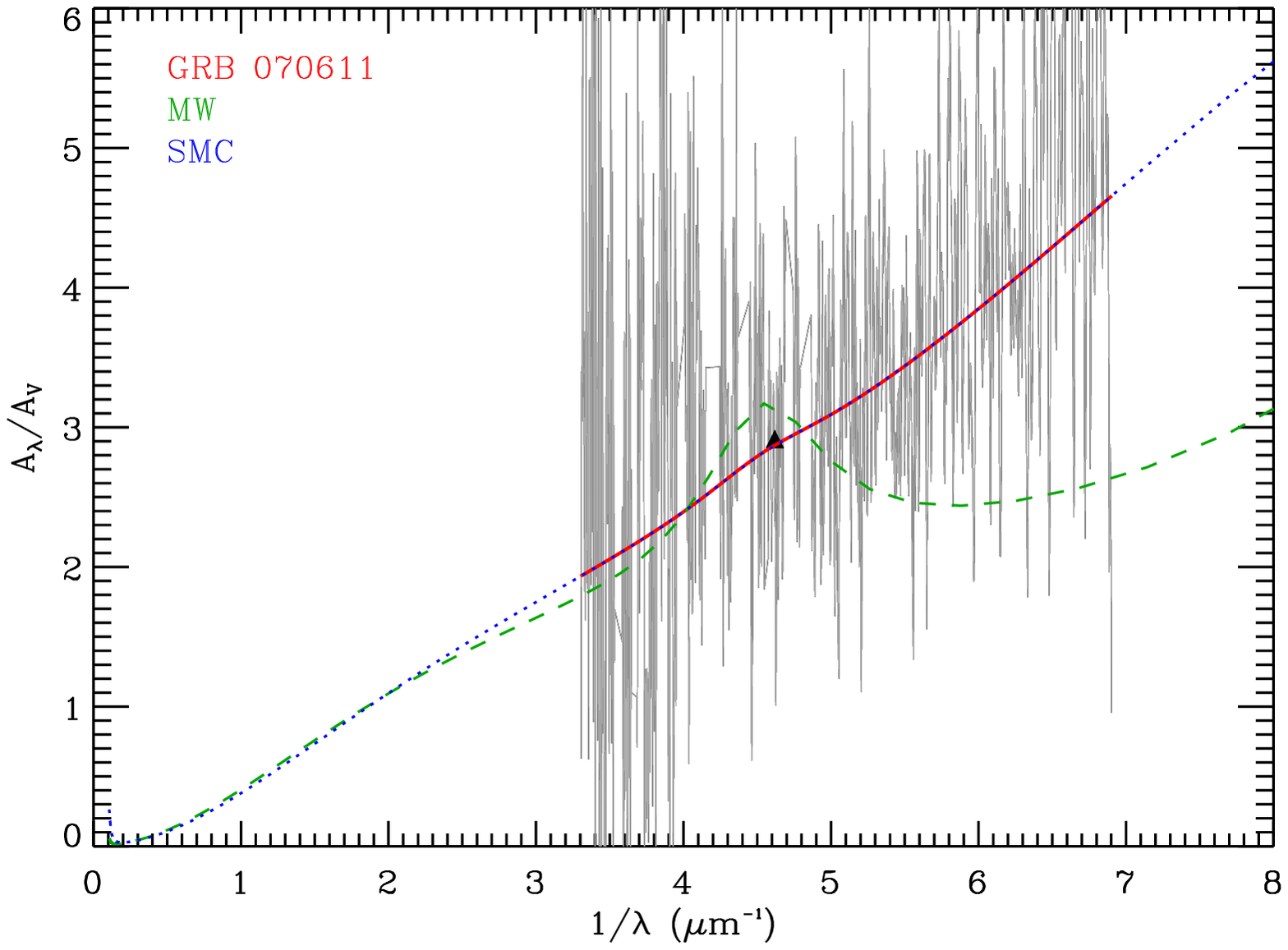}}  &
   {\includegraphics[width=0.9\columnwidth,clip=]{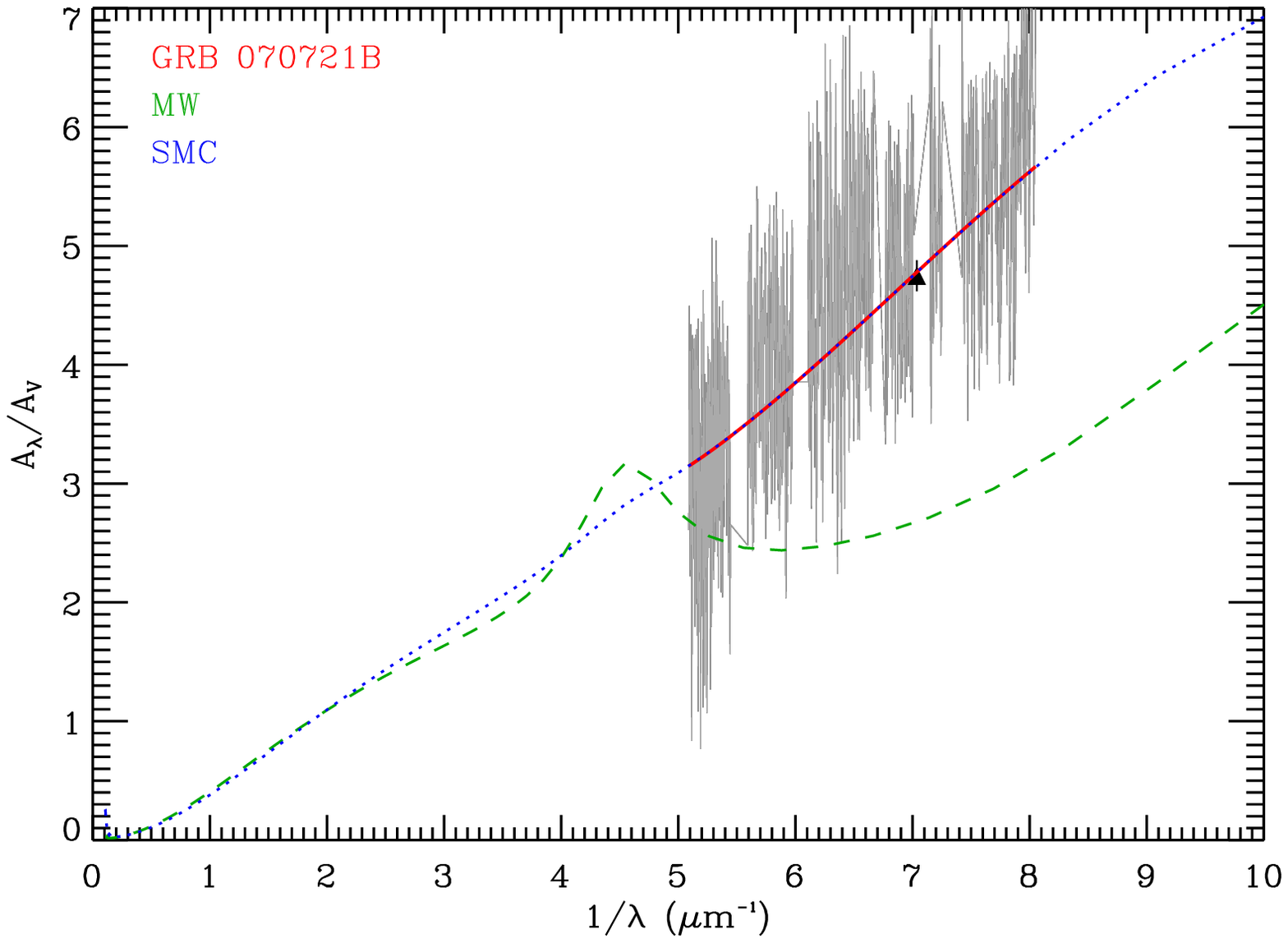}}  \\
   {\includegraphics[width=0.9\columnwidth,clip=]{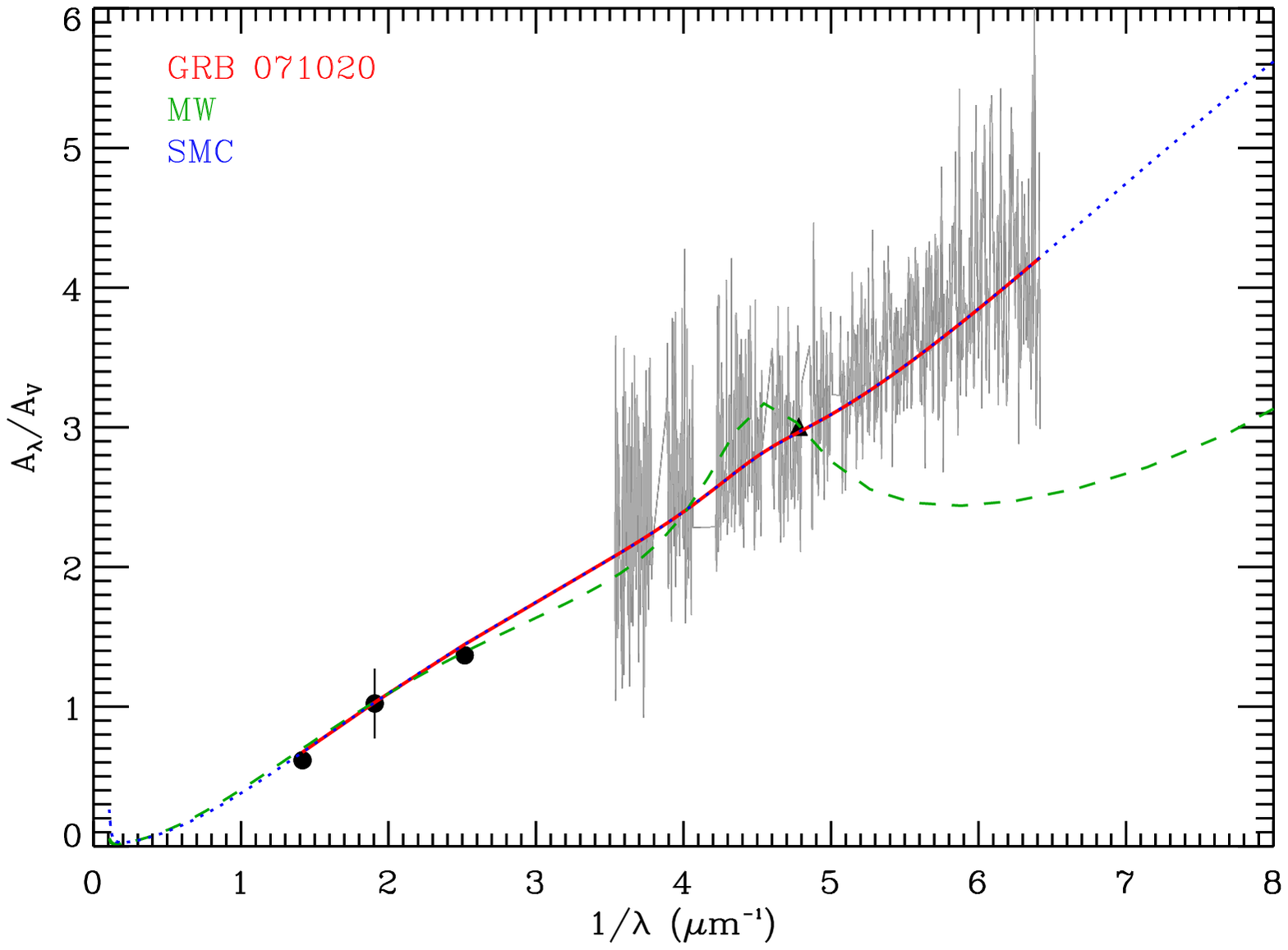}} &
   {\includegraphics[width=0.9\columnwidth,clip=]{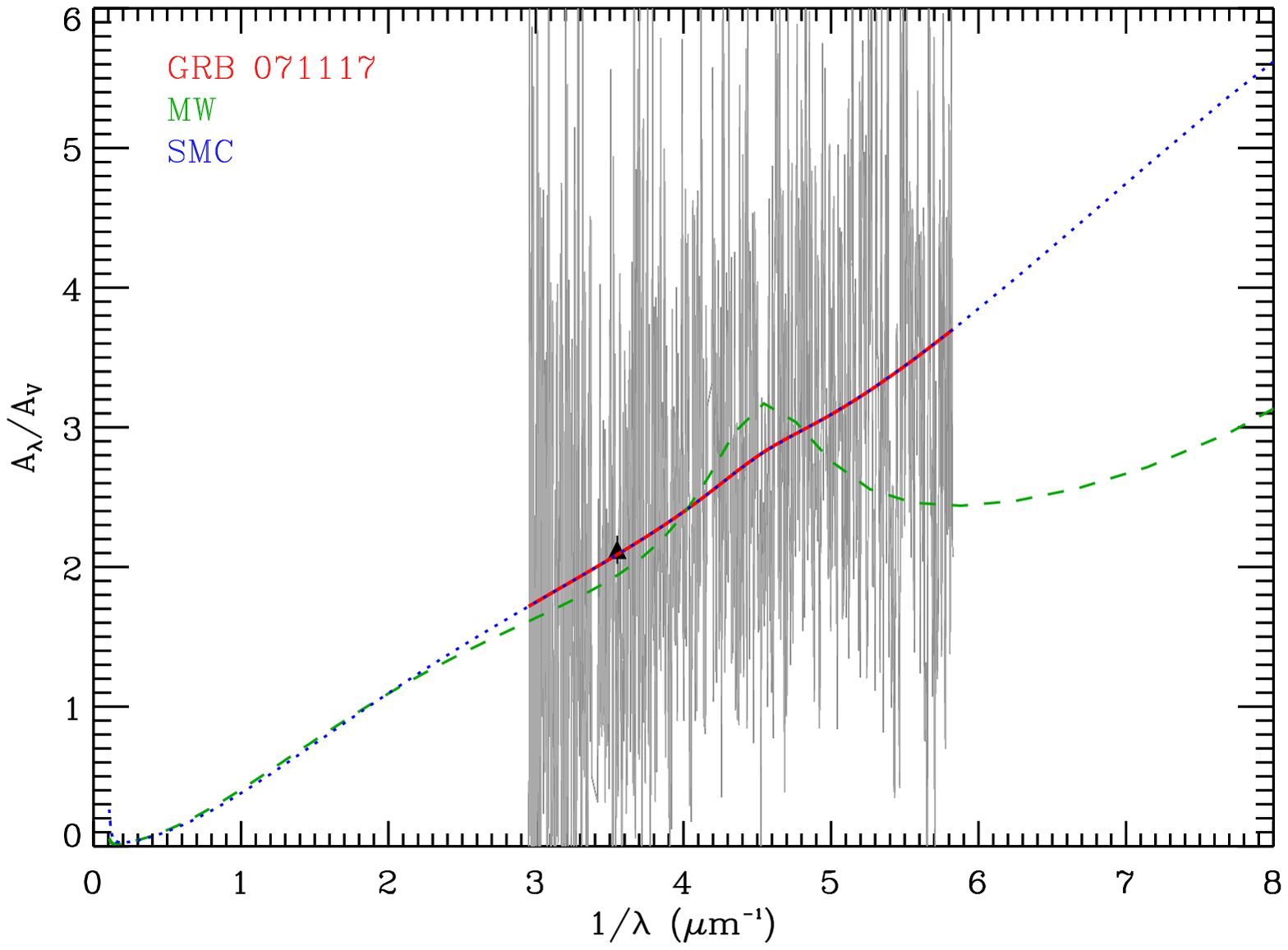}}  \\
   {\includegraphics[width=0.9\columnwidth,clip=]{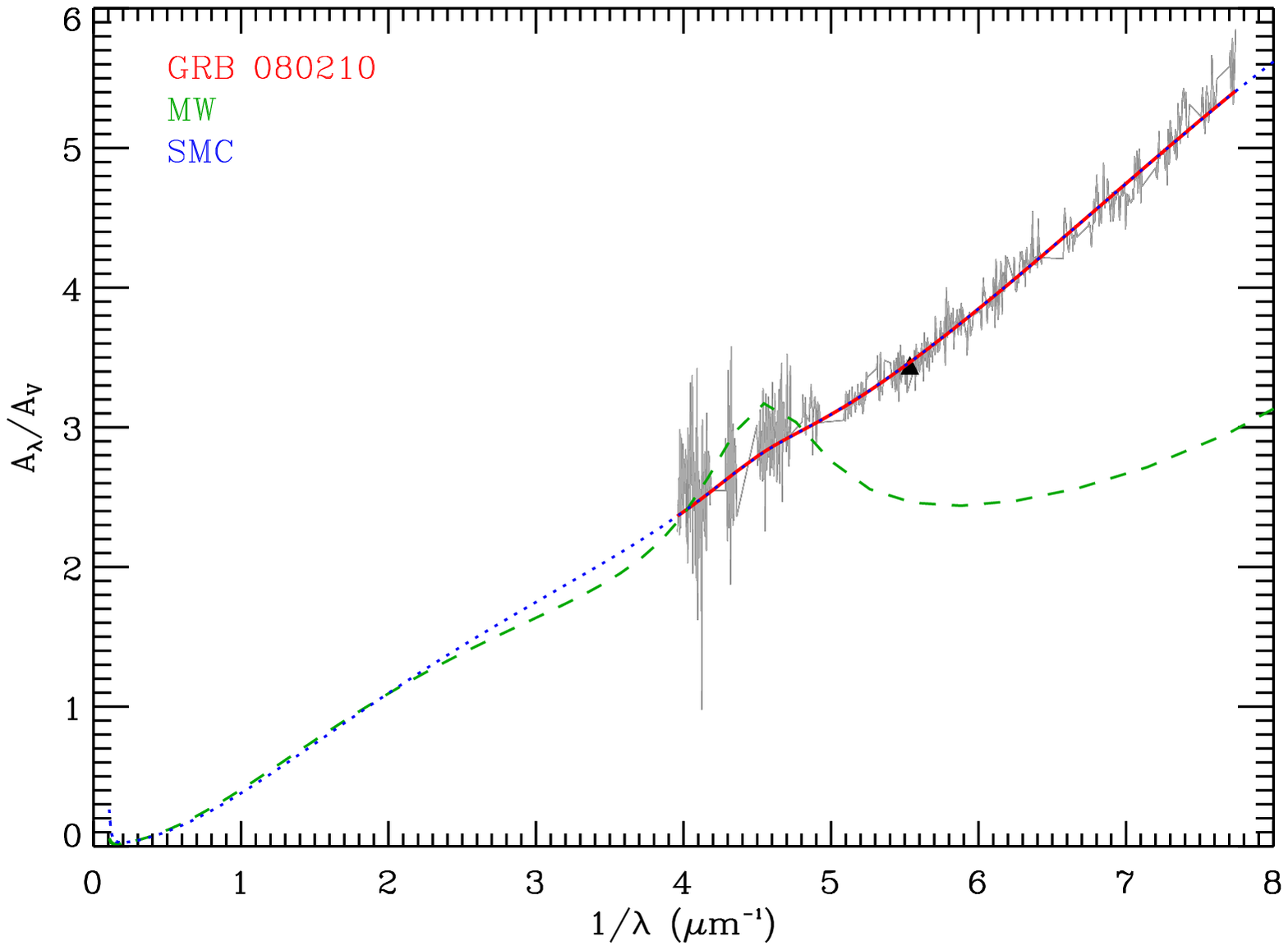}}  &
        {\includegraphics[width=0.9\columnwidth,clip=]{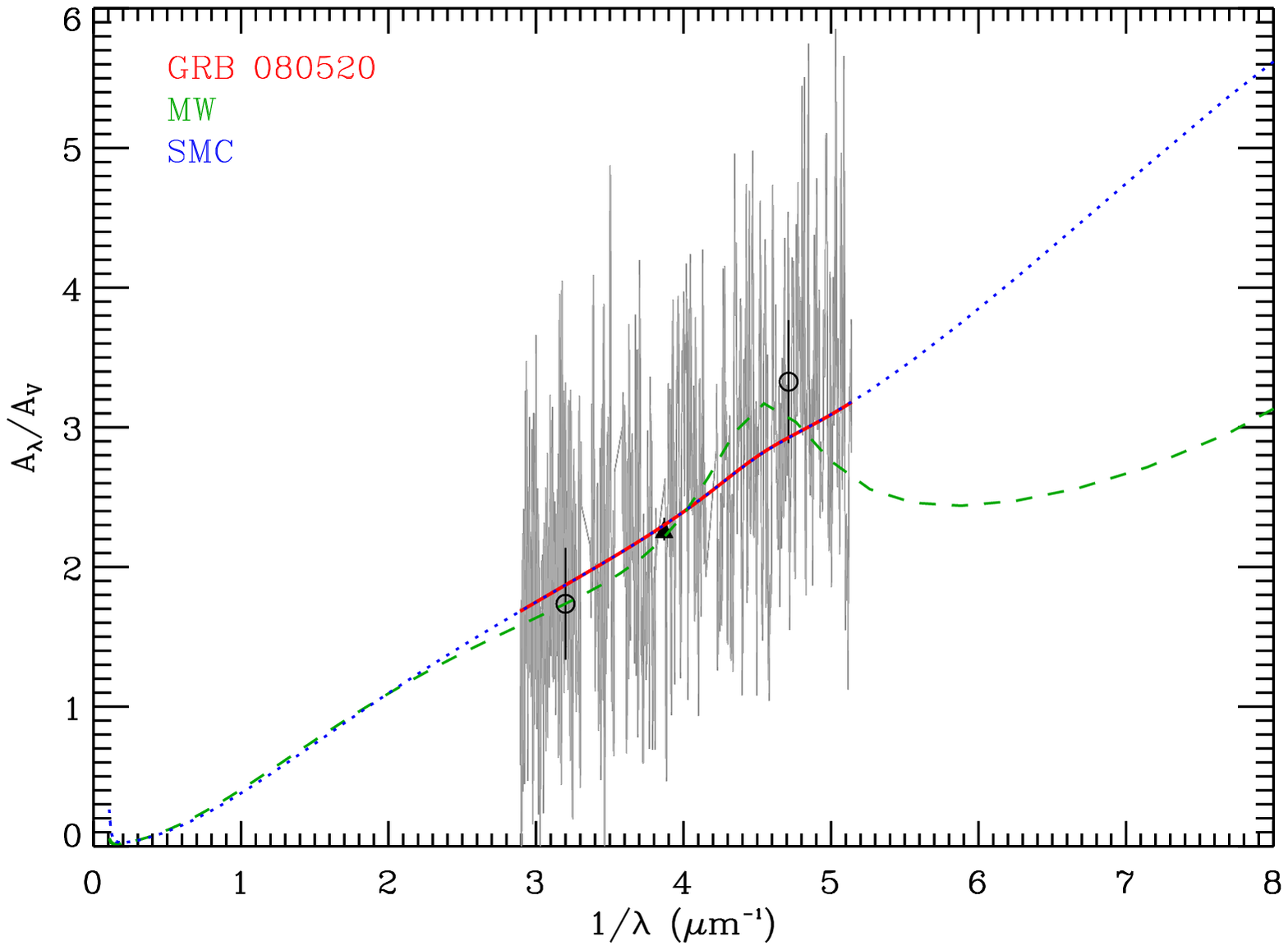}}  \\
                {\includegraphics[width=0.9\columnwidth,clip=]{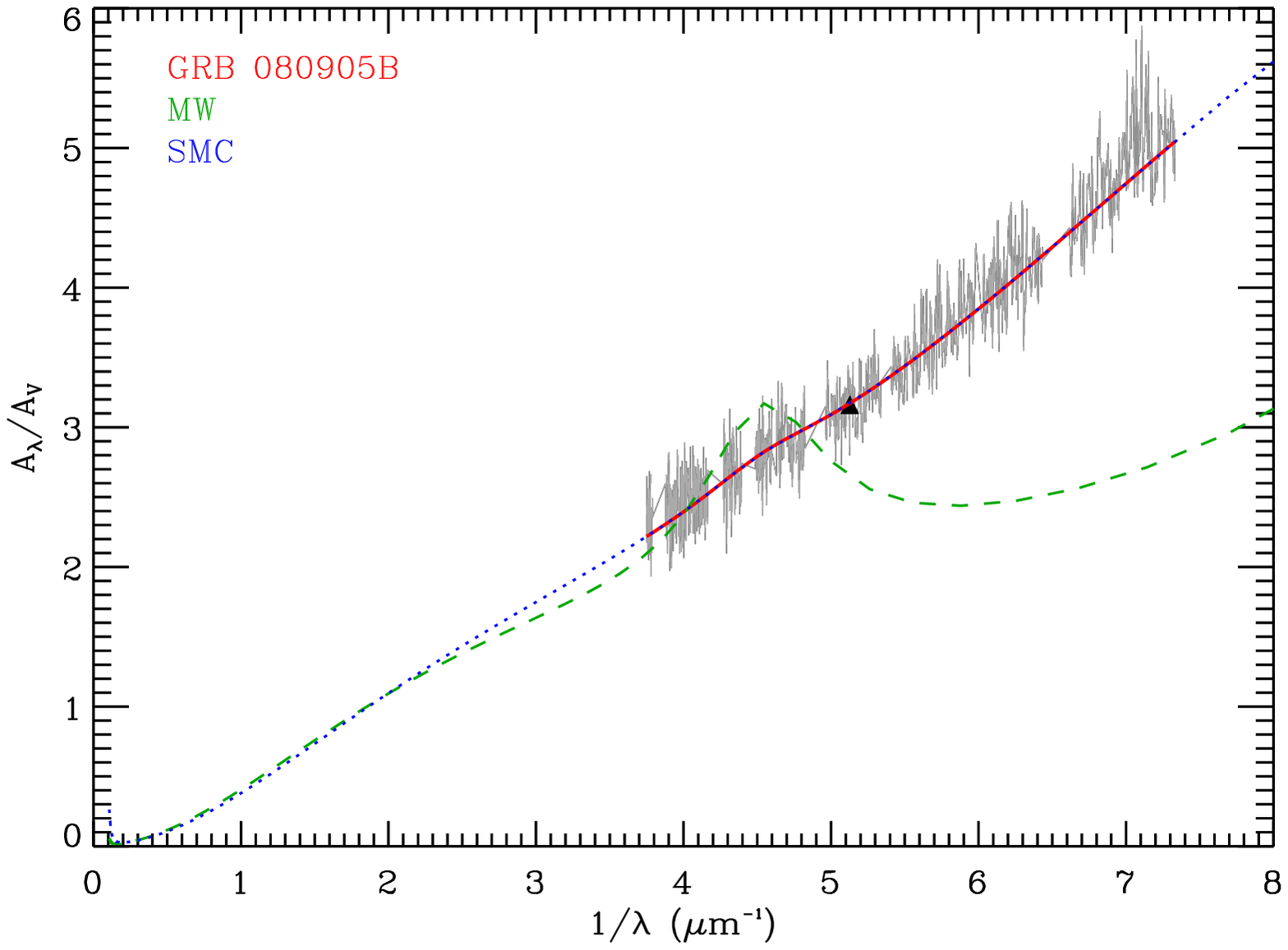}}  &
            {\includegraphics[width=0.9\columnwidth,clip=]{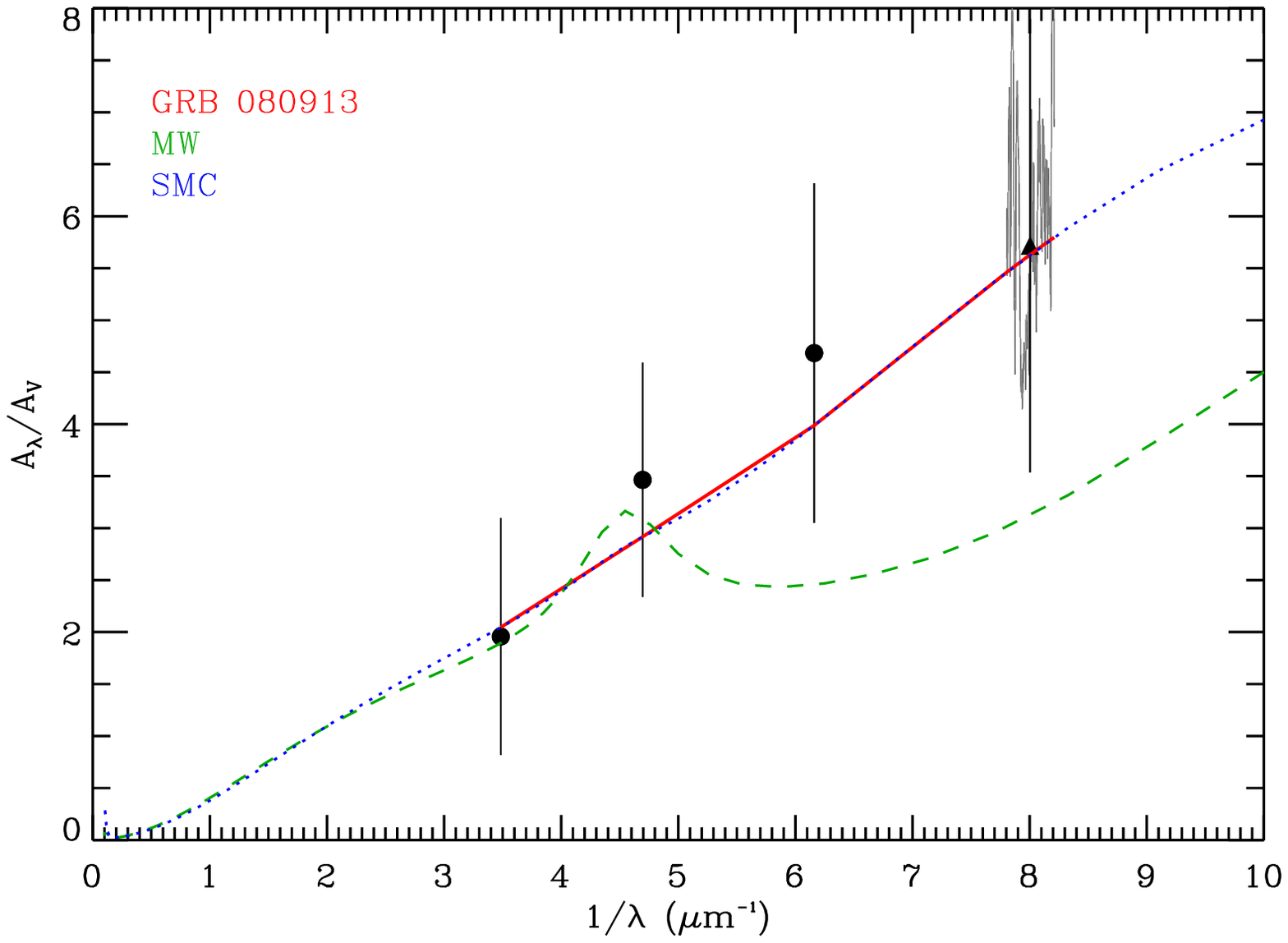}} \\
            \end{tabular}
\caption{(continued)}
   \end{figure*}
   \clearpage
   
\addtocounter{figure}{-1}
  \begin{figure*}
  \begin{tabular}{c c}
          {\includegraphics[width=0.9\columnwidth,clip=]{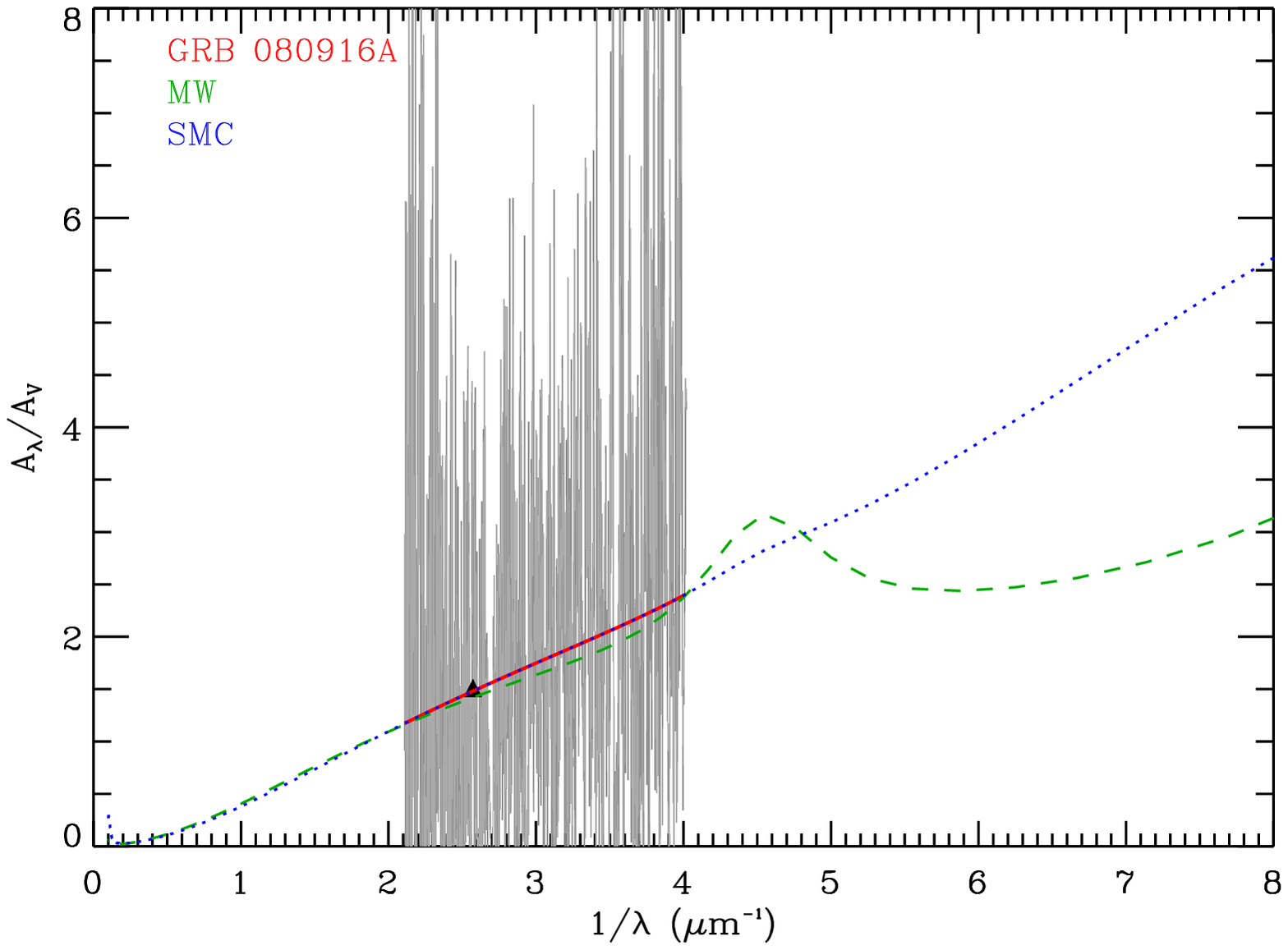}} &
     {\includegraphics[width=0.9\columnwidth,clip=]{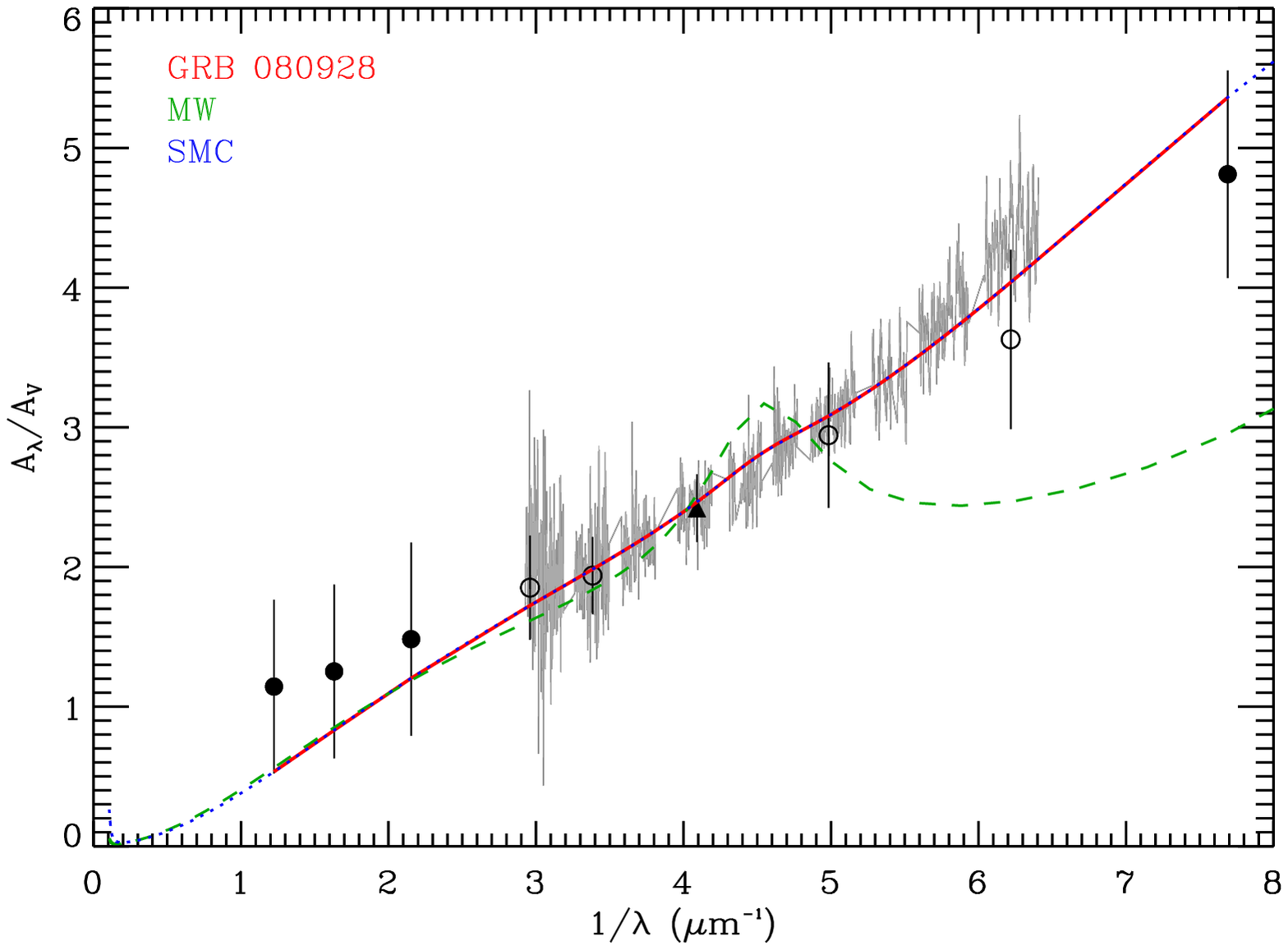}}  \\
    {\includegraphics[width=0.9\columnwidth,clip=]{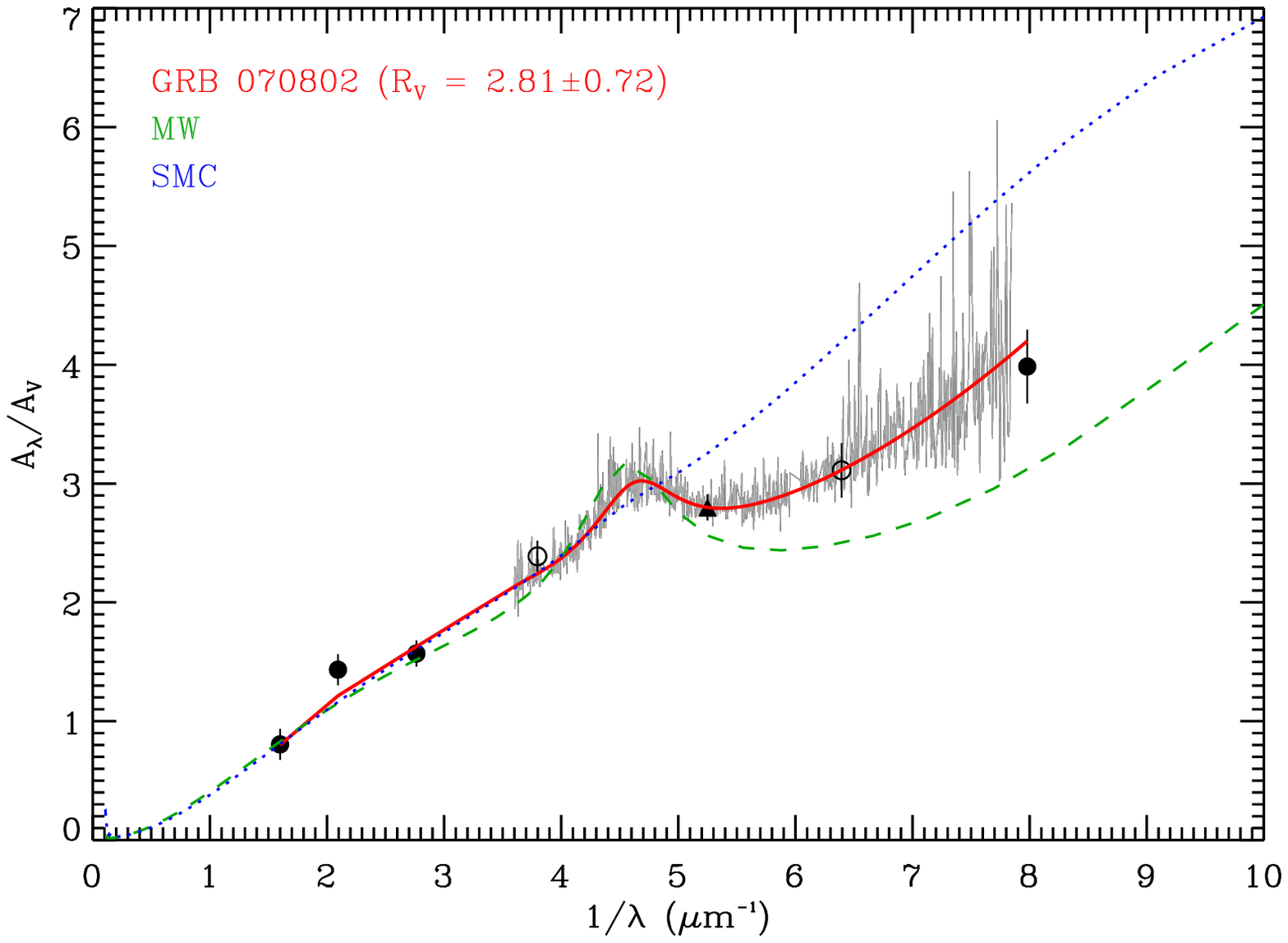}}  &
   {\includegraphics[width=0.9\columnwidth,clip=]{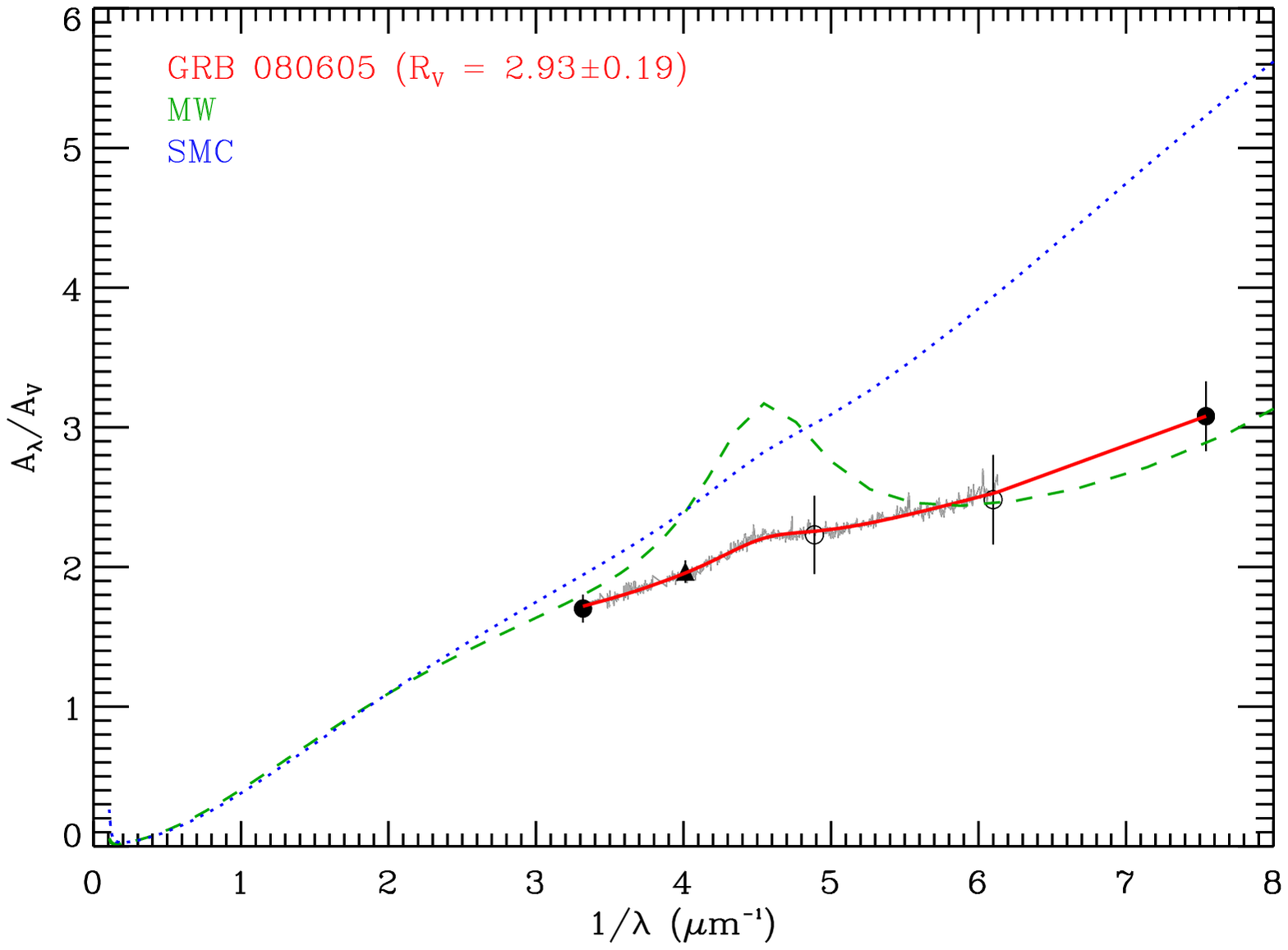}}  \\
   {\includegraphics[width=0.9\columnwidth,clip=]{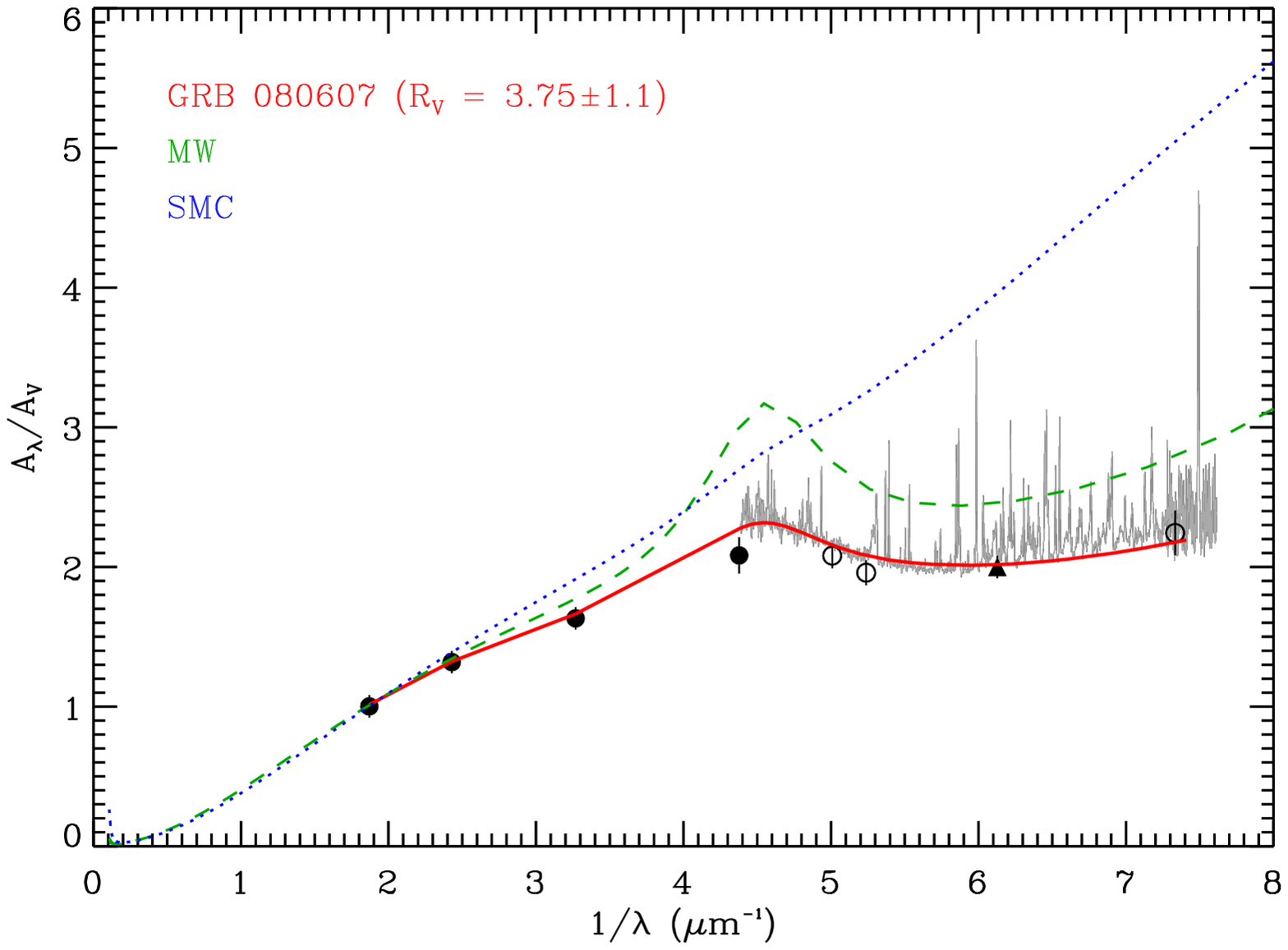}} &
   {\includegraphics[width=0.9\columnwidth,clip=]{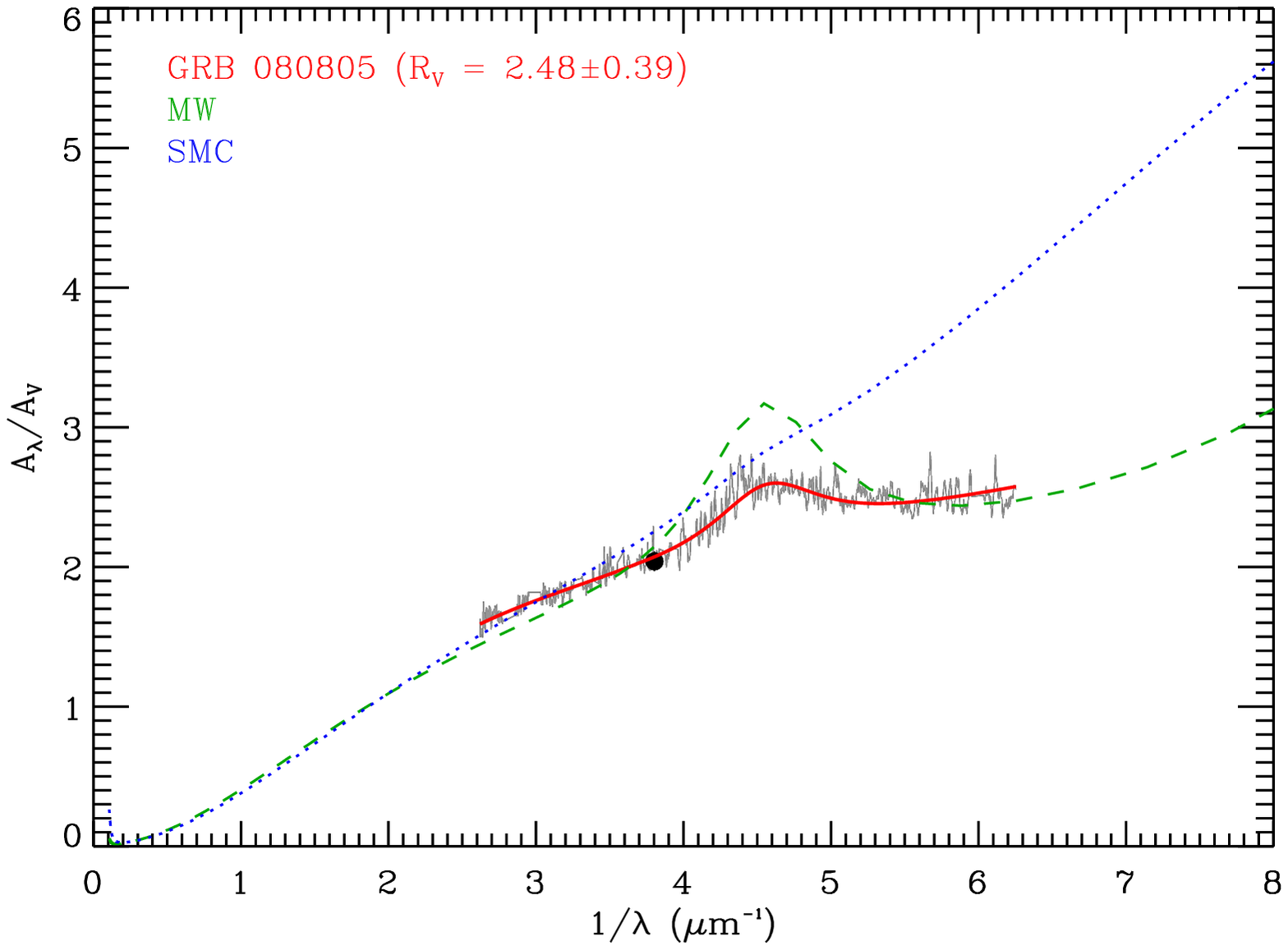}}  \\
\end{tabular}
\caption{(continued)}
   \end{figure*}
\end{appendix}

\end{document}